\documentclass[a4paper, 12pt]{article} 

\usepackage{fullpage}
\usepackage[utf8]{inputenc}
\usepackage[T1]{fontenc}
\usepackage[UKenglish]{babel}
\usepackage{lmodern}

\usepackage{nicefrac}
\usepackage{mathtools}
\usepackage{amsfonts, dsfont}
\usepackage{amsthm, stmaryrd, mathabx}
\usepackage{verbatim, hyperref}
\usepackage{ulem}
\usepackage[dvipsnames]{xcolor}
\usepackage{subcaption}
\usepackage{soul}
\usepackage{natbib}
\usepackage{makecell}
\usepackage[export]{adjustbox}

\usepackage{graphics}
\usepackage{graphicx}
\usepackage{svg}
\usepackage[font=footnotesize]{caption}
\graphicspath{ {./pics5/} }
\usepackage{wrapfig, lipsum}

\usepackage{floatrow}
\usepackage{subcaption}

\usepackage[plain]{algorithm}
\usepackage{algorithmic}

\theoremstyle{remark}

\renewcommand{\tilde}{\widetilde}
\renewcommand{\L}{\mathcal{L}}		%likelihood
\newcommand{\dif}{\mathrm{d}}
\newcommand{\Cinit}{C_{\text{init}}}				
\newcommand{\Cpers}{C_{\text{pers}}}	
\newcommand{\obs}{\text{obs}}	
\newcommand{\U}  {{\mathcal U}}

\begin{document}
\title{Mechanistic-statistical model for the expansion of ash dieback}
\author{Coralie Fritsch$^1$, Marie Grosdidier$^2$, Anne Gégout-Petit$^1$, Benoît Marçais$^3$}

\footnotetext[1]{Université de Lorraine, CNRS, Inria, IECL, F-54000 Nancy, France}
\footnotetext[2]{INRAE, BioSP, 84000 Avignon, France}
\footnotetext[3]{Université de Lorraine, INRAE, UMR1136 Interactions Arbres/Microorganismes, F-54000 Nancy, France}

%%%%%%%%%%%%%%%%%%%%%%%%%%%%%%%%%%%%%%%%%%%%%%%%%%%%%%%%%%%%%%%%%%%%%%
\maketitle
%%%%%%%%%%%%%%%%%%%%%%%%%%%%%%%%%%%%%%%%%%%%%%%%%%%%%%%%%%%%%%%%%%%%%%

\begin{abstract}

\textit{Hymenoscyphus fraxineus} is an invasive forest fungal pathogen that induces severe dieback in European ash populations. The spread of the disease has been closely monitored in France by the forest health survey system. We have developed a mechanistic-statistical model that describes the spread of the disease. It takes into account climate (summer temperature and spring rainfall), pathogen population dynamics (foliar infection, Allee effect induced by limited sexual partner encounters) and host density. We fitted this model using available disease reports. We estimated the parameters of our model, first identifying the appropriate ranges for the parameters, which led to a model reduction, and then using an adaptive multiple importance sampling algorithm for fitting. The model reproduces well the propagation observed in France over the last 20 years. In particular, it predicts the absence of disease impact in the south-east of the country and its weak development in the Garonne valley in south-west France. Summer temperature is the factor with the highest overall effect on disease spread, and explains the limited impact in southern France. Among the different temperature indices tested, the number of summer days with temperatures above 28°C gave the best qualitative behavior and the best fit. In contrast, the Allee effect and the heterogeneity of spring precipitation did not strongly affect the overall expansion of \textit{H.~fraxineus} in France and could be neglected in the modeling process. The model can be used to infer the average annual dispersal of \textit{H.~fraxineus} in France.

\end{abstract}

\tableofcontents

\section{Introduction}
\medskip
Biological invasions significantly impact ecosystems, with both major ecological and economic impact \citep{mollot_effects_2017, pysek_scientists_2020}. In particular, invasive pathogenic microorganisms are important threats for forest ecosystems and the problem has increased over the last decades \citep{santini_biogeographical_2013, desprez-loustau_evolutionary_2016, eschen_updated_2023, ghelardini_drivers_2016}. This has been associated with the globalization trends, and in particular intensified international trade and increased movement of ornamental plants worldwide \citep{santini_biogeographical_2013, sikes_import_2018}. Evaluating the potential distribution of an invasive forest pathogen is necessary to assess the threat it poses to a particular area \citep{paap_invasion_2022}. This represents a challenge for organisms that are still spreading and therefore do not occupy all the areas in which they could develop \citep{bebber_range-expanding_2015}. Consequently, it is difficult to understand the factors that shape the niche for invasive organisms, as only the realized niche, which is constrained by dispersal, is available, while the potential niche remains unknown. A good knowledge of the dispersal of invasive forest pathogens is therefore necessary to properly assess their impact and also to design adequate management strategies \citep{hudgins_predicting_2017}. Dispersal is strongly affected by landscape heterogeneity, both in terms of host availability and climate. Indeed, host density is a key feature in explaining disease emergence and spread \citep{ward_variable_2022, keesing_impacts_2010}. To properly manage invasive pathogens, it is therefore essential to assess how dispersion, host distribution and climate jointly shape their impact \citep{meentemeyer_landscape_2012}.
%\commentcora{C'était pas en Pologne la première fois? Je raconte des bêtises dans mes exposés alors. Benoit : Nouveau développement, cf https://theconversation.com/forets-et-parasites-invasifs-et-si-on-se-trompait-de-suspect-210995}
Ash dieback is a typical example of a severe epidemic induced by an invasive fungal pathogen. The causal agent, \textit{Hymenoscyphus fraxineus} (anamorph \textit{Chalara fraxinea}), was first reported in the Baltic region at the end of the 20th century \citep{kowalski_chalara_2006, agan_traces_2022}. The disease has spread to western and southern Europe over the last 30 years, causing severe dieback in affected ash stands \citep{coker_estimating_2019, marcais_ash_2022}. 

The life cycle of \textit{H. fraxineus} takes place mainly on ash leaves, which are infected by ascospores in late spring and summer \citep{gross_hymenoscyphus_2014}. Leaves remain symptomless for most of the summer \citep{cross_fungal_2017} until late August, when necrotic lesions develop. After leaf fall, the pathogen survives in forest litter on the rachis, which encompasses the petiole and the midrib of the compound leaf \citep{gross_hymenoscyphus_2014}. It produces pseudosclerotial plates on colonized rachises, which are melanized tissues offering protection against desiccation and microbial competitors. \textit{H. fraxineus} is able to survive and produce apothecia on pseudoclerotial rachises for several years \citep{gross_longevity_2013, kirisits_ascocarp_2015, laubray_hymenoscyphus_2024}. The majority of inoculum is produced by fruiting bodies, i.e. apothecia, which develop on pseudosclerotial rachis, although apothecia can occasionally be observed on other tissues infected by \textit{H. fraxineus} such as shoots or the collar of young trees \citep{kowalski_teleomorph_2009, kirisits_ash_2012, baxter_introduction_2023}. Asexual spores, i.e. conidia, are considered to have a limited role in disease dispersal \citep{gross_hymenoscyphus_2014, marcais_ash_2022}. Their main role is that of spermatia, \textit{H. fraxineus} being a heterothallic fungus \citep{gross_reproductive_2012}. The encounter of sexual partners on the forest floor can be a limiting factor at low ash densities or low infection levels, leading to a significant Allee effect \citep{laubray_evidence_2023}. Apothecia formation takes place from spring to midsummer, depending on location in Europe. Dispersal is by airborne ascospores, with an average dispersal distance of around 1 to 2 km \citep{timmermann_ash_2011, grosdidier_tracking_2018}. Apothecia and ascospore production, as well as the infection process, require humid conditions and are favored by rainy weather during the apothecia production period \citep{dvorak_detection_2016, dvorak_vertical_2023, havrdova_environmental_2017, chumanova_predicting_2019}. Ash dieback is caused by colonization of twigs from infected leaves in late summer and autumn \citep{gross_hymenoscyphus_2014}. At this stage of the disease, the limiting environmental condition is temperature. \textit{H. fraxineus} survives poorly at temperatures of 36°C and above \citep{hauptman_temperature_2013}.  Consequently, transfer of the pathogen from leaves to shoots may be limited after hot summers, and ash dieback is of limited severity in the Mediterranean climate of southeastern France \citep{grosdidier_higher_2018}. Nevertheless, \textit{H. fraxineus} can infect and reproduce on ash leaves without inducing significant shoot mortality \citep{marcais_ability_2023}. Infected ash trees then behave as healthy carriers. This occurs under specific micro/meso climatic conditions, notably after summer heat waves, on isolated trees or in hedgerows \citep{grosdidier_landscape_2020}. 

In this article, we aim to develop a spatio-temporal model describing the expansion of \textit{H. fraxineus}, taking into account climate data (in time and position) and ash tree density. By estimating the model parameters, we can measure the impact of local conditions on epidemic expansion. In particular, we try to identify the temperature index that gives the best qualitative and quantitative results among a set of temperature indices related to the intensity of heat during the summer. 

We are developing a mechanistic-statistical model comprising three nested sub-models: (1) a reaction-diffusion model describing the expansion of leaf infection by \textit{H. fraxineus} and the subsequent colonization of leaf debris in the litter (rachis); (2) a stochastic model describing the development of dieback symptoms as a function of leaf spore infection; (3) a stochastic model describing observation data as a function of symptoms and quadrats visited. We use a Bayesian framework to estimate the model parameters. The model and statistical inference methods we use are consistent with previous work by \cite{roques2011a} on pine processionary moth expansion processes, and by \cite{abboud2019dating,abboud2023forecasting} on the invasion of a phytopathogenic bacterium.

\medskip

In Section~\ref{sec.data}, we describe the data used to fit the model while in Section~\ref{sec.model}, we describe the three sub-models of our mechanistic-statistical model. 
Section~\ref{sec.stat.inf} is devoted to the statistical inference method used for parameter estimation. We first proceed to a rough estimation of the parameters leading to a model reduction (more technical details about this step are given in Appendix). Then we use an adaptive multiple importance sampling algorithm to estimate the posterior distribution of the parameters of the reduced model.  This method is repeated for different temperature indices. Section~\ref{sec.results} presents the numerical results (parameters estimation and model dynamics) only for the best temperature index (the comparison and discussion of the different temperature indices is postponed in Appendix). The model validation is also discussed in this section. Finally, we discuss in Section~\ref{sec.discussion} the biological information provided by the parameter estimation, the temperature index selection as well as the dynamics of this model.

\section{Data description and assumptions}
\label{sec.data}

\subsection{Disease prevalence}
%\commentcora{c'est quoi les emerald ashes?}
%\commentbenoit{ emerald ash borer = un insecte qui attaque les frêne en Amérique du Nord et est présent à l'est de l'Europe (Russie / Belarus). Du coup, c'est très surveillé par le DSF qui s'assure qu'il nest pas arrivé chez nous. J'ai rajouté le nom latin }
Since the first report of ash dieback in France in 2008 \citep{ioos_rapid_2009}, the disease progress has been tightly monitored by the forest health survey system (Département de la Santé des Forêts, DSF). The DSF implements a database which stores records of health problems observed in France by a network of foresters trained for the diagnosis of damages induced by abiotic, entomological or microbial causes.  Altogether, from 2008 to 2023, about 6000 health problems were recorded on ash, with numbers ranging from 100 to 750 per year. Each record include a date, a geographic location, a host, a list of observed problems and an associated prevalence as well as a brief description of the affected ecosystem. A specific effort was done by the DSF to monitor ash dieback (so called observation strategy), with, in particular, reports of  ash dieback absence. We collapsed reports made the same year on the same location and removed reports made in 2021-23 on emerald ash borer (\textit{Agrilus planipennis}) absence as it wasn't possible to conclude for them on \textit{H. fraxineus} presence / absence. Altogether we have 2840 reports of ash dieback presence, 2428 reports of ash dieback absence and 743 reports of a problem different without any reference to ash dieback. We assumed that ash dieback was absent  in these last reports, as is not reported and the reports are usually far from any previous report of ash dieback (median distance of 50 km). The symptoms induced  by \textit{H. fraxineus} are no very specific and a reliable diagnostic needs confirmation in the laboratory. Thus, new reports of ash dieback in area previously disease-free were usually confirmed by a qPCR test in laboratory, using the protocol of \cite{ioos_rapid_2009} (69\% of the case in a previously disease-free $16\times 16$ km$^2$ quadrat). Figure~\ref{fig.observations} show the infection data from 2008 to 2023. Data are grouped by $16\times 16$ km$^2$ quadrat.

\subsection{Ash density and meteorological data}
The model's dynamics depends on ashes density, rainfall during the spring and heat during the summer. Even though in the long term ash dieback leads to tree mortality, we neglected it in our model, assuming that the host density is constant in time and uniform within each quadrat. This is an acceptable assumption as data shows that the volume of ash did not yet significantly decrease in the parts of France with the oldest presence of ash dieback \citep{gomez-gallago_semi-mechanistic_2022}. Figure~\ref{fig.ash.density} shows the basal area of ashes in France on each quadrat. The data is computed from the database of the french forest inventory from 2006 to 2018 (https://inventaire-forestier.ign.fr/dataifn/). 
Informations on meteorological data collection are also given in Appendix \ref{append:data}. 
The main ascospore production period in north-east France is from the 15 of June to the 15 of July \citep{grosdidier_tracking_2018}. However, in the Pyrénées, apothecia may be observed as early as beginning of May \citep{marcais_ability_2023}.
For our estimation, precipitation during the period of apothecia production was computed as June rainfall that is represented in Figure~\ref{fig.rainfall}. 
Concerning the temperature impact on the colonization by \textit{H. fraxineus}, we tested several temperature indices. 
Figure~\ref{fig.NjS28} shows the evolution of the index which gives the best qualitative and quantitative results: the number of days during July and August with the maximal temperature above 28$^\circ$C.  
 Both meteorological and ash density data are grouped by the same $16\times 16$ km$^2$ quadrat grid used for ash dieback reports by the DSF.

\section{Model description}
\label{sec.model}

The full model is composed of three nested sub-models. The first one describes the dynamics of leaf infection (production and diffusion of spores, leaf infection). The second model represents the development of crown symptoms on ashes, i.e. shoot mortality and dieback, depending on leaf infection given by the first model. Finally, the third model describes the statistical model for data.

\subsection{Reaction-diffusion model for the leaf infection}
\label{reaction_diffusion_model}
\begin{figure}[H]
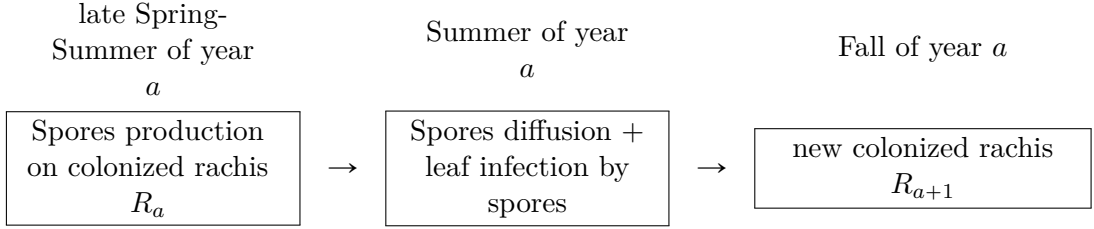

\begin{align*}
\begin{array}{ccccc}
\textrm{\begin{minipage}{2.7cm}\center \small late Spring- Summer of year $a$\end{minipage}}
&&
\textrm{\begin{minipage}{2.7cm}\center \small Summer of year $a$\end{minipage}}
&&
\textrm{\begin{minipage}{2.45cm}\center \small Fall of year $a$\end{minipage}}
\medskip
\\
\textrm{\fbox{\begin{minipage}{3.5cm}\center \small Spores production  \\ on colonized rachis $R_a$\end{minipage} }} 
&
\rightarrow
&
\textrm{\fbox{\begin{minipage}{3.5cm}\center \small Spores diffusion +\\ leaf infection by spores\end{minipage}}} 
  & 
  \rightarrow
  & \textrm{\fbox{\begin{minipage}{4.2cm} \center \small new colonized rachis $R_{a+1}$\end{minipage}}}
\end{array}
\end{align*}
\caption{\label{fig.cycle.chalarose} Reproduction cycle of \textit{Hymenoscyphus fraxineus}.}
\end{figure}

We detail here the reaction-diffusion model describing the dynamics of leaf infection by \textit{H. fraxineus}. The cycle of rachis colonization follows three steps, represented in Figure~\ref{fig.cycle.chalarose}: production of spores on colonized rachises on the forest floor during spring, spores diffusion and infection of leaves during the summer,  and colonization of the newly produced rachises by \textit{H. fraxineus} during fall. 

We denote by $R_a(x)$ the quantity of rachises colonized by \textit{H. fraxineus} at the spring of the year $a$ in location~$x$.

%----------------------
%----------------------

\medskip
\noindent{\textbf{Late spring and summer of the year $a$: spores production}}\\
The pathogen produces apothecia on colonized rachises, which then produces spores (asco\-spores).
We assume that the production of apothecia depends on humidity, and is subject to an Allee effect under a quantity of colonized rachises $r$ as suggested by \cite{laubray_evidence_2023}. The Allee effect is induced by a limited encounter of sexual partners at low rachis density on the forest floor. According to \cite{laubray_evidence_2023}, we propose to encode it in the model by reducing the ascospores production rate when the quantity of colonized rachises $R_a(x)$ is below a threshold $r$. Here, as it is usual to model Allee effect, the ascospores production is quadratic in $R_a(x)$ under $r$. % ; in this it is multiplied by $R_a(x)$ / $r$. 
The spores quantity $\nu_a(x)$ produced in location $x$ during late spring and summer of the year $a$ is then given by
\begin{align}
\label{eq.nu_a}
\nu_a(x) =
\begin{cases}
	 H_a(x)\,R_a(x) & \text{ if } R_a(x)\geq r 
\\[1em]
	H_a(x)\,\frac{R_a(x)^2}{r} & \text{ if } R_a(x)< r 
\end{cases}
\end{align}
with
\begin{align}
\label{eq.H_a}
H_a(x) = \beta_0+\beta_1\,h_a(x)
\end{align}
which depends linearly on the rainfall $h_a(x)$ in June of year $a$ at location $x$ (see Figure~\ref{fig.rainfall}) to account for the effect of the humidity condition on spore production. We therefore expect a positive parameter $\beta_1$.

\medskip
\noindent{\textbf{Late spring / summer of the year $a$: spores diffusion}}\\
The spores quantity $\nu_a(x)$ produced by rachises of the forest floor throughout late spring / summer diffuses at long range \citep{grosdidier_tracking_2018} and infects the leaves. We assume that \textit{H. fraxineus} release spores uniformly during the $\tau=60$ days diffusion period (June and July). The dynamics of the spores quantity $w_a(t,x)$ located in $x$ at time $t$ is then given by the following reaction-diffusion equation\begin{align}
\label{eq.diffusion}
	\frac{\partial w_a(t,x)}{\partial t} = D\,\Delta w_a(t,x) + \frac{\nu_a(x)}{\tau}
\end{align}
with $w_a(0,x)=0$, $\Delta$ is the Laplacian operator and $D$ is the diffusion coefficient.
At the end of the diffusion period of year $a$, the spores quantity is then $w_a(\tau,.)$.

\medskip
\noindent{\textbf{Fall of year $a$: colonization of newly produced rachises}}\\
In the fall, infected leaves are shed resulting in the production of newly colonized rachises on the forest floor.  The local amount of infected leaves shed depends on both the ashes density $d$ (represented in Figure \ref{fig.ash.density}) and on the level of leaf infection assumed to be linearly dependent on the spores quantity $w_a(\tau,x)$ at the end of the diffusion period. We denote by $\chi_a(x)$ the quantity of colonized rachis produced in the location $x$ at the end of the summer of year $a$. It is assumed that, above a saturation threshold $S$, the increase in spore load does not increase infection because all leaves are already infected. Then, $\chi_a(x)$ can be expressed as
\begin{align}
\label{eq.rachis_infection}
	\chi_{a}(x) = (w_a(\tau,x)\wedge S)\,d(x).
\end{align} 

It is known that both in artificial conditions (laboratory, container in nursery) and forest litter, colonized rachises can produce apothecia during several years \citep{gross_longevity_2013, kirisits_ascocarp_2015, laubray_hymenoscyphus_2024}. We assumed that colonized rachises fully persist during at least two years. The total colonized rachis quantity $R_{a+1}$ on the forest floor thus depends on the quantity produced during two successive falls and is given by
\begin{equation}
\label{eq:R_a_plus_one}
    R_{a+1}(x) = \chi_{a}(x) + \chi_{a-1}(x)\,.
\end{equation}

Note that we do not account for a poor survival of \textit{H.~fraxineus} at temperature above 35$^{\circ}$C in rachises on the forest floor. In shaded forest conditions, the effect of high temperatures at soil level is supposed to be low and negligible compared to the temperature impact on the pathogen survival in the crown.

The initialization of the model has then to be made for two initial years and is made from observation data. We discuss it at Section~\ref{symptoms.model}. 

%\medskip

%Note that a conversion coefficient should occurs in the conversion of spores quantity $w_a$ to new infected rachis $\chi_a$ in \eqref{eq.rachis_infection}. We expressed here the spore quantity in a measure unit such that this coefficient equals one, which is included in parameters $\beta_0$ and $\beta_1$ in~\eqref{eq.H_a}. 

\medskip

%\commentcora{J'ai déplacé ici les conditions aux bords qui était décrites juste après l'équation \eqref{eq.diffusion}. En fait, ça va avec l'éq \eqref{eq.diffusion} (choix de modélisation), mais il y a un choix numérique de simuler ça sur un domaine (d'où le fait de déplacer ça ici). Je pense que les 2 se tiennent, je voulais surtout sortir la partie technique numérique de la description du modèle.}
\noindent{\textbf{Numerical scheme}:} Numerically, we use the same spatial discretization for the model as for the rainfall $h_a$, temperature $T_a$ and ash dieback observation data (see Section~\ref{sec:stat_model}), that is a discretization at the scale of one quadrat of $16\times 16$ km$^2$. The model is computed by a Crank–Nicolson scheme. We consider the Neumann boundary conditions $\nabla w_a(t,.) \cdot \mathbf{n}  = 0$, where $\mathbf{n}$ denotes the normal to the boundary. However, note that, for convenience, we numerically consider a rectangular domain, where the density of ashes is zero outside France (that is, we neglect the countries bordering France). The boundary condition is then on the bounds of this rectangular domain and not  directly on the bounds of France. Neglecting spore coming from the borders may be justified as at the simulation onset, all areas close to Germany and the Flemish part of Belgium were already affected by ash dieback, which is taken into account by the initialization. Others borders of France (Wallonia part of Belgium, SW part of Switzerland, Italy, Spain) were affected by ash dieback after the French neighboring part. This condition is of course met when the border is maritime. At the end of the section, Table~\ref{tab.notation} gives the notations of the principal modeled quantities.

\subsection{Model for the symptoms development}
\label{symptoms.model}

The development of crown dieback is caused by the transfer of the pathogen from infected leaves to the shoots before leaf shedding. The infection then remains latent and becomes perceptible during the winter following the infection. We assume that an ash tree without symptom, in location $x$, develops dieback observable the year $a$, with a probability linearly depending on the pathogen quantity infecting its leaves in the year $a-1$, that is proportional to  $\chi_{a-1}(x)$. Reports made in December of the year $n$ were counted as occurring in the year $n+1$ as most of them concern young stand where symptoms can be seen quicker. 
Making the assumption that pathogen quantities by tree are uniform in each quadrat, this probability is proportional to $\chi^i_{a-1}/d(i)$ where $\chi^i_a=\int_{\omega_i}\chi_{a}(x)\,\dif x$ is the total pathogen quantity in the quadrat $\omega_i$ and $d(i)$ the density of ashes in the quadrat $\omega_i$.
Moreover, this also depends on the temperature during the summer of the year $a-1$. \textit{H. fraxineus} is negatively impacted by high temperatures \citep{hauptman_temperature_2013, grosdidier_higher_2018}. It was shown that the transfer from infected leaves to shoots, and thus the shoot mortality and dieback symptoms are hampered when the summer temperatures are too high \citep{marcais_ability_2023}.  
Finally, an ash tree of the quadrat $\omega_i$ develops observable symptoms the year $a$ with probability
%\commentcora{On a transformé la densité terrière en ``vraie'' densité ($d\to d\times 100/(\pi 0.125^2$) en supposant que le rayon d'un frêne est 125cm.  Comme on n'a plus la double binomiale, le rayon du frêne supposé n'impact plus les résultats :-), mais par contre dans ce cas, il faut faire la transo $(d/\alpha, r/\alpha, \alpha \beta_0, \alpha \beta_1)$ avec $\alpha\approx 2037.18$}
\begin{equation}
\label{eq.q.tilde}
	\tilde q_a^i = \frac{\chi^i_{a-1}f(T^i_{a-1})}{r_S\,d(i)}\wedge 1,
\end{equation}
where $r_S$ is the parameter tuning the proportionality and where $T^i_a$ is a temperature index quantifying heat during the summer of the year $a$. We will consider several temperature indices. For instance, one of them is the number of days in July and August of year $a$ for which the maximal daily temperature is over 28$^{\circ}$C on the quadrat $\omega_i$ (see Section~\ref{sec:comparison_temp} for the description of the different indices, and Figure~\ref{compare.temp} for their corresponding results).
We will consider that $f$ is of the form
\begin{align}
\label{eq.fct_temp}
 	f(T) = \left[1-\frac{T}{\gamma}\right]^\kappa_+ \,,
\end{align} 
where $\gamma$ is the threshold of the temperature variable from which the symptoms development is no more possible and the power $\kappa>0$ describes how is the impact of low (but positive) values of $T$ (namely the impact of few days with temperature getting over 28$^{\circ}$C for the selected model). 

Moreover, if a tree develops symptoms, the symptoms persist from one year to the following one with a probability that we denote by $\Cpers$. Therefore, the probability that a tree of quadrat $\omega_i$ has dieback symptoms during the year $a$ is
\begin{align}
q_a^i 
	%&= \PP(Y_a^i=1) \nonumber \\
	%&=  \PP(Y_a^i=1 \cap Y_{a-1}^i=1 \cap \text{symptoms persistence})\nonumber\\
	%&\quad + \PP(Y_a^i=1 \cap \overline{Y_{a-1}^i=1 \cap \text{symptoms persistence}})\nonumber\\
	& = \Cpers\, q_{a-1}^i+ (1-\Cpers\, q_{a-1}^i)\,\tilde q_a^i \,.
\label{eq.q}
\end{align}
%where $\bar A$ is the complementary event of $A$.

Now, it is natural to consider that conditionally to the pathogen leaves infection rate (or equivalently the colonized rachis quantity), the symptoms development of one tree does not depends on the symptoms development of the other trees.

\begin{figure}
\begin{center}
\includegraphics[width=7cm, trim = 0cm 1.78cm 0cm 0cm, clip=true]{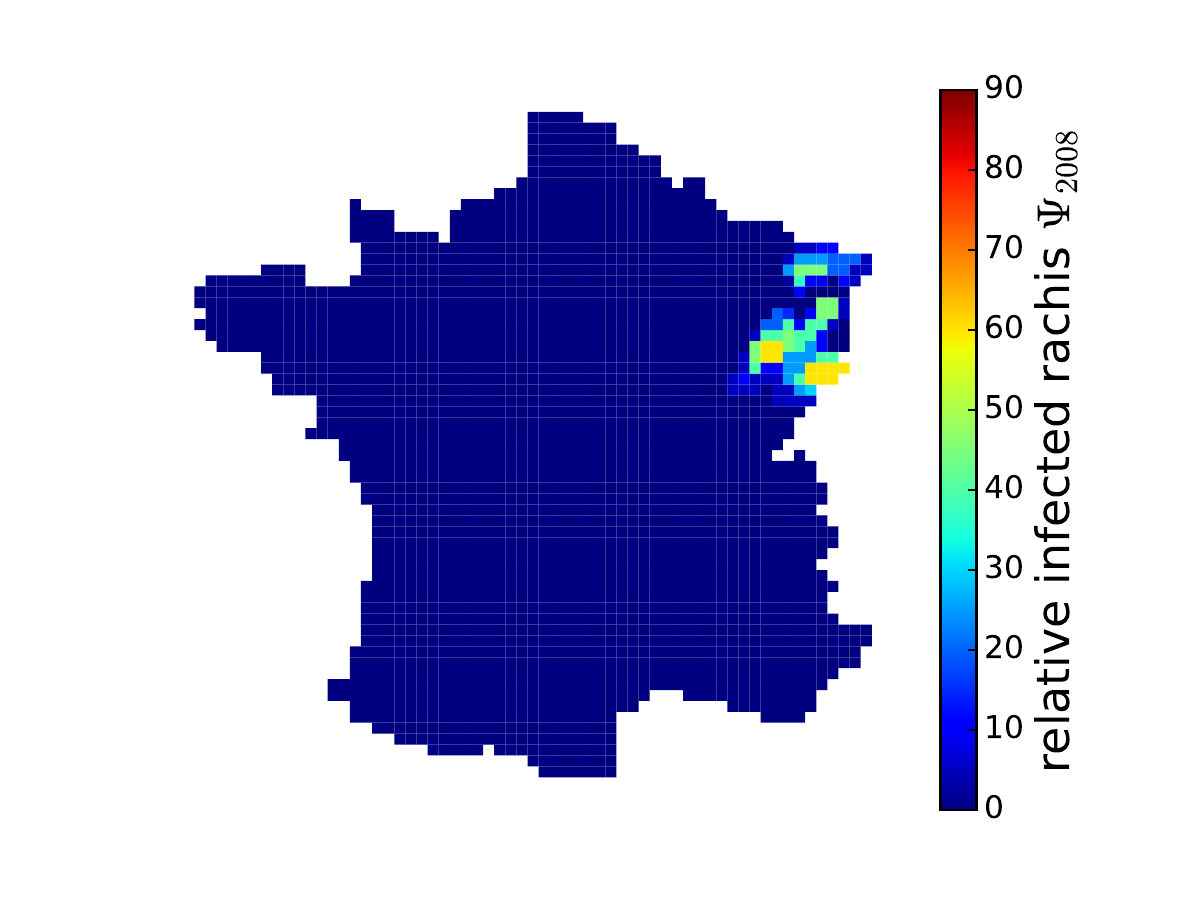}
\includegraphics[width=7cm, trim = 0cm 1.78cm 0cm 0cm, clip=true]{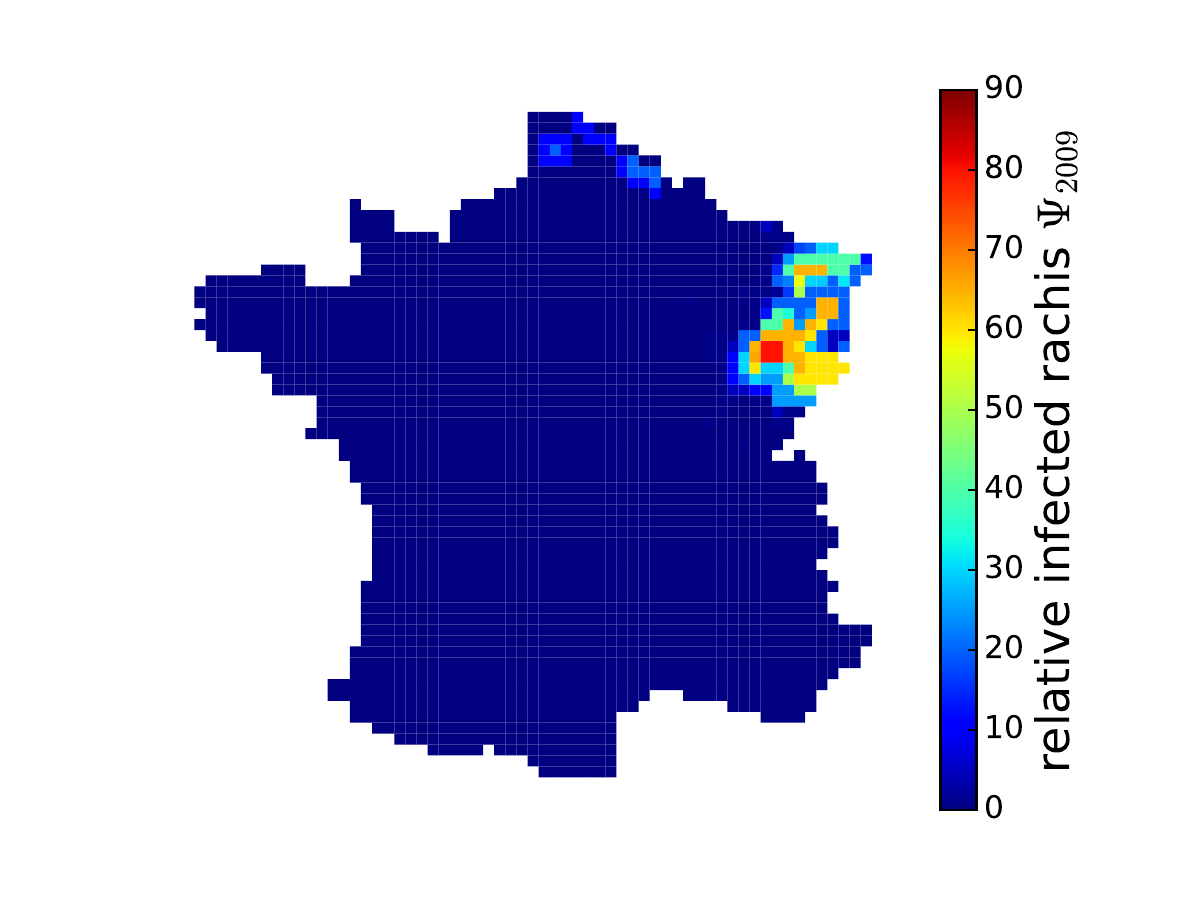}
\end{center}
\caption{\label{fig.initial.infected.rachis} Prevalence intensity of ash dieback symptoms for the years 2008 (left) and 2009 (right) used to compute initial colonize rachis quantities. }
\end{figure}

\medskip
\noindent{\textbf{Initialisation of the model}}\\
The new colonized rachis quantity $\chi_{2007}$ and $\chi_{2008}$ for the two first years had to be imputed from exogenous information. They were derived from the average local prevalence intensity of ash dieback observed by the DSF in each quadrat in 2008 and 2009 and denoted by $\Psi^i_{2008}$ and $\Psi^i_{2009}$, using the INLA package of R to obtain a smoothed map. In 2009, we assumed that ash dieback also appeared in northern France due to pathogen invasion from the flemish region of Belgium and we attributed low prevalence values to quadrats adjacent to that border (5\%). Both quantities $\Psi^i_{2008}$ and $\Psi^i_{2009}$ are given in Figure~\ref{fig.initial.infected.rachis}. We assume that $\chi_{2006}$ can be neglected, meaning that the pathogen was almost absent in 2006 and then that ash diebacks were very limited in 2007 ($q^i_{2007}\approx 0)$. Then we approximate $q_{2008}$ and $q_{2009}$ up to a multiplicative constant $\Cinit$ 
\begin{equation}
\label{eq.initialisation}
    q^i_{2008} \approx \Cinit\,\Psi^i_{2008} \, , \qquad q^i_{2009} \approx \Cinit\,\Psi^i_{2009}\,,
\end{equation}
and then from \eqref{eq.q}, we directly impute the colonized rachis quantities by
\begin{equation*}
\chi^i_{2007}  \approx \Cinit\,\Psi^i_{2008} \times \frac{r_S\,d(i)}{f(T^i_{2007})}, \qquad 
\chi^i_{2008}  \approx \Cinit\,\frac{\Psi^i_{2009} -\Cpers\, \Psi^i_{2008}}{(1-\Cpers\, \Cinit \, \Psi^i_{2008})}\times \frac{r_S\,d(i)}{f(T^i_{2008})}.
\end{equation*}

%\begin{align*}
%\chi^i_{2007} & \approx \Cinit\,\Psi^i_{2008} \times \frac{r_S\,d(i)}{f(T^i_{2007})} \\
%\chi^i_{2008} & \approx \Cinit\,\frac{\Psi^i_{2009} -\Cpers\, \Psi^i_{2008}}{(1-\Cpers\, \Cinit \, \Psi^i_{2008})}\times \frac{r_S\,d(i)}{f(T^i_{2008})}.
%\end{align*}

%$\chi^i_{2007}$ and $\chi^i_{2008}$  by
%\[
%	\chi^i_{a-1} =\frac{ \tilde q_a^i\times r_S\,d(i)}{f(T^i_{a-1})}  = \frac{q_a^i -\Cpers\, q_{a-1}^i}{(1-\Cpers\, q_{a-1}^i)}\times \frac{r_S\,d(i)}{f(T^i_{a-1})}.
%\]

%\begin{align*}
%\chi^i_{2007} & \approx   q_{2008}^i\times \frac{r_S\,d(i)}{f(T^i_{2007})}  \approx \Cinit\,\Psi^i_{2008} \times \frac{r_S\,d(i)}{f(T^i_{2007})} \\
%\chi^i_{2008} & = \frac{q_{2009}^i -\Cpers\, q_{2008}^i}{(1-\Cpers\, q_{2008}^i)}\times \frac{r_S\,d(i)}{f(T^i_{2008})}  
%		\approx \Cinit\,\frac{\Psi^i_{2009} -\Cpers\, \Psi^i_{2008}}{(1-\Cpers\, \Cinit \, \Psi^i_{2008})}\times \frac{r_S\,d(i)}{f(T^i_{2008})} 
%\end{align*}
%and we initialize the quantities of active colonized rachis of the quadrat $\omega_i$ in 2009 to $R^i_{2009} = \chi^i_{2007} + \chi^i_{2008}$.

\subsection{Statistical model for the observation data}
\label{sec:stat_model}

For each $a\in \llbracket 2008, 2023\rrbracket$, we denote by $I_a$ the set of indices of observed quadrats during the year $a$ and for each $i\in I_a$, we set $\obs_a(i)$ the number of observed plots in quadrat $i$ during the year $a$.
The observed data are given by $(p_a^k(i))_{a\in \llbracket 2008, 2023\rrbracket,i\in I_a,k\in \llbracket 1, \obs_a(i)\rrbracket}$ where $p_a^k(i)$ is the proportion of infected trees for the $k$-th observed plot from quadrat $i$ in year $a$.
As noticed in Section~\ref{symptoms.model}, reports made in December are counted as occurring in the following year.
At each plot, around $m=30$ trees are observed so that $m \,\obs_a(i)$ is the number of observed trees in quadrat $i$.

We assume that each tree is observed at most once during all the observation period, such that the different observations are independent. As most stands are observed only once and since trees are not marked in the few stands observed more than one time, this assumption is reasonable.
Therefore, we can derive the number $N_a^i$ of infected trees observed for the quadrat $i$ in the year $a$, through a binomial mechanism from the quantities $q_a^i$ given by the model of symptoms (Eq. \eqref{eq.q}) by
\begin{align}
\label{bin_law_obs}
	N_a^i \sim \text{Bin}(m\,\obs_a(i), q_a^i).
\end{align} 
Moreover, conditionally on the process $(q_a^i)_{a,i}$, it is natural to consider that  variables $(N_a^i)_{a,i}$ are independent. 

Letting  $p_a(i)$ the observed proportion of infected trees for the quadrat $i$ in the year $a$ defined by
\begin{align}
\label{def.p_a_i}
	p_a(i)
	=\frac{1}{\obs_a(i)}\sum_{k=1}^{\obs_a(i)} p_a^k(i),
\end{align}
then the number of infected trees observed in the quadrat $\omega_i$ is $(m\,\obs_a(i)\,p_a(i))_{a,i}$; it is a realisation of $(N_a^i)_{a,i}$.
The data $(p_a(i))_{a,i}$ are represented in Figure~\ref{fig.observations}.

%\subsection{Notations}
\label{sec:notation}

\begin{table}[]
    \centering
\begin{tabular}{|c|c|c|}
\hline
Notation & designation & Equation \\
\hline
$a$ & year & \\
$x$ & location & \\
$i$ & index of quadrat $\omega_i$ & \\
\hline
$R_a(x)$ & quantity of rachises colonized by H. fraxineus at the spring
 & Fig.\ref{fig.cycle.chalarose},\eqref{eq:R_a_plus_one}\\
$\nu_a(x)$ & spores quantity produced in $x$ during late spring and summer & \eqref{eq.nu_a}\\%, \eqref{eq.diffusion} \\
$\omega_a(t,x)$ & spores quantity at time $t$ in June and July & \eqref{eq.diffusion} \\
$\chi_a(x)$ & colonized rachis produced at the end of the
summer & \eqref{eq.rachis_infection} \\
$\tilde q_a^i$ & probability that a tree develops observable
symptoms  & \eqref{eq.q.tilde} \\
$q_a^i$ & probability that a tree has dieback 
symptoms  & \eqref{eq.q} \\
$p_a(i)$ & observed proportion of trees with symptoms  & \eqref{def.p_a_i} \\
\hline
\end{tabular}
    \caption{Notations for the modeled quantities}
    \label{tab.notation}
\end{table}

\section{Statistical Inference }
\label{sec.stat.inf}

\subsection{Parameters}
\label{sec.identifiability}
The model is composed of the following ten parameters: 
\begin{itemize}
\item \textbf{Rachis colonization model:} the diffusion coefficient $D$ (Eq.~\eqref{eq.diffusion}), the spores production parameter $\beta_0$, the humidity impact measurement parameter $\beta_1$ (Eq.~\eqref{eq.H_a}), the Allee effect parameter $r$ (Eq.~\eqref{eq.nu_a}), the colonized rachis saturation quantity $S$ (Eq.~\eqref{eq.rachis_infection}), the multiplicative constant $\Cinit$ for colonized rachis initialization (Eq.~\eqref{eq.initialisation});

\item \textbf{Symptoms development model:} the tuning parameter $r_S$ of the symptoms development Bernoulli parameter (Eq.~\eqref{eq.q.tilde}), the threshold $\gamma$ of  the temperature variable which control the possibility of transfer of the infection from the leaves to the shoots, the parameter $\kappa$ of the temperature function $f$ (Eq.~\eqref{eq.fct_temp}), and the probability of symptoms persistence $\Cpers$ (Eq.~\eqref{eq.q}).
\end{itemize}

\medskip

It is easy to see that, for all $\alpha>0$, the parameters $\theta=(D,\beta_0,\beta_1,r,\gamma,\kappa,S,\Cinit,r_S,\Cpers)$ and $\tilde \theta=(D,\beta_0,\beta_1,\alpha r,\gamma,\kappa, \alpha S,  \Cinit,\alpha r_S,\Cpers)$ give the same model dynamics. In fact, changing the coefficient $\alpha$ just means a change of the measure unit of the colonized rachis. We arbitrarily set the parameter $r_S=1000$ in the following in order to set the rachis measure unit.

\subsection{Likelihood}
\label{sec:likelihood}

We use data from years 2008 to 2019 to estimate our model parameters and we keep data from 2020 to 2023 to measure the model's forecast capabilities.
%For $a=2008, \dots, 2019$, let $I_a$ be the set of indices of quadrats observed the year $a$.
By assumption of independence between observations, conditionally on the process $(q_a^i)_{a,i}$, and by \eqref{bin_law_obs},  the likelihood of a set of parameters $\theta=(D,\beta_0,\beta_1,r,\gamma,\kappa,S,\Cinit,r_S=1000,\Cpers)$ is given by 
\begin{align}
\L(\theta)
	&= 
		\prod_{a=2008}^{2019} \prod_{i\in I_a}
			\dbinom{m\,\obs_a(i)}{m\,\,\obs_a(i)\,p_a(i)} \,
			\left(q_a^i(\theta)\right)^{m\,\,\obs_a(i)\,p_a(i)}\,
			\left(1-q_a^i(\theta)\right)^{m\,\obs_a(i)\,(1-p_a(i))}
\label{eq.logvrai}
\end{align}
where $p_a(i)$ is the function of the observations $(p_a^k(i))_{1\leq k\leq \obs_a(i)}$ defined by \eqref{def.p_a_i} and $(q_a^i(\theta))_{a,i}$ is the process of the dieback symptoms probabilities defined by \eqref{eq.q} for the model described in Sections~\ref{reaction_diffusion_model} and \ref{symptoms.model} with the set of parameters $\theta$. At a first look, the likelihood seems very simple but the processes $(q_a^i(\theta))_{a,i}$ are derived from the reaction diffusion process so that it is impossible to have an explicit formula for the optimal $\theta$. It is why we choose a bayesian approach to optimise the likelihood.

\subsection{Bayesian method for the parameters estimation} 
 %\subsection{Bayesian optimisation of the likelihood} 
 \label{sec:bayesian_method}
 In this section, we describe the process that leads to the parameters estimation. We have chosen to compute the posterior distribution of the parameter $\theta$ by the Adaptive Multiple Importance Sampling (AMIS) algorithm \citep{cornuet2012a} described in Section \ref{sec:AMIS}. Even if the main aim of AMIS is to overcome the difficulty to choose a proposal distribution as close as possible to the posterior distribution, authors strongly require that a significant part of the computing effort be spent on the initialization stage. That is crucial to design an efficient importance sampling algorithm and we have first proceeded to an initial rough estimation of the parameters. The purpose of this stage is twofold: to see if some parameters can be dropped or fixed to a constant and also to have a proposal not so far from the posterior. The results of this procedure are described in the following sections, while the technical details are postponed in Appendix. 
%\subsection{A first naive approach leading to model reduction} 

 \subsubsection{First estimation of the parameters:  model reduction and parameter initialization} 
\label{sec:first_est_model_reduction} 

It is well known that the initialization of the algorithms is one of the main difficulties in the parameters estimation by iterative methods. The process used for the first rough estimation of the parameters as well the numerical results are described in Appendix~\ref{sec:appendix.first.estimation}. 
This approach leads to a model reduction setting four parameters to specific values.
In particular, it stands out that the probability of the symptom persistence $\Cpers$ is equal or very close to 1. This is consistent with field observation. Ash dieback usually result in severe mortality of at least small and medium shoots and those can be observe on the tree during several years with thus persistent symptoms. Moreover, the parameters $\beta_1$ seems to be negligible with respect to $\beta_0$ in \eqref{eq.H_a}, meaning that the rainfall variations have a minor impact on the long range dispersal of \textit{H.~fraxineus}. Furthermore, the Allee effect parameter $r$ is negligeable in the more likely zone of parameters. Finally, on our first estimates, it appears that the posterior distribution of the parameter $\gamma$ is, for the temperature index considered in Section~\ref{sec.results}, almost degenerated in a value close to, but strictly larger than $21.5$ (see Section~\ref{sec:degenerated_dist_gamma}).

After the first numerical tests, we then set the probability of symptoms persistence ($\Cpers$) to one, the Allee effect parameter ($r$) and impact of the rainfall ($\beta_1$) to zero and the threshold $\gamma$ of the temperature index to 21.50001:
\begin{align}
\label{eq:model_reduction}
	\Cpers=1, \qquad \beta_1 = 0, \qquad r=0, \qquad \gamma=21.50001.
\end{align}

In addition, this first step gives a rough estimation of the parameters $D$, $\beta_0$, $\kappa$, $S$ and $\Cinit$ for the initialization of the AMIS algorithm, described in the next section.

\subsubsection{Adaptive Multiple Importance Sampling Algorithm}
\label{sec:AMIS}

Following the model reduction \eqref{eq:model_reduction}, the posterior distribution of the set of parameters $\theta = (D,\beta_0,\kappa,S,\Cinit)$ is computed by the AMIS algorithm described by Algorithm~\ref{algo.AMIS} \citep{cornuet2012a}. The principle of the algorithm is the following: at each step 1) we sample $N$ sets of parameters according to a normal proposal distribution $\mathcal{N}(m_1,\Sigma_1)$; 2) we compute the weight associated to each parameter (included parameters samples of previous steps) with respect to the likelihood function, the prior density and the proposal density; 3) we update the proposal distribution according to the parameters weights.
This algorithm then provides weighted parameters sets describing the posterior distribution of $\theta$.

This algorithm has the advantage to allow the parallelization of the likelihood computations of the sampled parameters at each iteration step and to recycle all simulated parameters (which is particularly important when the likelihood is costly to compute). Moreover,  the efficiency of the AMIS algorithm for banana shape target (as suggested by our first estimation step for the 2-dimensional posterior distribution of $(D,\beta_0)$, see Figure~\ref{fig.grid.D.beta0.likeli}) has been illustrated by \cite{cornuet2012a}.
The main difficulty is to determine the initial proposal distribution (namely the mean $m_1$ and the covariance matrix $\Sigma_1$). 

 %--------------------------
\begin{algorithm}
\begin{center}
\begin{minipage}{10.5cm}
\small
\begin{algorithmic}
\STATE initialization of the mean $m_1$ and the covariance matrix $\Sigma_1$ 
\FOR {$k=1 \cdots N_L$}
  \STATE $\theta_{ki}\sim \mathcal{N}(m_k,\Sigma_k)$, $i=1 \cdots N$
  \COMMENT{new parameters sample}
	\STATE  $\tilde w_{li}\leftarrow \frac{L(\theta_{li}) \pi (\theta_{li})}{\frac{1}{k}\sum_{\ell=1}^k g_{m_\ell,\Sigma_\ell}(\theta_{li})}$ $l=1 \cdots k$, $i=1 \cdots N$
	\COMMENT{weights}
	\STATE  $w_{li}\leftarrow \tilde w_{li} / \sum_{\ell=1}^k\sum_{\iota=1}^N\tilde w_{\ell \iota}$
	\COMMENT{normalized weights}
    \STATE $m_{k+1} \leftarrow \sum_{l=1}^k\sum_{i=1}^N w_{li} \, \theta_{li}$ 
    \COMMENT{proposal distribution update}
    %\STATE $\Sigma_{k+1} \leftarrow  \sum_{l=1}^k\sum_{i=1}^N w_{li} \, (\theta_{li}-m_l)\,(\theta_{li}-m_l)^t$ {\color{red}FAUX, corrigé le 25/10/22}
     \STATE $\Sigma_{k+1} \leftarrow  \sum_{l=1}^k\sum_{i=1}^N w_{li} \, (\theta_{li}-m_{k+1})\,(\theta_{li}-m_{k+1})^t$
    %\STATE $\begin{cases}m_{k+1} \leftarrow \sum_{l=1}^k\sum_{i=1}^N w_{li} \, \theta_{li}\\
    	%			\Sigma_{k+1} \leftarrow  \sum_{l=1}^k\sum_{i=1}^N w_{li} \, (\theta_{li}-m_l)\,(\theta_{li}-m_l)^t\end{cases}$
\ENDFOR
\end{algorithmic}
\end{minipage}
\end{center}
\caption{Adaptive Multiple Importance Sampling Algorithm for parameter estimation of the reduced model.  The function $g_{m,\Sigma}$ is the density of a normal distribution $\mathcal{N}(m,\Sigma)$ with mean $m \in \mathbb{R}^5$ and covariance matrix of size $(5, 5)$.  For $\theta = (D,\beta_0,\kappa,S,\Cinit)$, $L(\theta)=\mathcal{L}(D,\beta_0,\beta_1=0,r=0,\gamma=21.50001,\kappa,S,\Cinit,r_S=1000,\Cpers=1)$ is the likelihood function defined in Section~\ref{sec:likelihood} with the model reduction \eqref{eq:model_reduction}.
\label{algo.AMIS}}
\end{algorithm}
%--------------------------
 
Even though the first estimation of the parameters reveals a strong correlation between some parameters (in particular between $\beta_0$ and $D$, see Figure~\ref{fig.grid.D.beta0.likeli}), our prior information remains imprecise.  We then choose five independent uniform distributions for our prior as the dependency will be corrected by the algorithm:
\begin{align*}
	\pi(\theta)&=
	\U_{[0,\bar D]\times [0,\bar \beta_0] \times [0,\bar \kappa] \times [0,\bar S] \times [0,\widebar \Cinit]}
	(D,\beta_0,\kappa, S, \Cinit)
\end{align*}
with $\bar D$, $\bar \beta_0$, $\bar \kappa$, $\bar S$ and $\widebar \Cinit$ sufficiently large such that they don't impact the posterior distribution.

We initialize the proposal distribution by independent normal distribution (i.e. the covariance matrix $\Sigma_1$ is diagonal) whose the means and the variances are roughly estimated thanks to the first estimation step (see Section~\ref{sec:coarse_grid} for more details). The values of the mean $m_1$ and the covariance matrix $\Sigma_1$ are given in Table~\ref{table.init.prop.amis} for the temperature index used for the numerical results of the next section.

\begin{table}[H]
\begin{center}
\begin{tabular}{lll}
$m_D=18$ & $m_{\beta_0}=24$ & 
$m_{\kappa}=0.05$ \\ $m_S=90$ & $m_{\Cinit}=0.0085$
\\
$\sigma^2_D=0.4$ & $ \sigma^2_{\beta_0}=3$ &
$\sigma^2_{\kappa}=1.5 \times 10^{-5}$ \\ $\sigma^2_S=1$ &  $\sigma^2_{\Cinit}=1\times 10^{-8}$
\end{tabular}
\end{center}
\caption{\label{table.init.prop.amis} Initial values of the mean $m_1=(m_D, m_{\beta_0}, m_{\kappa}, m_S, m_{\Cinit})$ and the covariance matrix  $\Sigma_1 = \text{diag}(\sigma^2_D, \sigma^2_{\beta_0},  \sigma^2_{\kappa}, \sigma^2_S, \sigma^2_{\Cinit})$ of the AMIS Algorithm~\ref{algo.AMIS} for the temperature index T28.} 
\end{table}

The convergence of the algorithm is discussed in Appendix~\ref{sec:converenge_AMIS}.

\section{Numerical Results for the best temperature index (T28)}
\label{sec.results}

The method described in Section~\ref{sec:bayesian_method} have been repeated for several temperature indices.  The qualitative and quantitative behaviors of the model for these different temperature indices are discussed in Section~\ref{sec:comparison_temp} in Appendix. In this section, we only present the numerical results for the model where the temperature variable $T_a^i$ is the number of days of July and August of the year $a$ for which the temperature is over 28$^{\circ}$C on the quadrat $\omega_i$ (called the \textit{temperature index T28}). 
Among the temperature indices tested, T28 gives the best qualitative performance (in particular, it correctly explains the lack of expansion of ash dieback in south-east France, see Section~\ref{sec:evolution_chalara}) as well as the best quantitative performance (highest likelihood for the period 2008-2023 among the temperature indices tested, see Section~\ref{sec:comparison_temp}).

We run\footnote{Simulations are run on the babycluster of the Institut Élie Cartan de Lorraine.} Algorithm~\ref{algo.AMIS} with $N_L=60$ iterations, the samples size $N=10000$, the initialization of the posterior distribution given in Table~\ref{table.init.prop.amis} and the temperature index T28. We first present the computed posterior distributions. Then we plot the dynamics for the best parameter set. Finally we discuss the validation of the model.

\subsection{Posterior distributions of parameters}

\begin{figure}
\begin{center}
\includegraphics[height=4cm, trim = 1cm 0cm 1.cm 0cm, clip=true]{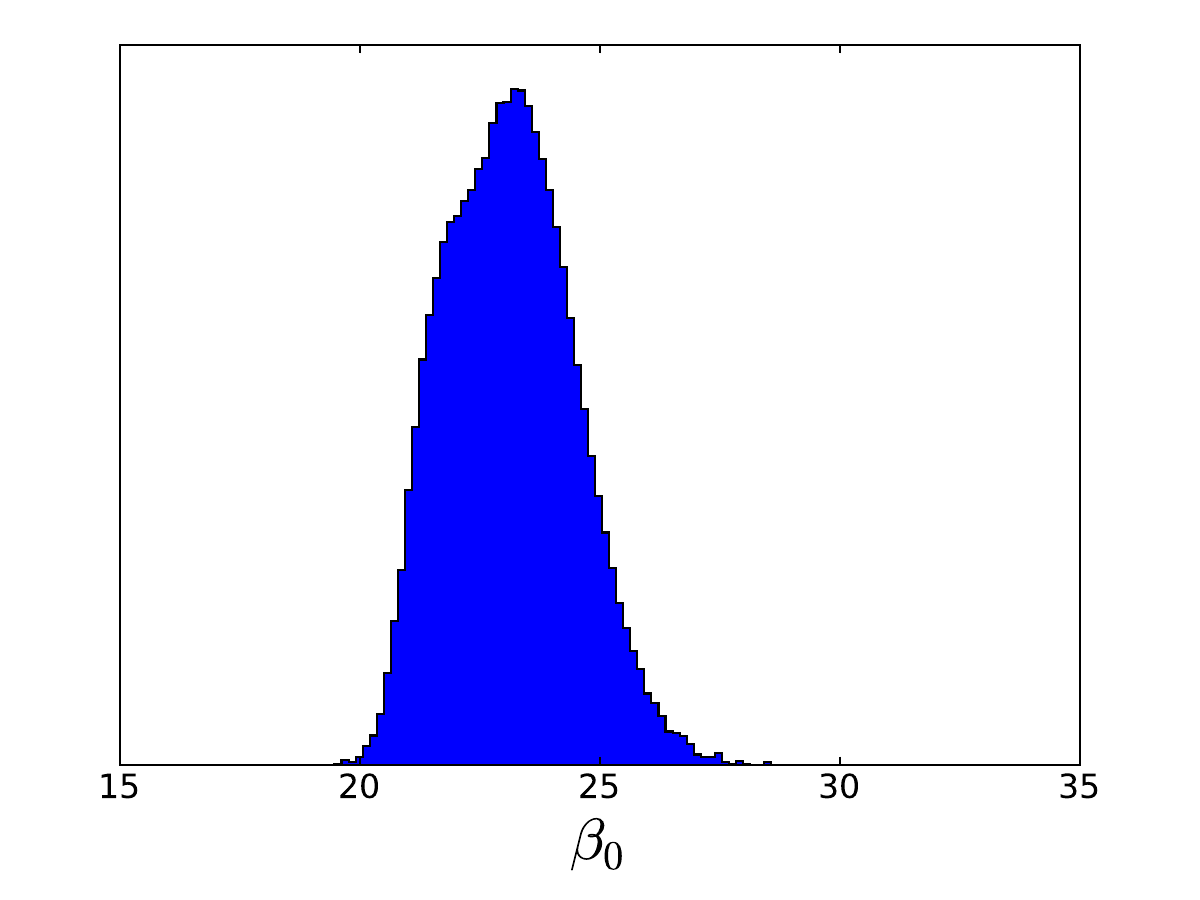}
\includegraphics[height=4cm, trim = 1cm 0cm 1.cm 0cm, clip=true]{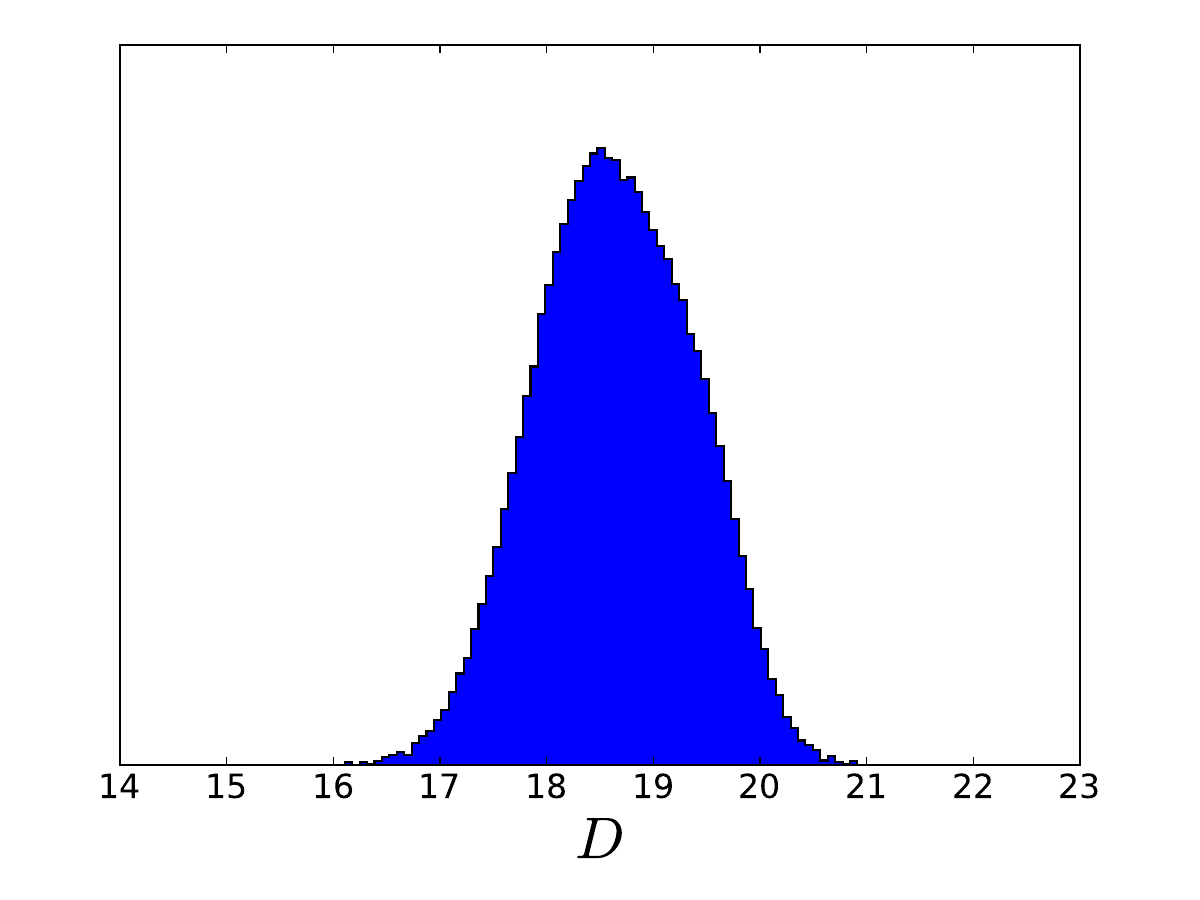}
\includegraphics[height=4cm, trim = 1.cm 0cm 1.cm 0cm, clip=true]{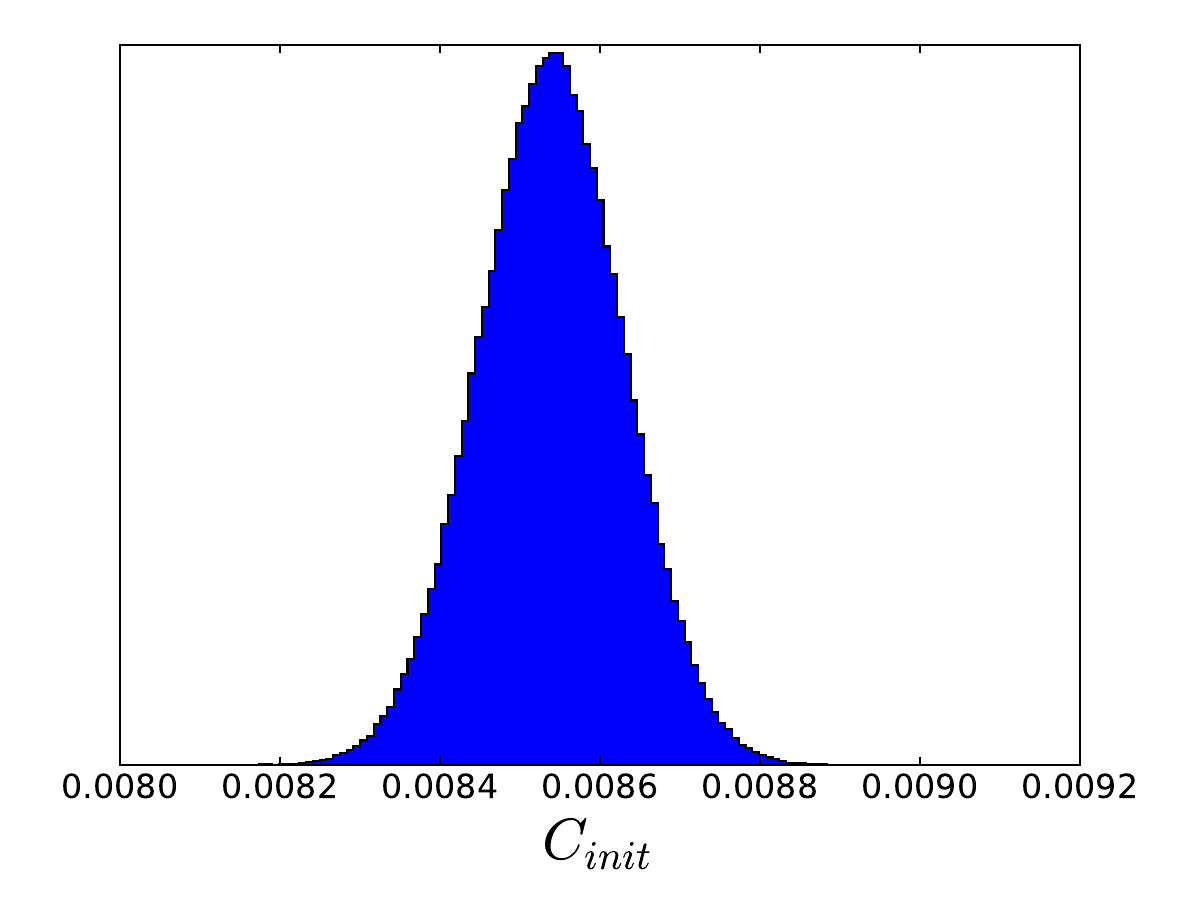}
\\
\includegraphics[height=4cm, trim = 1cm 0cm 1.cm 0cm, clip=true]{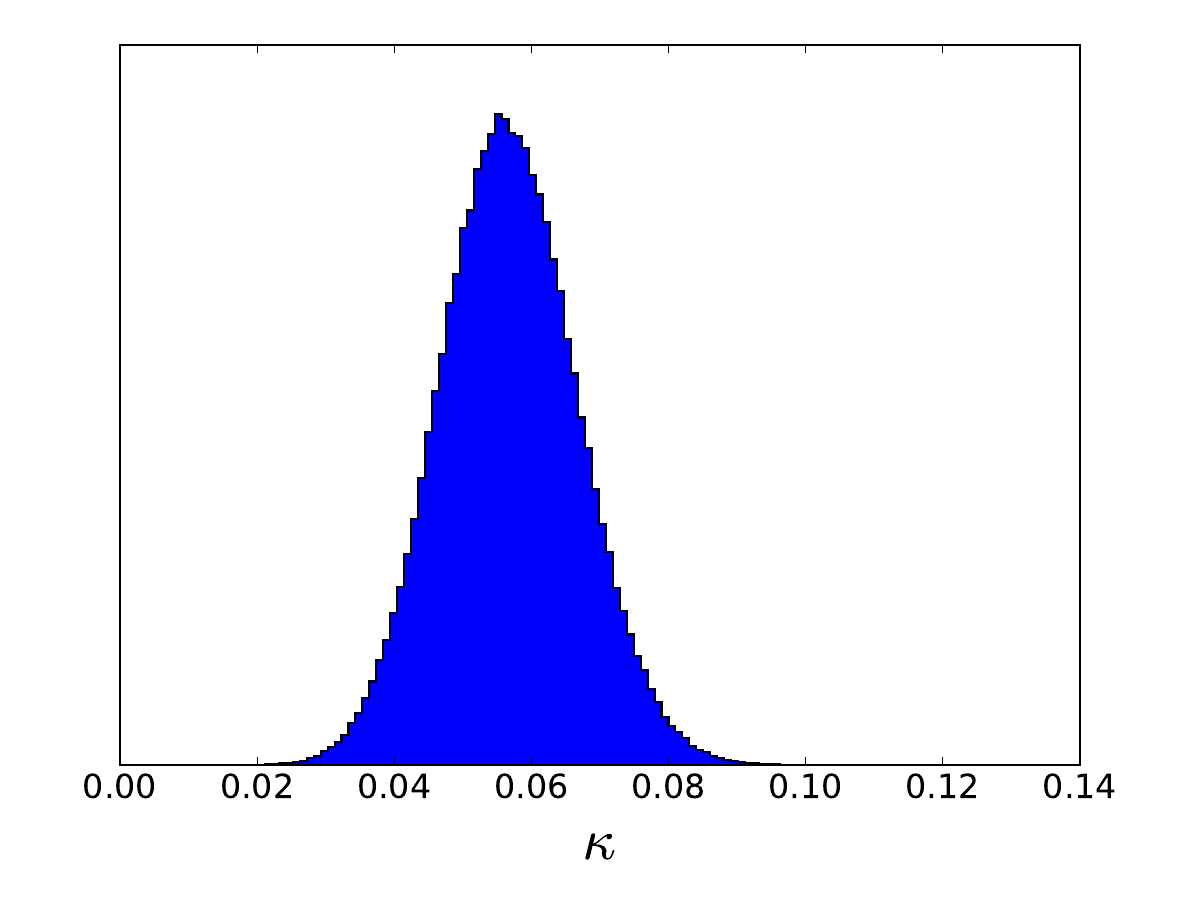}
\includegraphics[height=4cm, trim = 1cm 0cm 1.cm 0cm, clip=true]{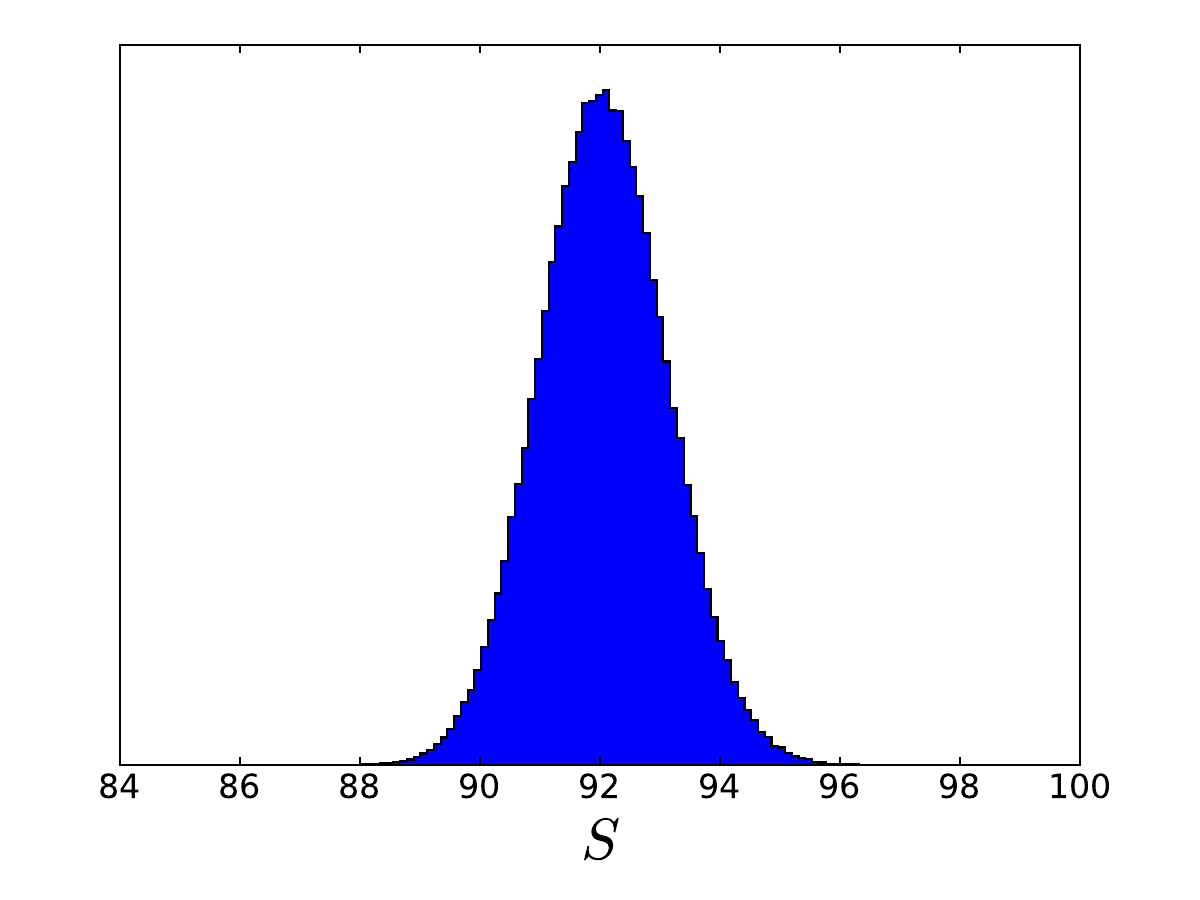}
\end{center}
\caption{\label{fig.posterior.distributions} Marginal posterior distributions of the five estimated parameters $\beta_1$, $D$, $\Cinit$, $\kappa$ and $S$.% estimated by Algorithm~\ref{algo.AMIS} with $N_L=60$ iterations, samples size $N=10000$,  the initialization given in Table~\ref{table.init.prop.amis} and the temperature index T28. 
}
\end{figure}

The one-dimensional posterior distributions of the five estimated parameters are presented in Figure~\ref{fig.posterior.distributions}.
They give the acceptable ranges for the five estimated parameters $\beta_1$, $D$, $\Cinit$, $\kappa$ and $S$.

\medskip

The correlation coefficients between estimated parameters are summarize in Table~\ref{table.correlation}. In addition, the 2-dimensional posterior distributions of the parameters $(\beta_0,D)$ and $(S,\kappa)$, for which the correlations are the highest, are presented in Figure~\ref{fig.2D.posterior.distributions}.
As already observed in the first estimation step (see Appendix~\ref{sec:appendix.first.estimation} and Figure~\ref{fig.grid.D.beta0.likeli}), there is a strong negative correlation between parameters $\beta_0$ and $D$, which can be interpreted as follows: increasing $\beta_0$ increases the source term $\nu_a(x)/\tau$ in \eqref{eq.diffusion} (amount of spores produced at $x$) and then speeds up the propagation of \textit{H.~fraxineus}, which is counterbalanced by decreasing the diffusion coefficient $D$. However, increasing $\beta_0$ and decreasing $D$ increases the local infection~/ colonization. 

Both parameters $S$ and $\kappa$ are also correlated. $S$ set the maximal spore load possible ($\chi_{a}(x)$ in \eqref{eq.rachis_infection}). This will determine the amount of foliar infection at a given environmental condition. $\kappa$ determines how high summer temperatures control the rate of leaf infection transfer to shoot infection. When the spore load and thus the leaf infection are higher (high $S$), there is a compensation by a stronger and more gradual control of the transfer of the pathogen from leaves to shoots (higher $\kappa$).

\begin{table}
\begin{center}
\begin{tabular}{c|cccc}
& D & $\beta_0$ & $\kappa$ & $S$\\
\hline
$\beta_0$ & \textbf{-0.97}\\
$\kappa$ & 0.06 & -0.04 \\
$S$ & 0.09 & -0.11 & \textbf{0.58}\\
$\Cinit$ & -0.09 & 0.08 & 0.02 & -0.14
\end{tabular}
\end{center}
\caption{\label{table.correlation} Correlation coefficients between estimated parameters.% by Algorithm~\ref{algo.AMIS} with $N_L=60$ iterations, samples size $N=10000$, the initialization given in Table~\ref{table.init.prop.amis} and the temperature index T28.
}
\end{table}

\begin{figure}
\begin{center}
\includegraphics[height=5cm, trim = 0cm 0.5cm 4.5cm 0.5cm, clip=true]{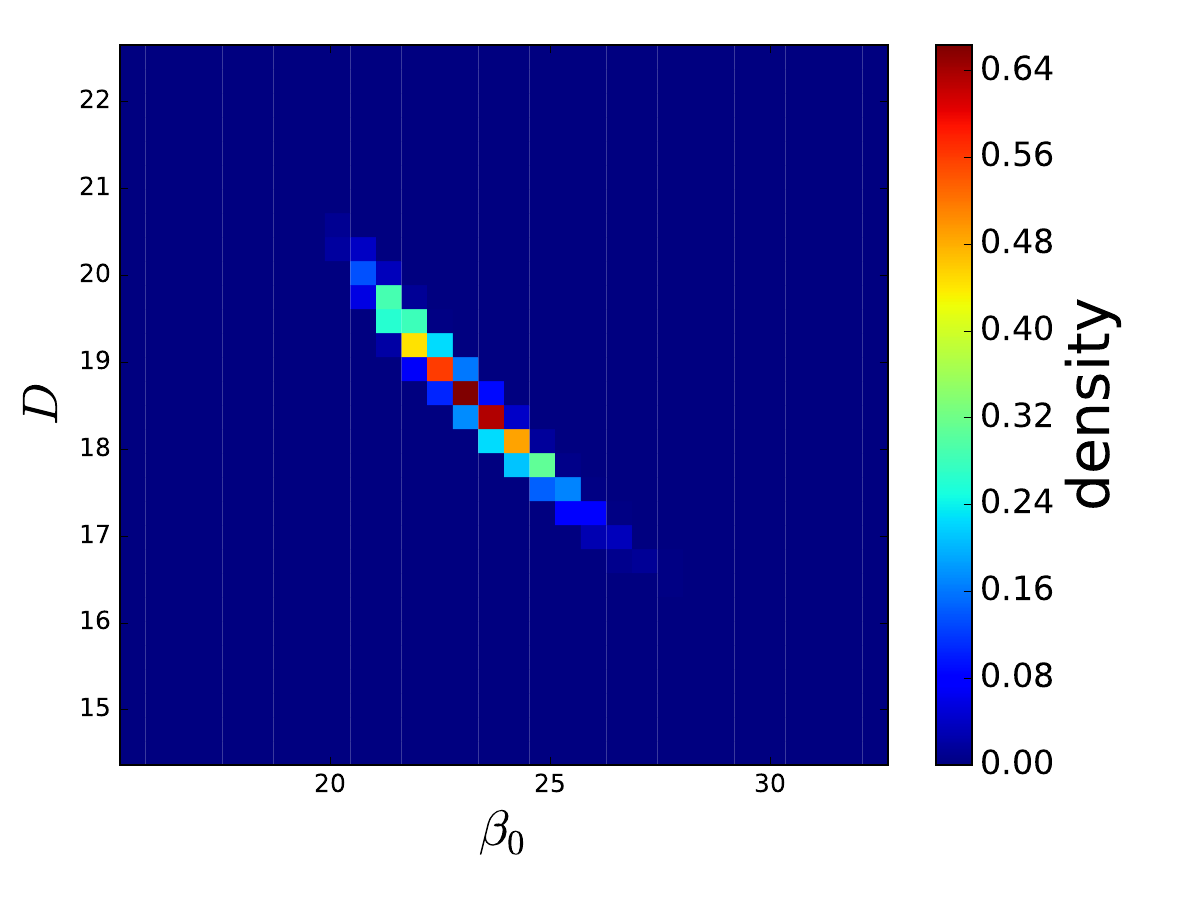}
\includegraphics[height=5cm, trim = 0cm 0.5cm 4.5cm 0.5cm, clip=true]{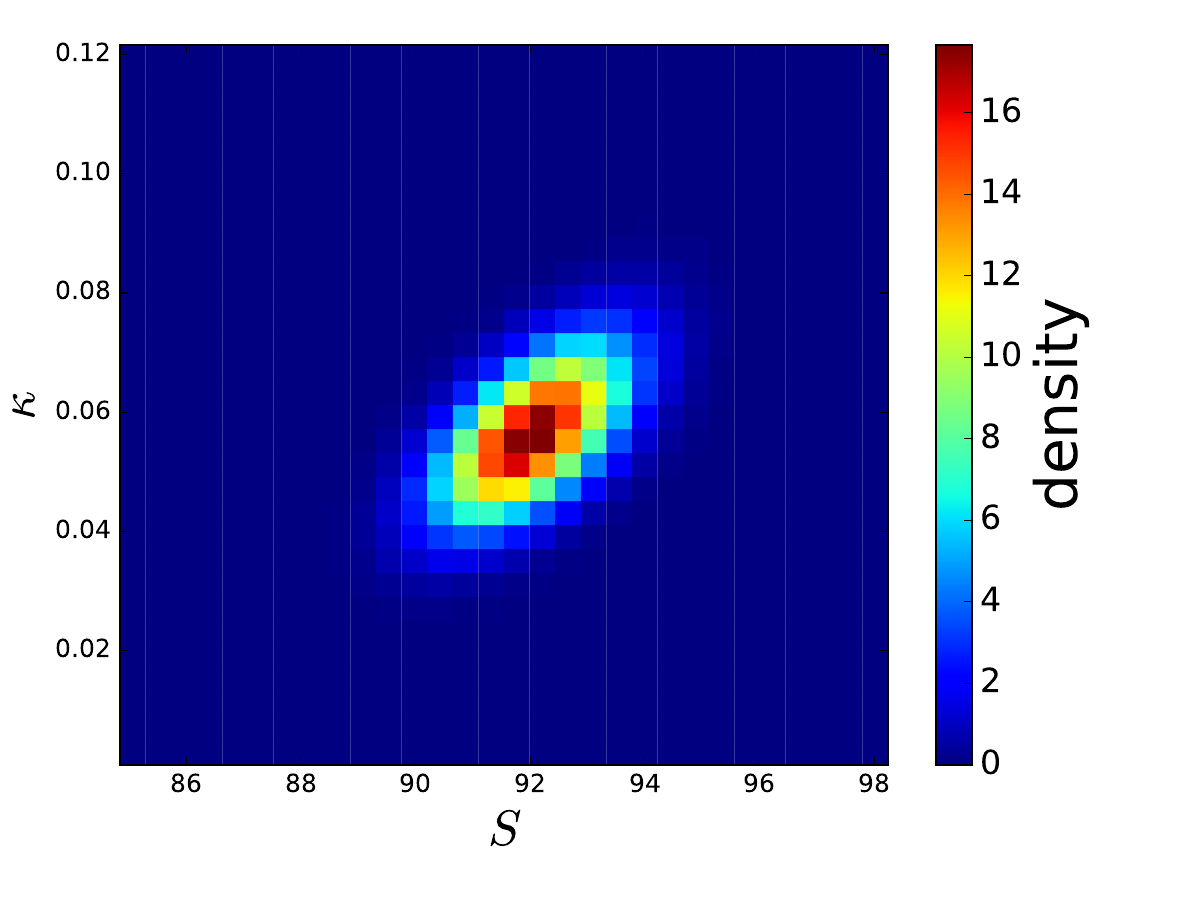}
\end{center}
\caption{\label{fig.2D.posterior.distributions} 2-dimensional posterior distributions of the parameters $(\beta_0,D)$ and $(S,\kappa)$.% estimated by Algorithm~\ref{algo.AMIS} with $N_L=60$ iterations, samples size $N=10000$,   the initialization given in Table~\ref{table.init.prop.amis} and the temperature index T28.
}
\end{figure}

%The parameters $\beta_0$ and $S$ are negatively correlated. Both parameters act as an infection threshold and then impact the local infection level. Increasing one parameter is counterbalanced by decreasing the other one.

%There is a low correlation between parameter $S$ and $\Cinit$.  {\color{red}Se justifie par les arguments de Section~\ref{sec.identifiability}? Corrélés car tous les 2 des para d'échelle / faiblement corrélés car $r_S$ fixé?}
%Note that, except $S$,  $\Cinit$ is not correlated to other parameters. It is in accordance with Figure~\ref{fig.parameters.convergence}, where the convergence of $\Cinit$ seemed independent of other parameters.

%No significant correlation is observed between others pairs of parameters. {\color{red} voir les vitesses de convergence des paramètres à la Figure~\ref{fig.parameters.convergence}}

\subsection{Predicted expansion of ash dieback for the best set of parameters}
\label{sec:evolution_chalara}

In this section, we present the model dynamics for the parameters set, among the parameters sampled by the AMIS algorithm, which maximises the likelihood. It corresponds to the parameter $\theta^{\max}=(D^{\max},\beta_0^{\max},0,0,21.50001,\kappa^{\max},S^{\max},\Cinit^{\max},1000,1)$ with
\begin{align}
\begin{matrix}
\label{para.max.like1}
D^{\max}=18.51 ; & \beta_0^{\max} = 23.26  ;  & \kappa^{\max} =  0.056 ;\\
S^{\max} = 92.0 ; & \Cinit^{\max}=0.00855.
\end{matrix}
\end{align}

\begin{figure}
\begin{center}
\begin{tabular}{cc}
\makecell{\includegraphics[height=4.2cm, trim = 3.2cm 1.8cm 5.4cm 0.5cm, clip=true]{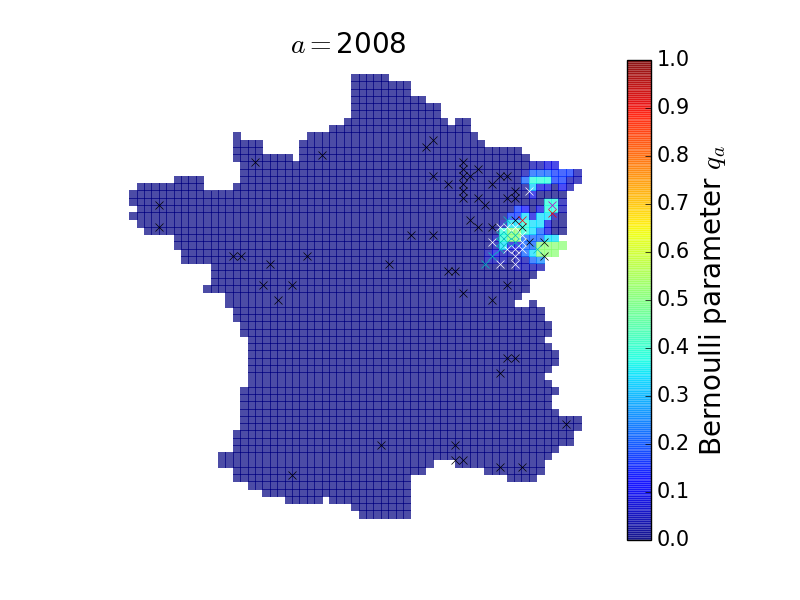}
\includegraphics[height=4.2cm, trim = 3.2cm 1.8cm 5.4cm 0.5cm, clip=true]{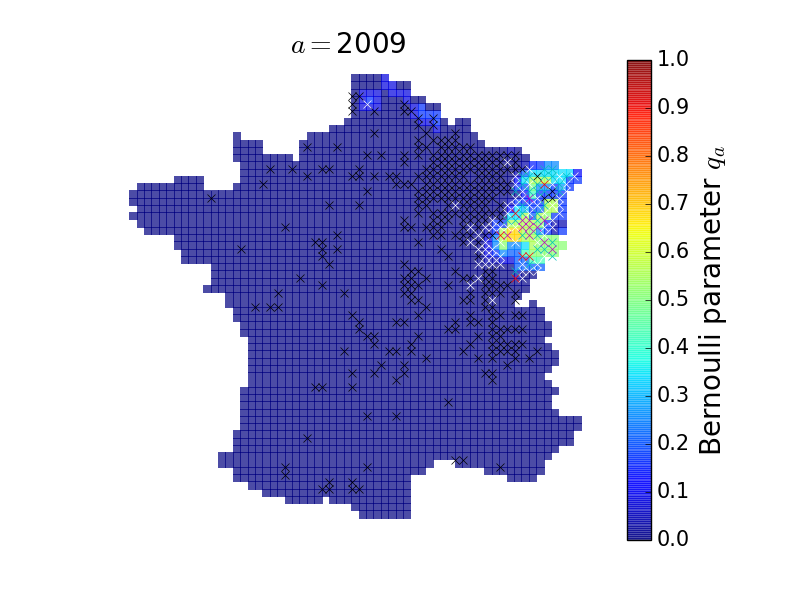}
\includegraphics[height=4.2cm, trim = 3.2cm 1.8cm 5.4cm 0.5cm, clip=true]{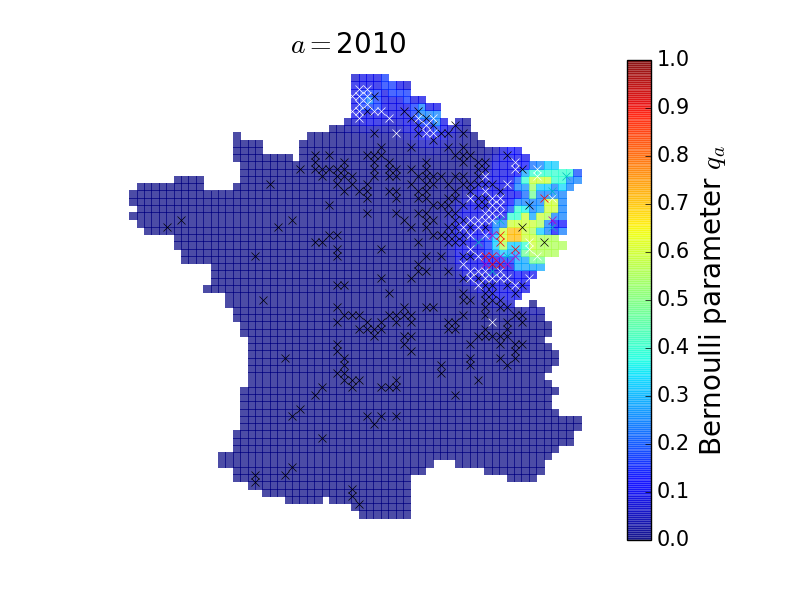}\\
\includegraphics[height=4.2cm, trim = 3.2cm 1.8cm 5.4cm 0.5cm, clip=true]{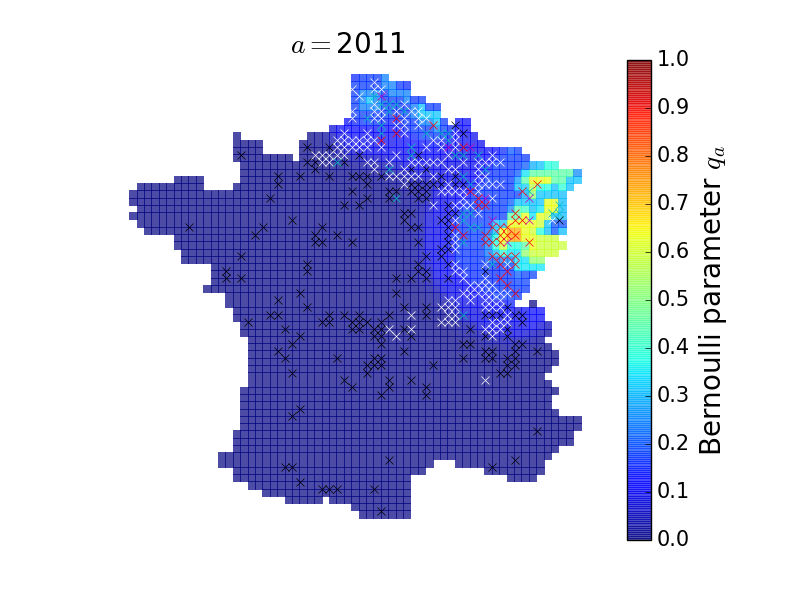}
\includegraphics[height=4.2cm, trim = 3.2cm 1.8cm 5.4cm 0.5cm, clip=true]{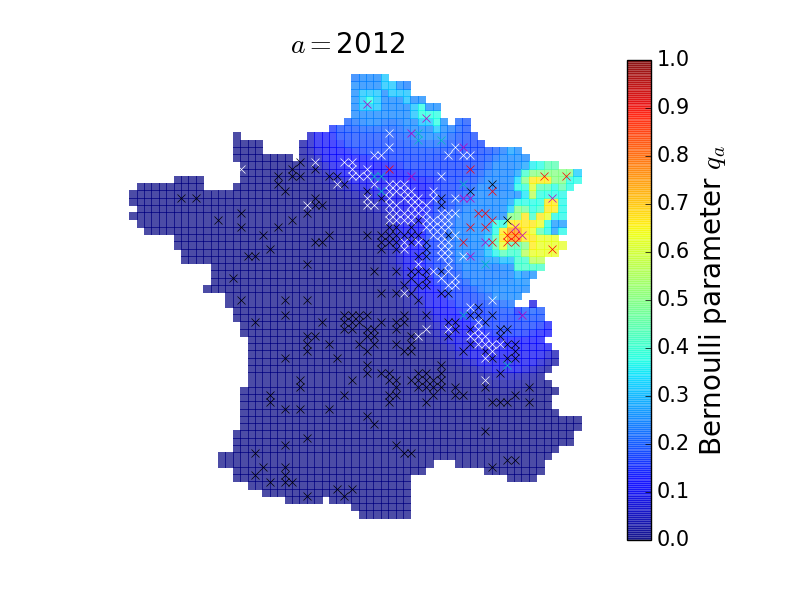}
\includegraphics[height=4.2cm, trim = 3.2cm 1.8cm 5.4cm 0.5cm, clip=true]{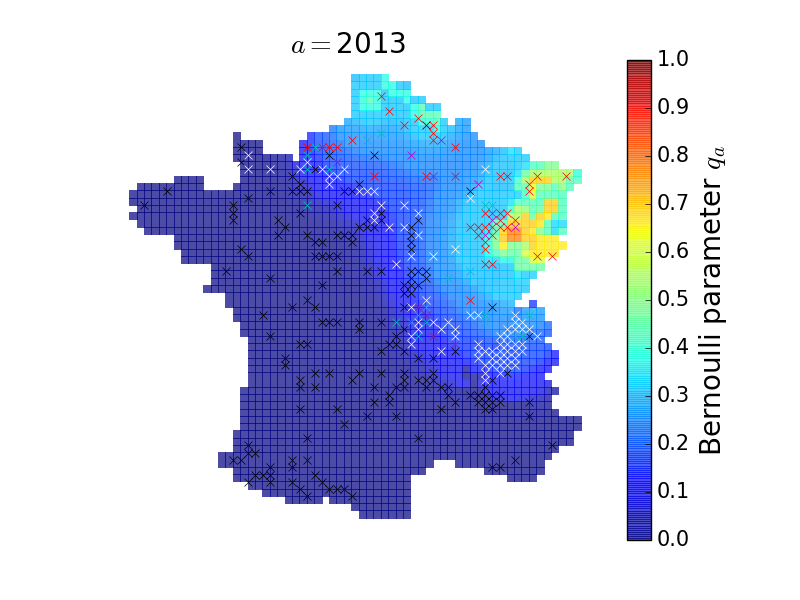}\\
\includegraphics[height=4.2cm, trim = 3.2cm 1.8cm 5.4cm 0.5cm, clip=true]{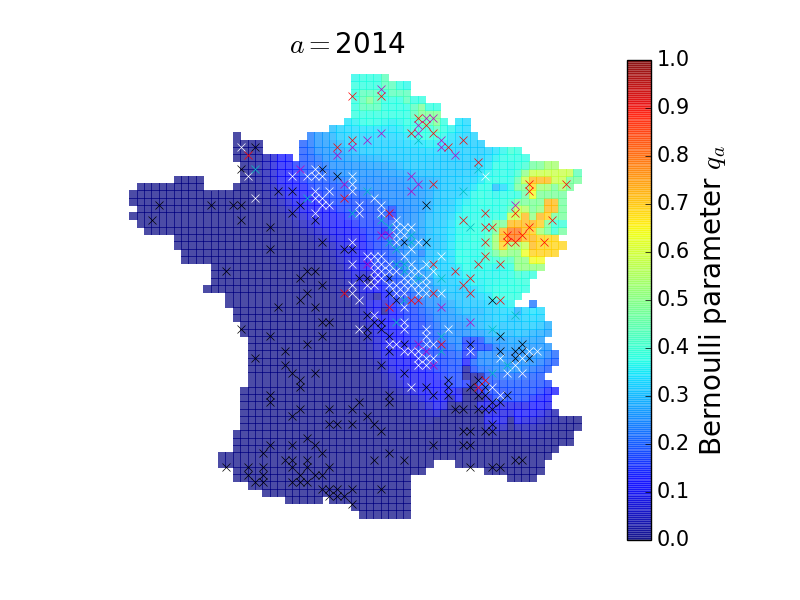}
\includegraphics[height=4.2cm, trim = 3.2cm 1.8cm 5.4cm 0.5cm, clip=true]{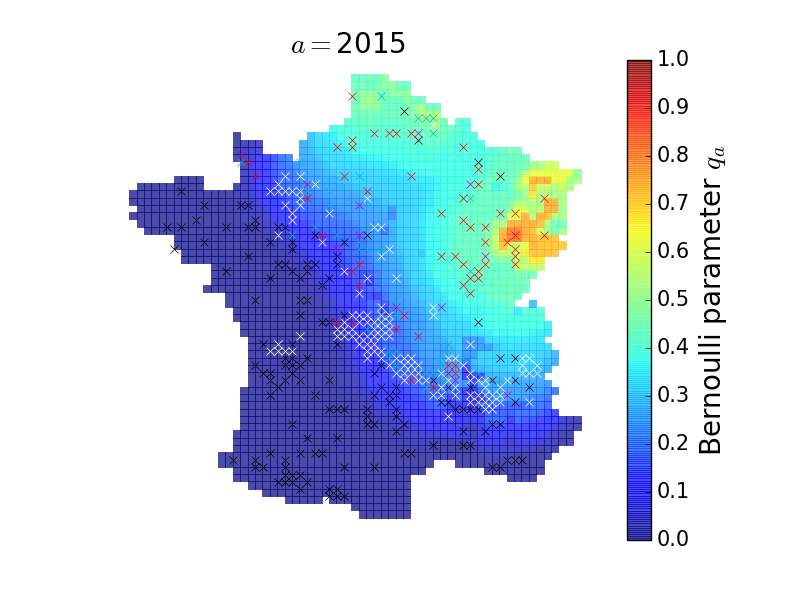}
\includegraphics[height=4.2cm, trim = 3.2cm 1.8cm 5.4cm 0.5cm, clip=true]{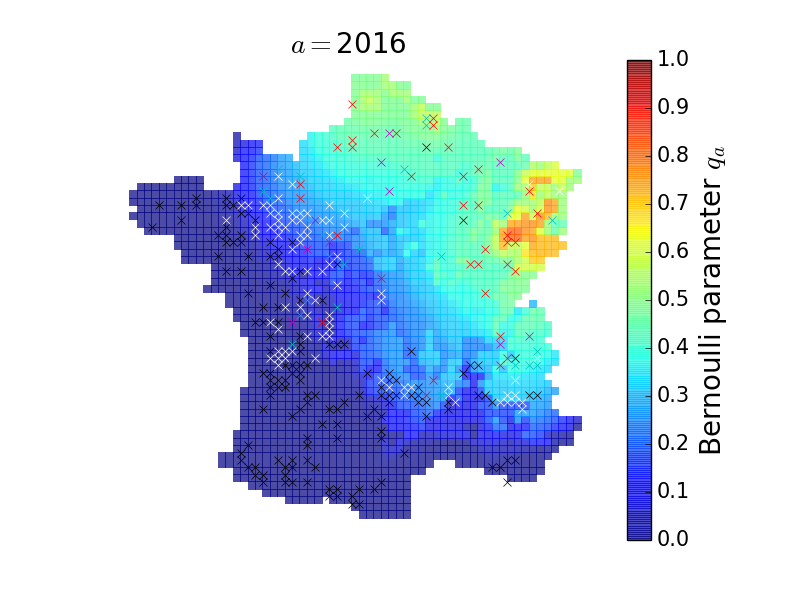}\\
\includegraphics[height=4.2cm, trim = 3.2cm 1.8cm 5.4cm 0.5cm, clip=true]{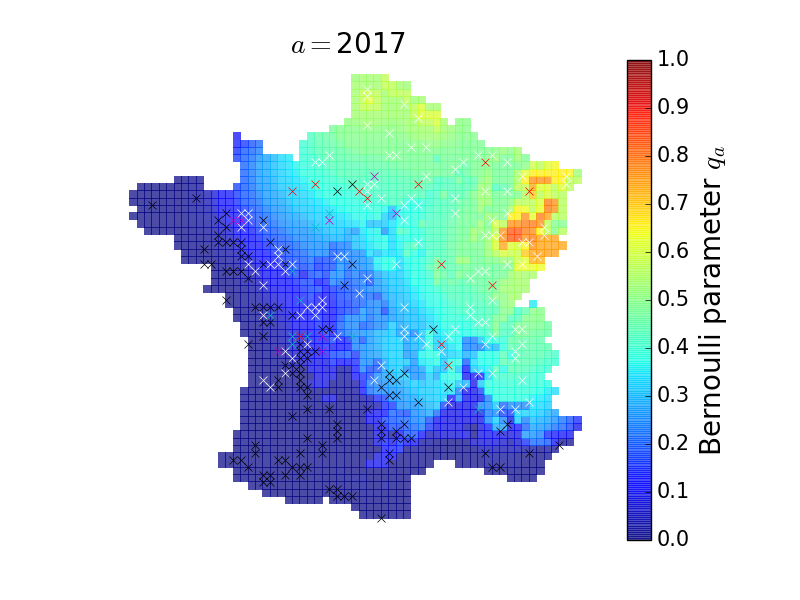}
\includegraphics[height=4.2cm, trim = 3.2cm 1.8cm 5.4cm 0.5cm, clip=true]{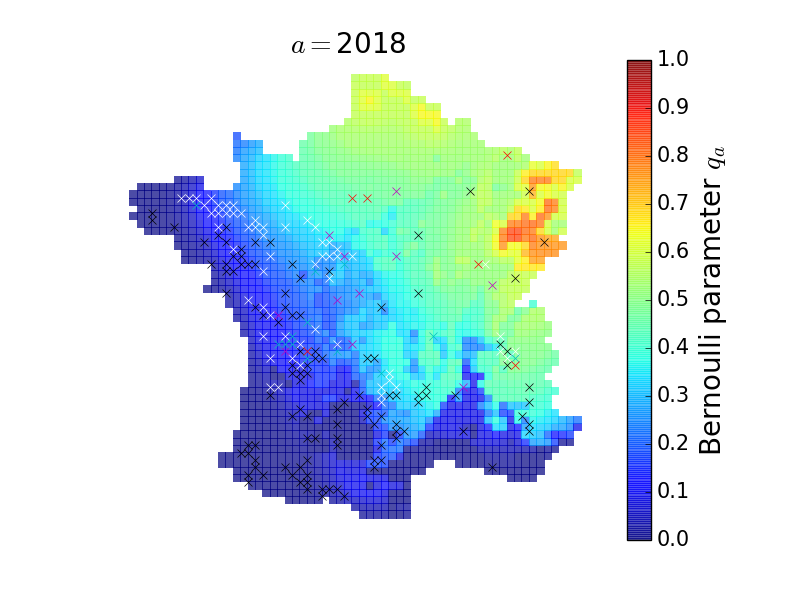}
\includegraphics[height=4.2cm, trim = 3.2cm 1.8cm 5.4cm 0.5cm, clip=true]{fig_AMIS_Njs28_231120135239_bernouilli_parameter_2019}}
&
\makecell{
\includegraphics[height=6cm, trim = 15.5cm 1.2cm 1.6cm 0.5cm, clip=true]{fig_AMIS_Njs28_231120135239_bernouilli_parameter_2019}\\
\includegraphics[height=3.5cm, trim = 12.5cm 8.8cm 1.6cm 0.5cm, clip=true]{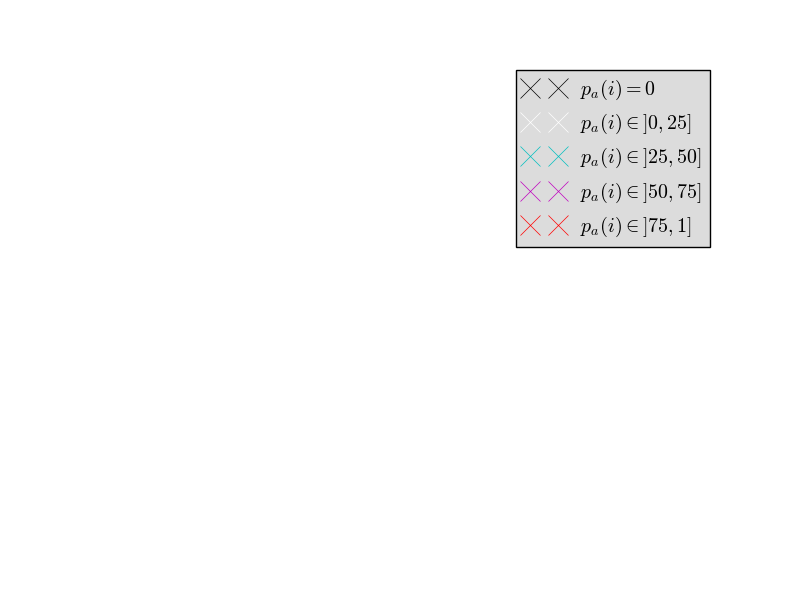}}
\end{tabular}
\end{center}
\caption{\label{fig.chalarose.dynamics} Evolution of the estimated probability that an ash tree of the quadrat $\omega_i$ develops observable symptoms (Bernoulli parameter in \eqref{eq.q}) for the set of parameters $\theta^{\max}$ given by \eqref{para.max.like1} which maximizes the likelihood for the temperature index T28. Crosses: observations data $p_a(i)$ (see \eqref{def.p_a_i}) from 2008 to 2019. Black crosses if no observed symptoms and white, cyan, magenta and red crosses for proportions of trees with dieback symptoms in $]0,0.25]$, $]0.25,0.5]$, $]0.5,0.75]$ and $]0.75,1]$ respectively.}
\end{figure}

Figure~\ref{fig.chalarose.dynamics} represents the evolution of the predicted probability that an ash tree develops observable symptoms (bernoulli parameter $q_a^i$  in \eqref{eq.q}) from $a=2008$ to $a=2019$, i.e. for the period used for the estimation, with the parameter set $\theta^{\max}$.
The front propagation, represented by the white crosses for the data on Figure~\ref{fig.chalarose.dynamics}, is generally well captured by our model. In some years, isolated reports were made well in advance of the predicted propagation front, in areas not contiguous to previously infected areas (2011 in central France, 2012-13 in the Cotentin, 2015 in the Charentes). They are often reports in ash plantation, in particular in the Cotentin, 2012-13. However, they usually remained isolated reports and led to a clear regional outbreak only in 2015 in the Charentes (central western France).
After 2015, ash dieback is still predicted to expand in western, and, to a lesser degree, in southwest France, while it expansion is predicted to stop in southeast France. Expansion is limited by high summer temperatures in this region. This fits well the observations as no further ash dieback expansion is observed in southeast France after 2015.

\begin{figure}
\begin{center}
\includegraphics[height=3.7cm, trim = 3.2cm 1.8cm 5.4cm 0.5cm, clip=true]{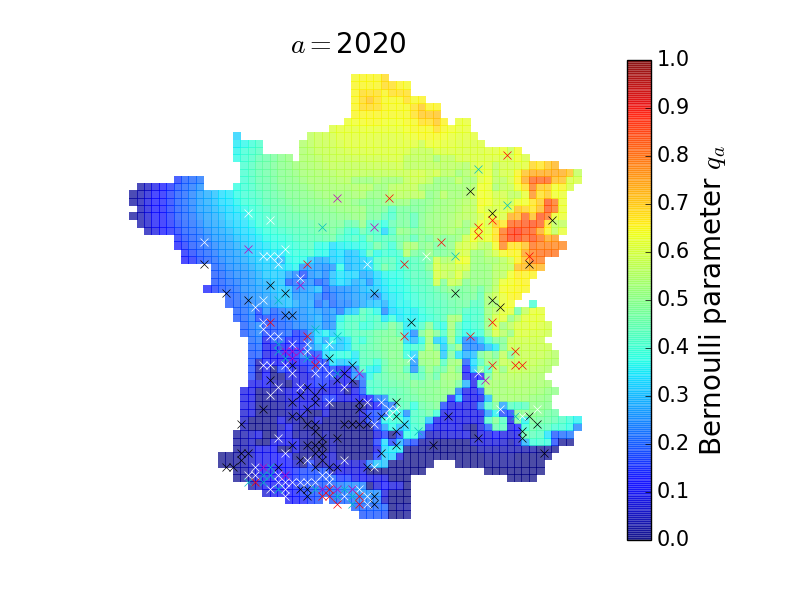}
\includegraphics[height=3.7cm, trim = 3.2cm 1.8cm 5.4cm 0.5cm, clip=true]{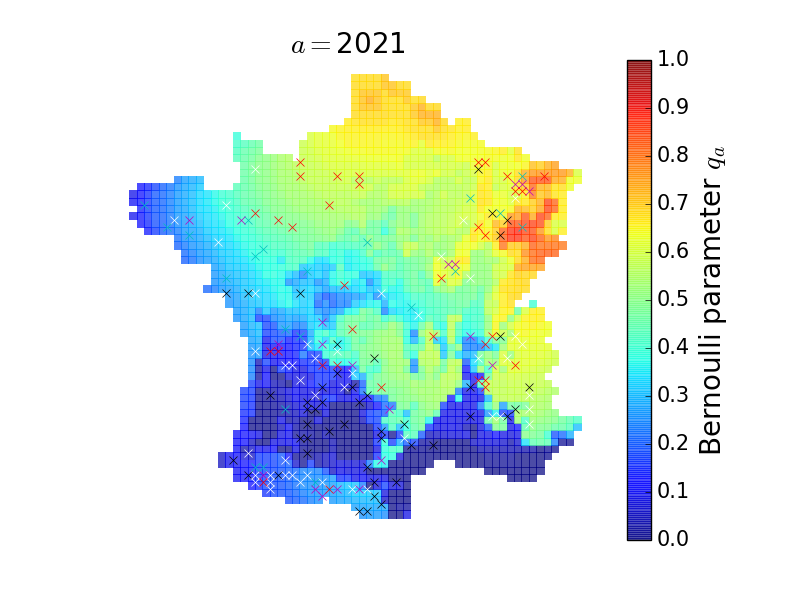}
\includegraphics[height=3.7cm, trim = 3.2cm 1.8cm 5.4cm 0.5cm, clip=true]{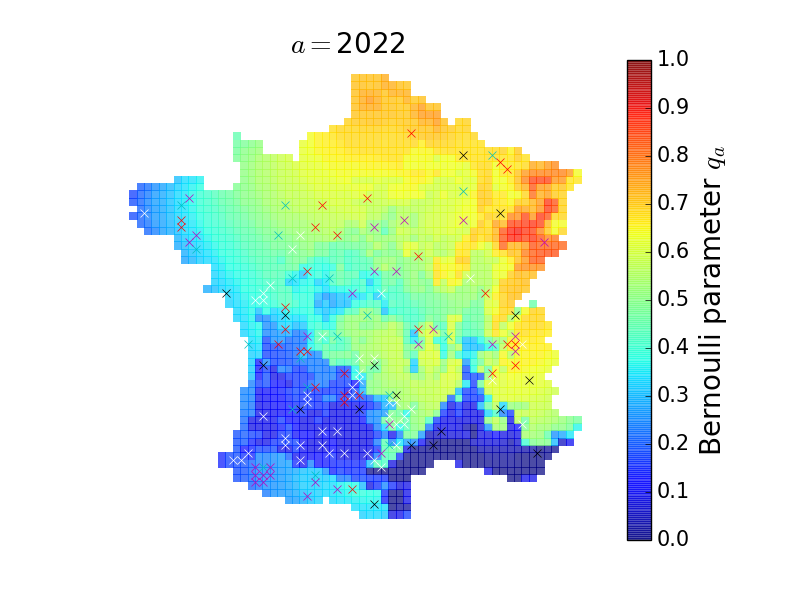}
\includegraphics[height=3.7cm, trim = 3.2cm 1.8cm 5.4cm 0.5cm, clip=true]{fig_AMIS_Njs28_231120135239_bernouilli_parameter_2023}
\includegraphics[height=3.7cm, trim = 15.5cm 1.2cm 1.cm 0.8cm, clip=true]{fig_AMIS_Njs28_231120135239_bernouilli_parameter_2023}
\end{center}
\caption{\label{fig.chalarose.dynamics.pred} Prediction of the Bernoulli parameter \eqref{eq.q} for the set of parameters $\theta^{\max}$ given by \eqref{para.max.like1} (fitted on data from 2008 to 2019 for the temperature index T28) from 2020 to 2023. Crosses: observation data from 2020 to 2023  (see Figure~\ref{fig.chalarose.dynamics} for the color coding).}
\end{figure}

Figure~\ref{fig.chalarose.dynamics.pred} represents the prediction of the probability that an ash tree will develops observable symptoms ($q_a^i$) for years 2020 to 2023 for the parameter $\theta^{\max}$. Note that this parameter was estimated with data from 2008 to 2019 and that we are in a context of prediction. We can see that the model fits well what is observed in 2020-2023. In particularly, it predicts that ash dieback will expand toward Spain in the eastern part of the Pyrénées throughout a relatively narrow corridor. Notice that no observation of ash dieback has been reported by the DSF in this area during the fitted period 2008-2019, ash dieback was observed for the first time in the Pyrénées in 2020. The corridor corresponds to an area of higher elevation where summer temperature remains cooler and ash density is significant (Figure~\ref{fig.ash.density}).
Report of ash dieback remained scarce in the Garonne valley and the ash dieback prevalence of the few reports available in the area in 2020-23 is limited (under 10\%).

\begin{figure}
\begin{center}
\begin{tabular}{cc}
 \makecell{\includegraphics[height=2.8cm, trim = 3.2cm 1.8cm 5.4cm 0.5cm, clip=true]{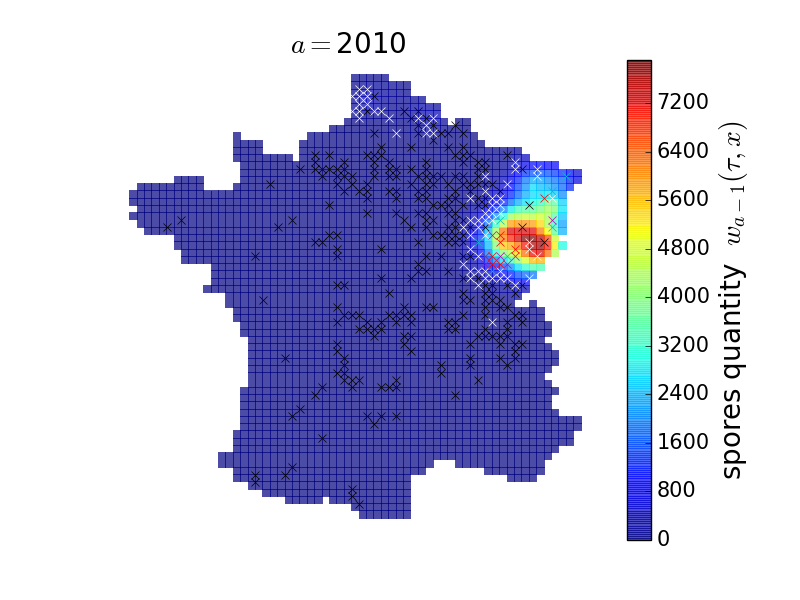}
\includegraphics[height=2.8cm, trim = 3.2cm 1.8cm 5.4cm 0.5cm, clip=true]{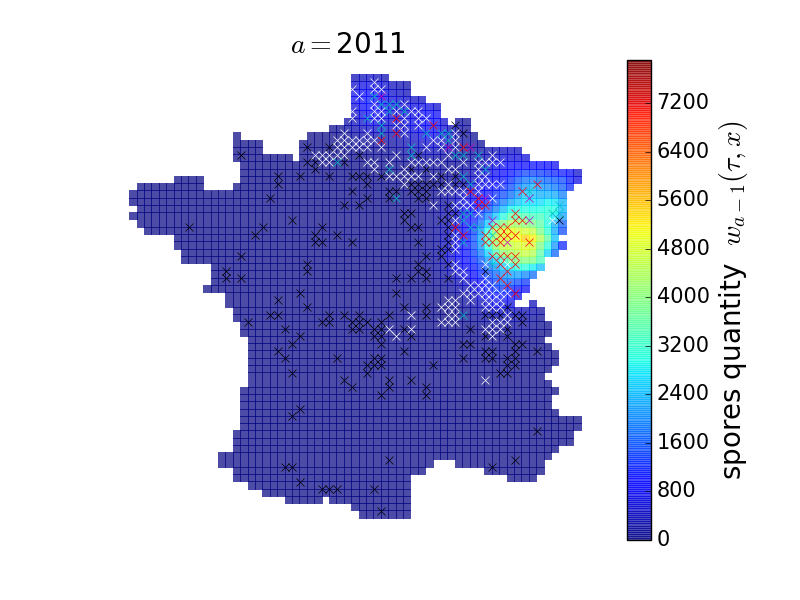}
\includegraphics[height=2.8cm, trim = 3.2cm 1.8cm 5.4cm 0.5cm, clip=true]{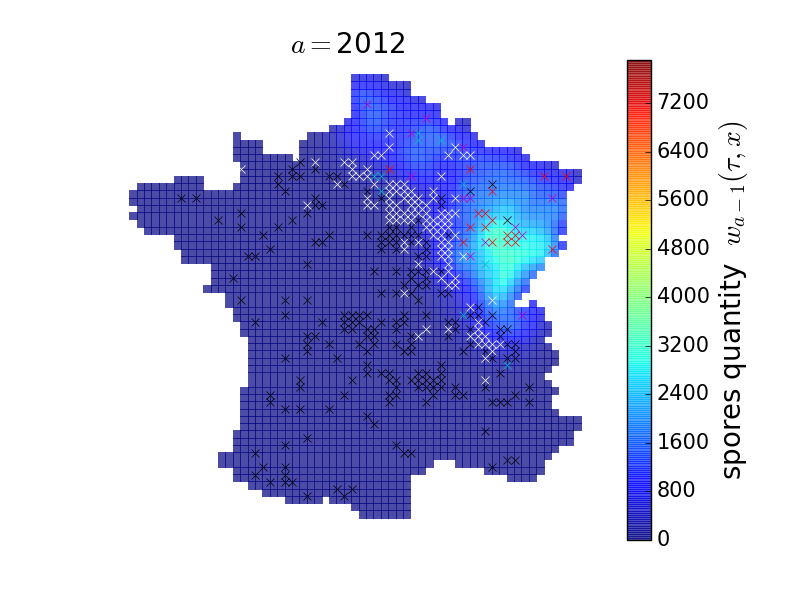}
\includegraphics[height=2.8cm, trim = 3.2cm 1.8cm 5.4cm 0.5cm, clip=true]{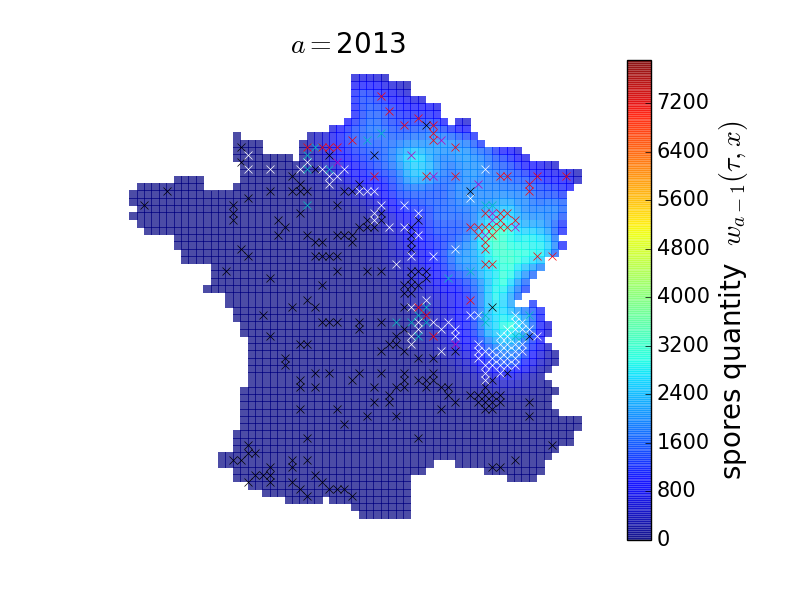}
\includegraphics[height=2.8cm, trim = 3.2cm 1.8cm 5.4cm 0.5cm, clip=true]{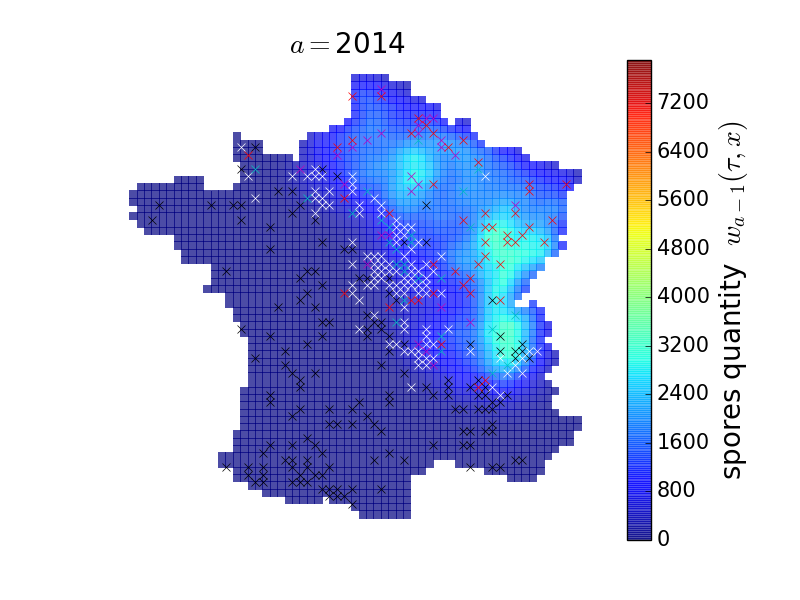}\\
\includegraphics[height=2.8cm, trim = 3.2cm 1.8cm 5.4cm 0.5cm, clip=true]{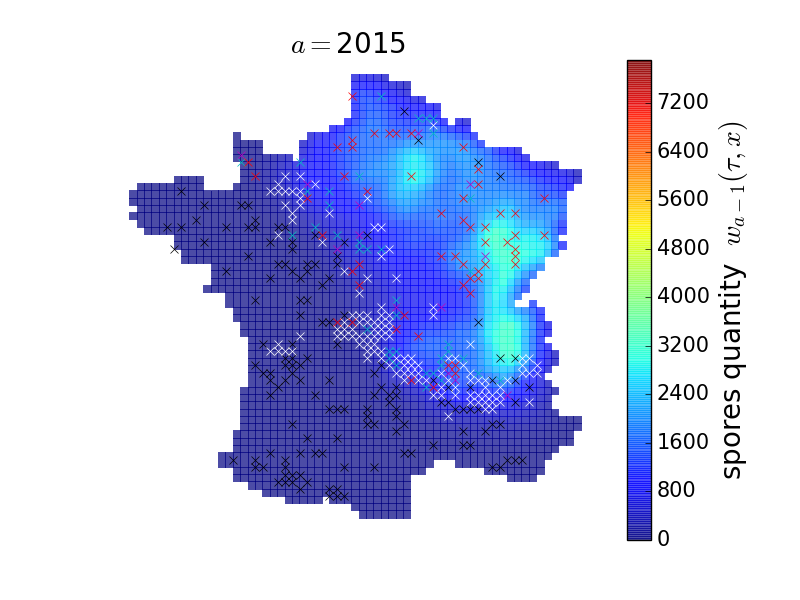}
\includegraphics[height=2.8cm, trim = 3.2cm 1.8cm 5.4cm 0.5cm, clip=true]{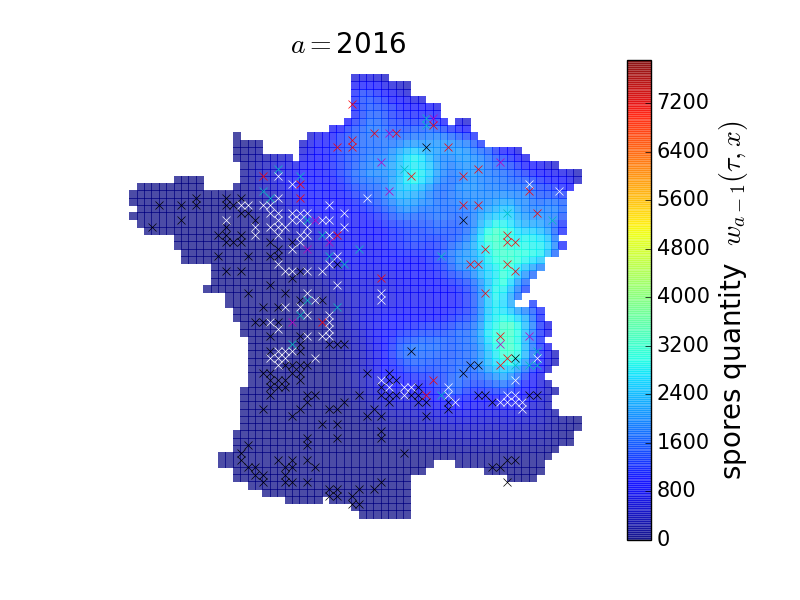}
\includegraphics[height=2.8cm, trim = 3.2cm 1.8cm 5.4cm 0.5cm, clip=true]{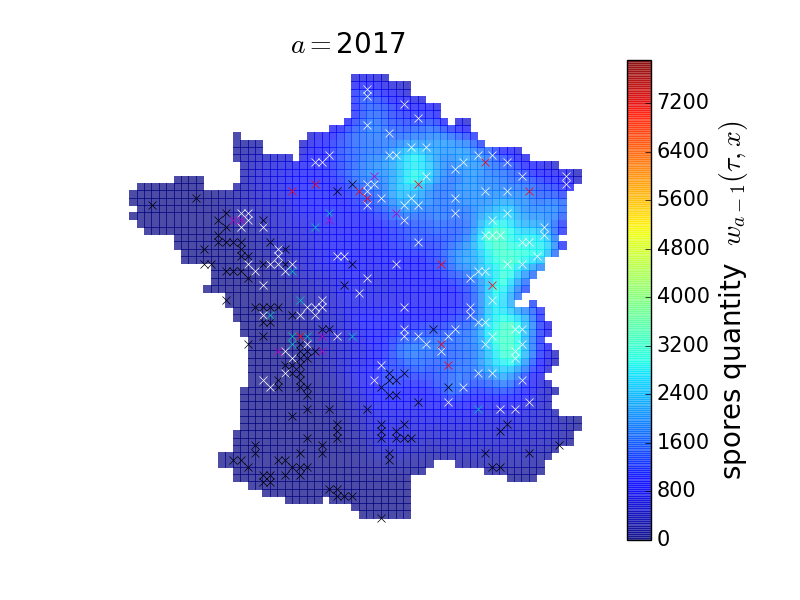}
\includegraphics[height=2.8cm, trim = 3.2cm 1.8cm 5.4cm 0.5cm, clip=true]{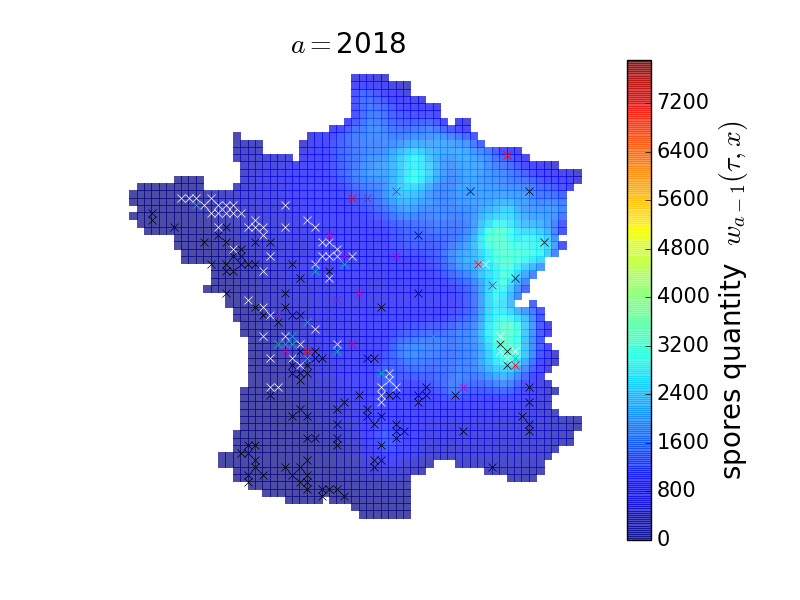}
\includegraphics[height=2.8cm, trim = 3.2cm 1.8cm 5.4cm 0.5cm, clip=true]{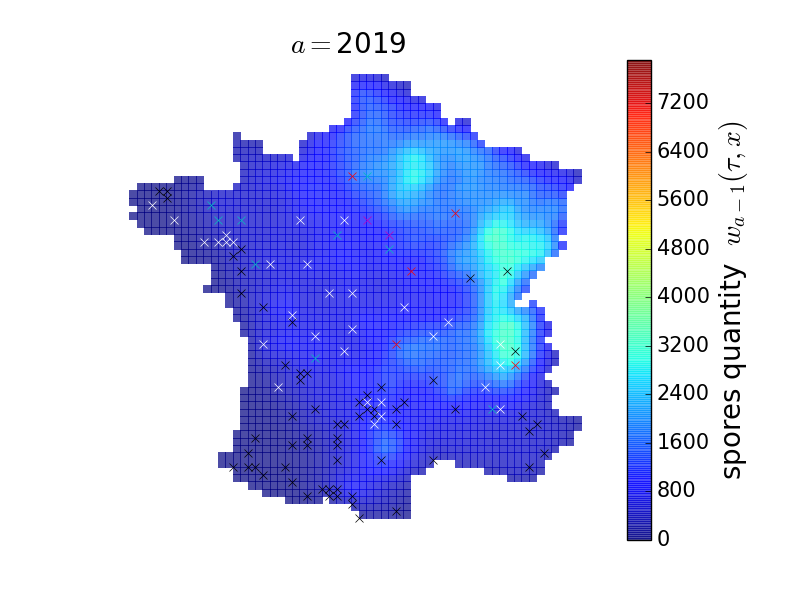}}
&
\includegraphics[height=5cm, trim = 15.5cm 1.2cm 1.cm 0.5cm, clip=true, valign=m]{fig_AMIS_Njs28_231120135239_spores_quantity_2010}
\end{tabular}
\end{center}
\caption{\label{fig.spores.dynamics} Evolution of the spore quantity $w_{a-1}(\tau,x)$, for the set of parameters $\theta^{\max}$ given by \eqref{para.max.like1}, from 2010 to 2019. Crosses: observation data from 2010 to 2019  (see Figure~\ref{fig.chalarose.dynamics} for the color coding).}
\end{figure}

\begin{figure}
\begin{center}
\begin{tabular}{cc}
 \makecell{\includegraphics[height=2.8cm, trim = 3.2cm 1.8cm 5.4cm 0.5cm, clip=true]{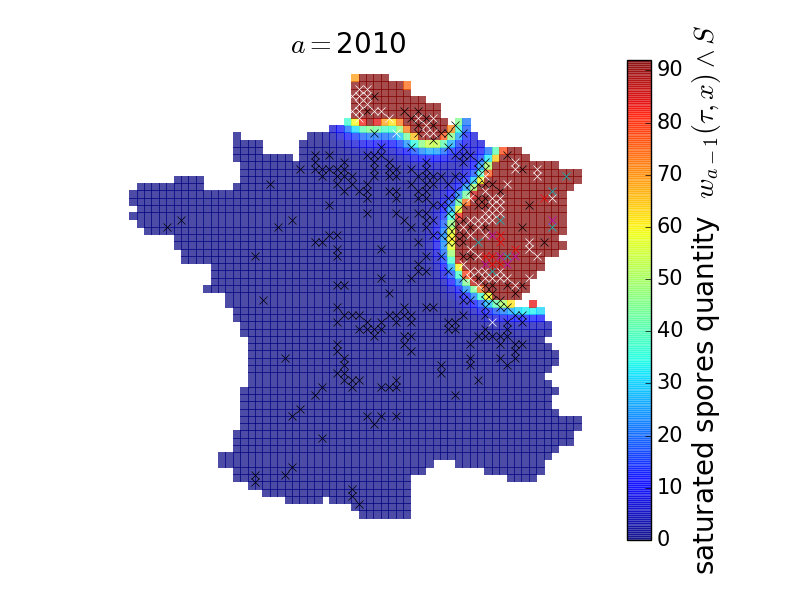}
\includegraphics[height=2.8cm, trim = 3.2cm 1.8cm 5.4cm 0.5cm, clip=true]{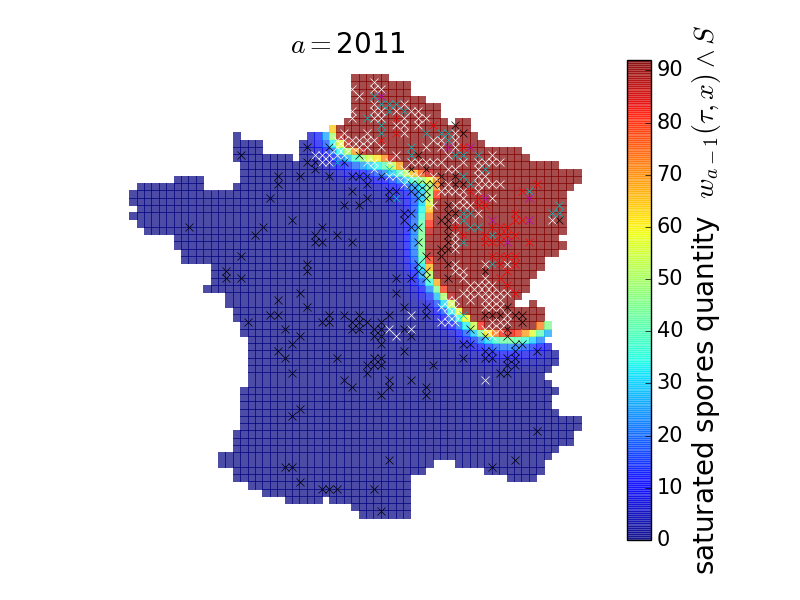}
\includegraphics[height=2.8cm, trim = 3.2cm 1.8cm 5.4cm 0.5cm, clip=true]{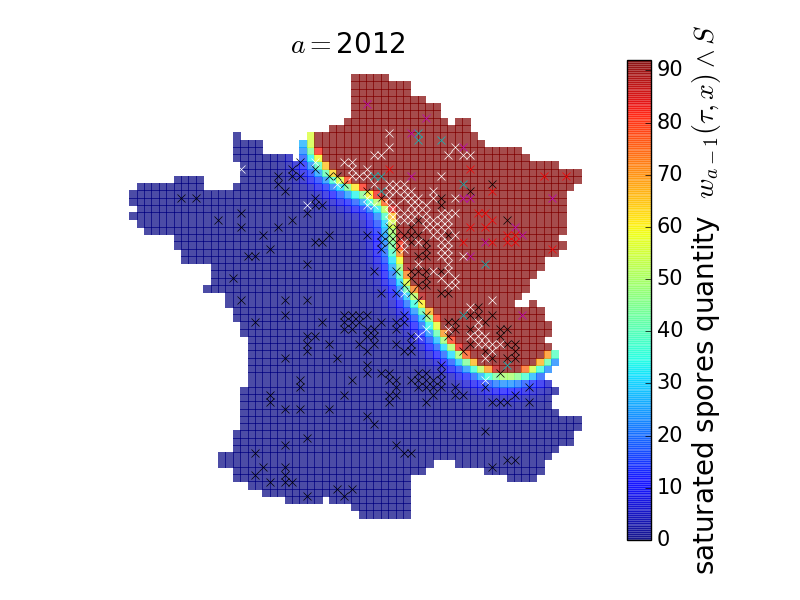}
\includegraphics[height=2.8cm, trim = 3.2cm 1.8cm 5.4cm 0.5cm, clip=true]{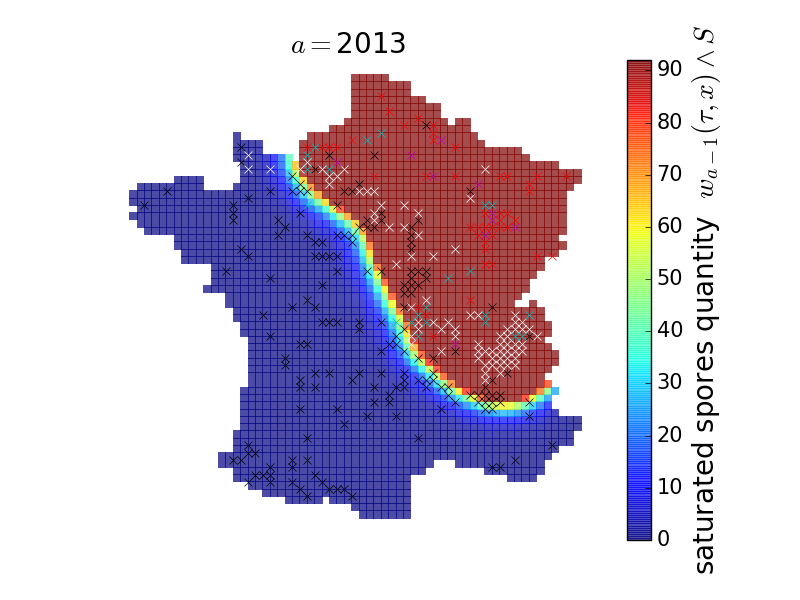}
\includegraphics[height=2.8cm, trim = 3.2cm 1.8cm 5.4cm 0.5cm, clip=true]{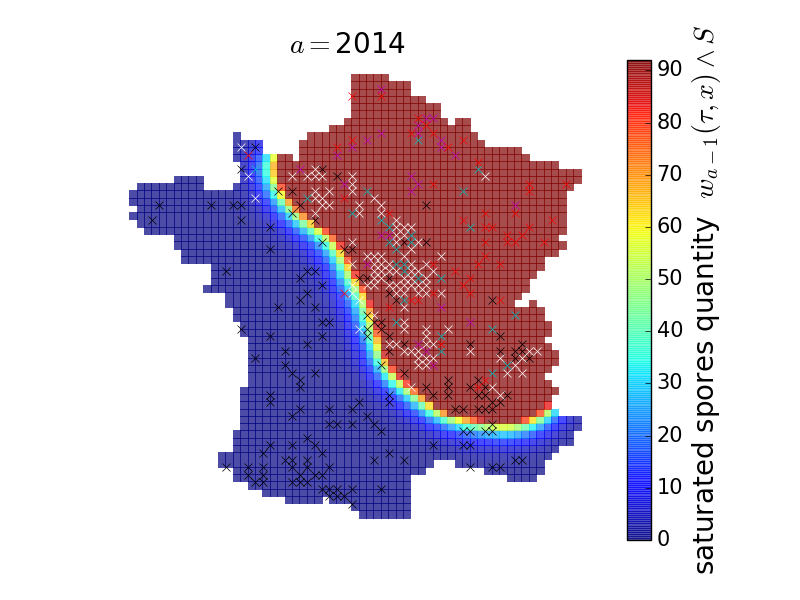}\\
\includegraphics[height=2.8cm, trim = 3.2cm 1.8cm 5.4cm 0.5cm, clip=true]{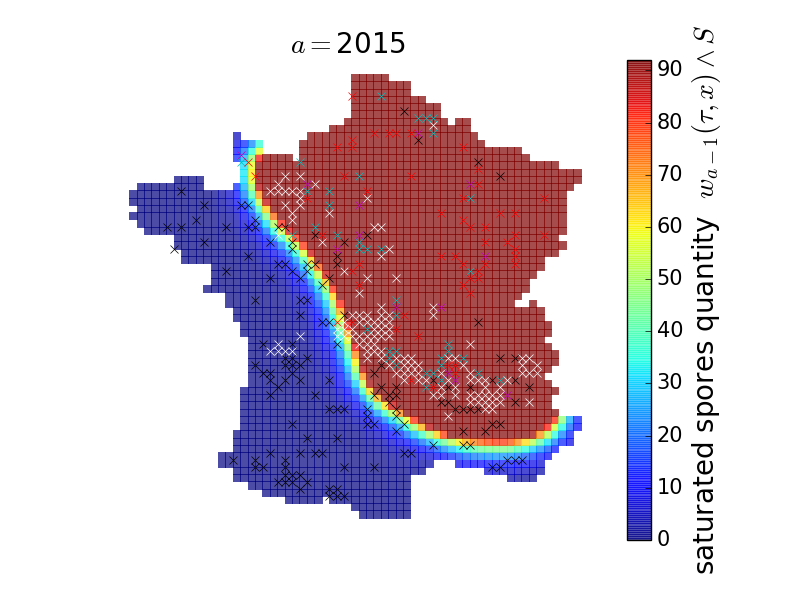}
\includegraphics[height=2.8cm, trim = 3.2cm 1.8cm 5.4cm 0.5cm, clip=true]{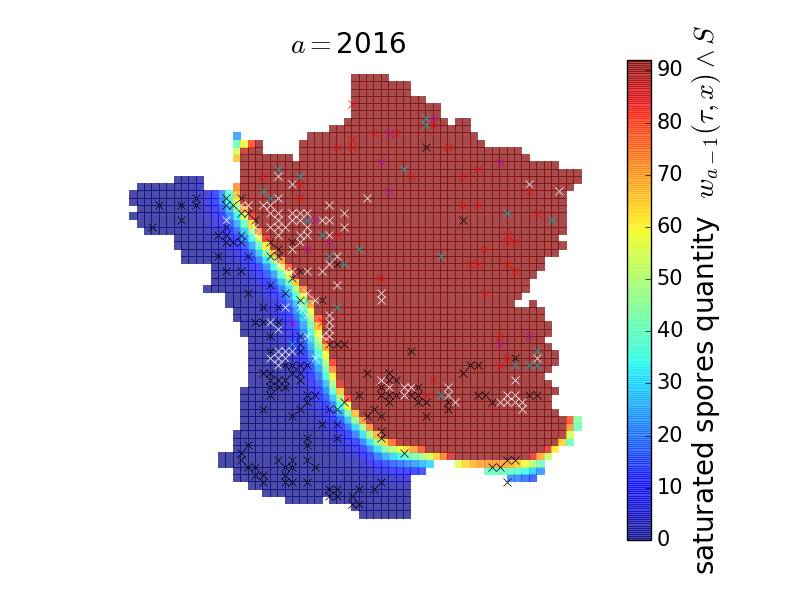}
\includegraphics[height=2.8cm, trim = 3.2cm 1.8cm 5.4cm 0.5cm, clip=true]{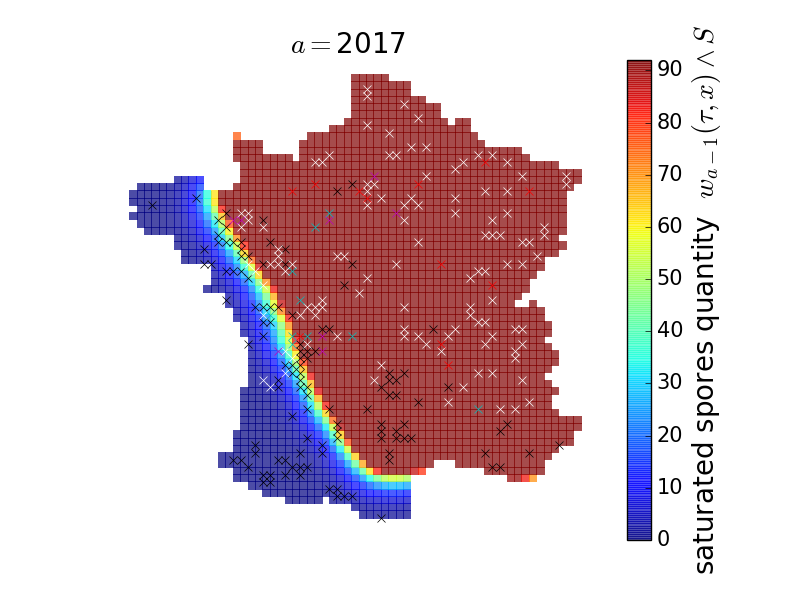}
\includegraphics[height=2.8cm, trim = 3.2cm 1.8cm 5.4cm 0.5cm, clip=true]{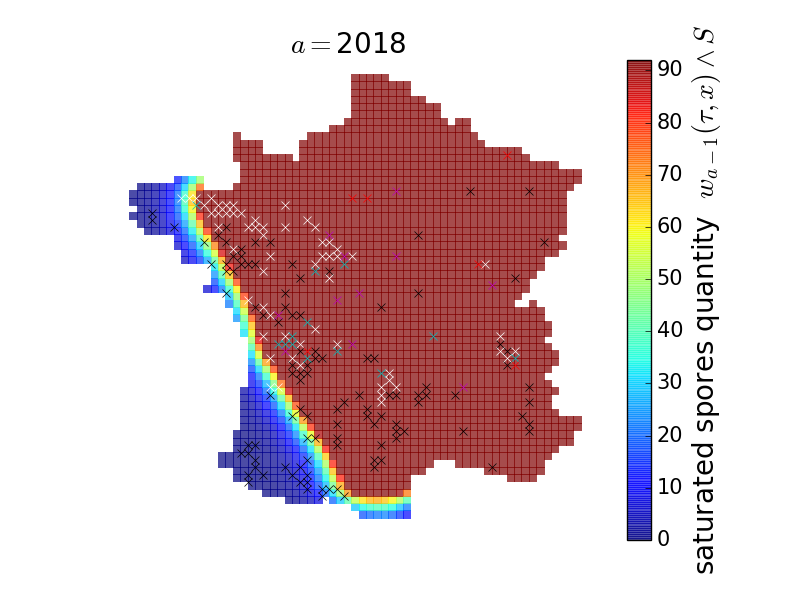}
\includegraphics[height=2.8cm, trim = 3.2cm 1.8cm 5.4cm 0.5cm, clip=true]{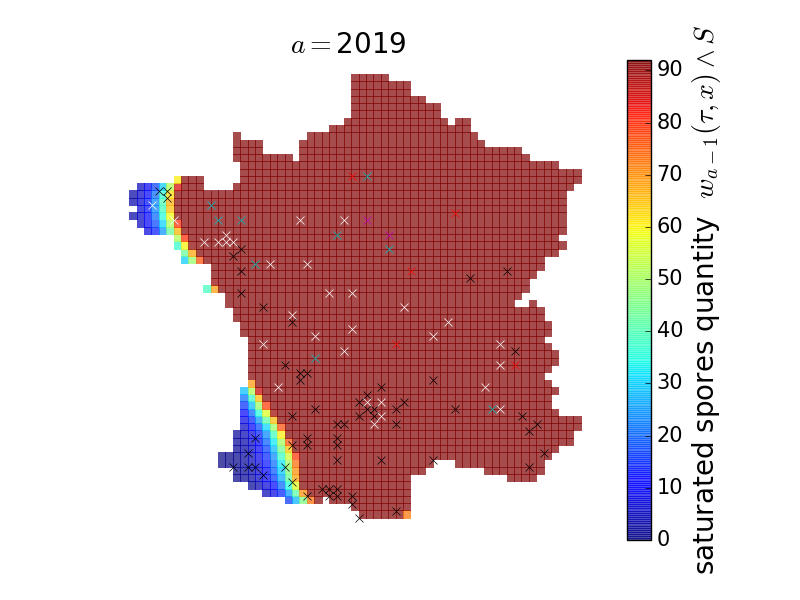}}
&
\includegraphics[height=5.5cm, trim = 15.5cm 0.5cm 1.cm 0.5cm, clip=true, valign=m]{fig_AMIS_Njs28_231120135239_spores_quantity_sat2010}
\end{tabular}
\end{center}
\caption{\label{fig.saturatedspores.dynamics} Evolution of the saturated spore quantity $w_{a-1}(\tau,x)\wedge S$, for the set of parameters $\theta^{\max}$ given by \eqref{para.max.like1}, from 2010 to 2019. Crosses: observation data from 2010 to 2019  (see Figure~\ref{fig.chalarose.dynamics} for the color coding).}
\end{figure}

Figures~\ref{fig.spores.dynamics} and \ref{fig.saturatedspores.dynamics} represent the dynamics of the spore quantity $w_{a-1}(\tau,x)$ and of the saturated spore quantity $w_{a-1}(\tau,x)\wedge S$ respectively, from 2010 to 2019. Note that, due to the delay between infection and the dieback visibility, graphs compare obervation data of the year $a$ and (saturated) spore quantities of the year $a-1$. We observe that, except on the propagation front, the zone is either infected and saturated or non infected ($w_a(\tau,x)=w_a(\tau,x)\wedge S\in \{0,S\}$), meaning that in infected zones, the rachis infection given by \eqref{eq.rachis_infection} is maximal and the spores quantity is non limiting. We also observe that the spore quantity decreases the first two years in east France, probably reflecting that the initial conditions over estimates the quantities of colonized rachis in 2007 and 2008 in order to compensate the neglected symptoms before 2008.
In addition, we can notice that, as nothing limits the propagation of spores in the model, they entirely colonize France. High summer temperatures only impact the development of the  dieback symptoms, as observed in Figure~\ref{fig.chalarose.dynamics}.

\subsection{Model validation}

%\commentanne{Après réflexion, je ne choisis pas le score de Brier car il est adapté à la proba d'événements ponctuels et pas tout à fait ici à une proba sur une parcelle. je m'explique : si on prédit correctement 50\% des arbres atteints de chalarose dans une parcelle, même si la prédiction est exacte, le score local de Brier vaut O,25. Alors qu'une prédiction de 0.5 pour un seul point veut dire que l'on n'a pas d'a priori sur la valeur. 
%Pour info à Benoît, les estimations initiales 2020-2022 pour le Brier score et le RMSE étaient bien éronnées comme soupçonné par Coralie, je 'ai pas triché sur les valeurs du tableau ci-après !}
%\subsubsection{Brier score}
%Dans la capture d'écran ci dessous, je print \\
%\texttt{année a : } $\sum_{i\in I_a} m\,\,\obs_a(i)$, $\frac{BS_a}{\sum_{i\in I_a} m\,\,\obs_a(i)}$, $(\sum_{y=2008}^a\sum_{i\in I_y} m\,\,\obs_y(i))^{-1}\sum_{y=2008}^a BS_y$\\
%où
%\begin{align*}
%BS_a = \sum_{\text{arbre observé}} (\text{observation}-q_a)^2 = \sum_{i\in I_a} \left[m\,\,\obs_a(i)\,p_a(i)(1-q_a^i)^2 + m\,\obs_a(i)\,(1-p_a(i))(0-q_a^i)^2\right]
%\end{align*}
%En d'autre terme, je print l'année $a$, le nombre d'arbres obervés de l'année, le brier score de l'année, le brier score de 2008 à $a$. 

%\includegraphics[height=4.2cm]{fig/validation/brier_score.png}

%Les paramètres sont estimés sur la période 2008-2019. En 2020 et 2021, les scores me semblent pas trop mal. En 2022 par contre, on est bcp plus élevé!
To complete Figures~\ref{fig.chalarose.dynamics} and~\ref{fig.chalarose.dynamics.pred}, we computed the Root Mean Square Error (RMSE) for each year $a$, for $a=2008$ to $a=2023$, given by
\[
\text{RMSE}(a) = \sqrt{\frac{\sum_{i \in I_a}\sum_{k=1}^{\obs_a(i)}(p_a^k(i)-q_a^i)^2}{\sum_{i \in I_a}\obs_a(i)}}
\]
where $p_a^k(i)$ (resp. $q_a^i$) is the observed Bernoulli parameter (resp.  the Bernoulli parameter estimated by the model of the previous section (that is for the set of parameter $\theta^{\max}$ given by \eqref{para.max.like1})).
It allows us to measure the accuracy of the model for each year. In addition, we also compute, for each year $a$, the RMSE of the period from 2008 to the year $a$ given by
\[
\text{RMSE}(2008 \to a)  = \sqrt{\frac{\sum_{y=2008}^a\sum_{i \in I_y}\sum_{k=1}^{\obs_y(i)}(p_y^k(i)-q_y^i)^2}{\sum_{y=2008}^a\sum_{i \in I_y}\obs_y(i)}}.
\]
Contrary to $\text{RMSE}(a)$, it gives a lower weight to the year with a low number of measures on the period.
They are given in Table~\ref{tab.chalarose.RMSE}.
Smaller is the RMSE, better is the prediction. We expect a model that fits better, at least, that a model that gives a purely random prevision. To check this, we simulated random imputations of the $q_a^i$ by a uniform distribution on $[0,1]$ and computed the empirical distributions of the RMSE$(a)$, for $a$ in the period 2008-2023, for this purely random model. 
Simulations of 100000 replica always give RMSE$(a)$ between 0.35 and 0.50. In addition, the first percentiles of the empirical distributions are between 0.42 and 0.52, clearly greater than the ones of our fitted model.
%Simulations of 1000 replica always give RMSE greater than 0.47 for 2020, (resp. 0.42 and 0.40 for 2021 and 2022).

\begin{table}[]
    \centering
\begin{tabular}{|c|c|c|c|}
\hline
Year $a$ & $\sum_{i \in I_a}\obs_a(i)$ & $\text{RMSE}(a)$ & $\text{RMSE}(2008 \to a)$ \\
\hline
2008 & 183 & 0.2222 & 0.2222 \\
2009 & 753 & 0.1465 & 0.1641 \\
2010 & 504 & 0.1898 & 0.1735 \\
2011 & 620 & 0.2387 & 0.1954 \\
2012 & 515 & 0.1940 & 0.1951 \\
2013 & 438 & 0.2621 & 0.2062 \\
2014 & 493 & 0.2739 & 0.2170 \\
2015 & 520 & 0.2720 & 0.2249 \\
2016 & 410 & 0.2437 & 0.2267 \\
2017 & 341 & 0.2956 & 0.2323 \\
2018 & 254 & 0.2484 & 0.2331 \\
2019 & 132 & 0.2207 & 0.2328 \\
2020 & 346 & 0.3071 & 0.2381 \\
2021 & 215 & 0.3286 & 0.2421\\
2022 & 166 & 0.3413 & 0.2455 \\
2023 & 111 & 0.3819 & 0.2487 \\
\hline
\end{tabular}
    \caption{Number of observed parcels in the year $a$, RMSE of the year $a$, and RMSE of the period from 2008 to the year $a$, with $a$ from 2008 to 2023, for the model fitted with data from 2008 to 2019.}
    \label{tab.chalarose.RMSE}
\end{table}

Note that our model is clearly better than the one of purely random prevision. We can also remark that years from 2008 to 2019 with many observations (column $\sum_{i \in I_a}\obs_a(i)$) are well fitted because they have a great contribution in the whole likelihood that as to be maximised.
Remember that data from 2020 to 2023 were not used in the fit, which explains a RMSE that gradually increases during the period, although it remains still clearly lower than the one of a random model. 

%{\color{red}Plutôt que de comparer les RMSE à 0.5, je suggère de les comparer à
%\[
%\sigma(a) = \sqrt{\frac{\sum_{i \in I_a}\obs_a(i)\mathbb{E}\left(\frac{N_a^i}{m}-q_a^i\right)^2}{\sum_{i \in I_a}\obs_a(i)}}
%=\sqrt{\frac{\sum_{i \in I_a}\obs_a(i)\frac{q_a^i(1-q_a^i)}{m}}{\sum_{i \in I_a}\obs_a(i)}}
%\leq \frac{0.5}{\sqrt{m}}=0.0913
%\]
%sauf qu'on est très mauvais. 
%}

\section{Discussion}
\label{sec.discussion}

We developed a complex mechanistic-statistical model describing the expansion of ash dieback, taking into account climate factors (temperature and rainfall), Allee effect, and heterogeneity in ashe tree density. We estimated the parameters of our model by first identifying appropriate ranges for parameters, which led to a model reduction. We then used an Adaptive Multiple Importance Sampling Algorithm for fitting the model. We tested several temperature indices.
The temperature index T28, which corresponds to the number of days in summer with a temperature exceeding 28°C, provided the highest forecast likelihood, in the sense that, for the parameter set maximizing the likelihood on the fitted period 2008-2019, it obtained the best likelihood for the period 2008-2023.
In addition the model with the temperature index T28 leads to good qualitative features. Specifically, for this index, the model well depicts the qualitative propagation behavior that has been observed in France over the past 20 years. In particular the model successfully reproduces the absence of disease spread in southeast France and the limited spread of the disease in southwest France, with a more significant development occurring once the disease reaches the Pyrénées mountains.  
Both the Allee effect and the heterogeneity of the rainfall in spring do not have a significant impact on the overall expansion of \textit{H.~fraxineus} in France and can be neglected in the modeling.  
Similar qualitative behaviors and likelihoods have been obtained with other indices, such as the number of days in summer with temperatures exceeding 24°C, 26°C or 30°C (indices T24, T26 and T30), not allowing selection of one of these indices (index T26 even leads to better likelihood for the fitted period 2008-2019). Nevertheless, these temperature indices lead to the same estimation range for parameters (excluding the parameters of the temperature impact function $f$), then leading to similar conclusions about the model dynamics.
However, our results suggest that the index corresponding to temperatures exceeding 35°C (index T35) should be excluded. This index does not accurately reproduce the propagation dynamics of the data, particularly failing to show the absence of disease in southeast France.

For the selected temperature index T28, the developed model adequately reproduces several key features of the ash dieback dynamics observed in France. First, the dynamics of the inoculum, i.e. the amount of infected rachises in the litter, show a quick and sharp transition from low to high levels within 1-2 years. This very abrupt transition, which has a significant impact on most local stands just a few years after the first observation of the disease in an area, corresponds to the observed dynamics in France \citep{grosdidier_landscape_2020}. 
Then the model well depicts the absence of dieback in many places, although it predicts the complete colonisation of France by \textit{H.~fraxineus}. This is in agreement with the prediction that the pathogen is able to complete its cycle everywhere in France \citep{marcais_ability_2023}. We do have indications that this prediction has some validity. A specific survey in southeast France (Avignon) during 2023 summer was able to locate ash trees with leaf infections by \textit{H.~fraxineus}, although shoot infections were absent (Marçais, unpublished). Ash dieback is still not reported around Avignon in accordance with our model which predicts no presence of disease in the area in 2023. \textit{H.~fraxineus} may thus be present much more widely in SE France than previously thought, but does not appear to induce any damage as predicted by our model. The model also adequately predicts that the impact of ash dieback will be very limited in the Garonne valley. In SW France, the apparent stop of ash dieback spread in 2015 and the colonization of the Pyrenées after 2019 is predicted by the model, even though only data from 2008-19 where used for model calibration. Following the widespread presence reported in the Pyrénées after 2020, the DSF conducted extensive surveys in the Garonne valley and confirmed that although the disease may occasionally be found there, its prevalence and severity remain very limited. Our model outputs are in agreement with the hypothesis of \cite{marcais_ability_2023} that undetected spread through foliar infection with limited induced dieback may explain the arrival in the Pyrénées in 2020, with a jump of about 200 km from the closest previous known disease location.

We thus confirm that summer high temperatures are a major climatic factor that affect the impact and spread of ash dieback, as was shown by \cite{hauptman_temperature_2013}, \cite{grosdidier_higher_2018} and \cite{marcais_ability_2023}. The  parameter $\gamma$ was estimated at 21.5 for the model with the temperature index T28, which means that ash dieback stops to be reported by the DSF in area where temperature exceeds 28°C for more than 21.5 days during the summer in most years. This occurs in France mainly in the SE, although during the period of 2018-2020, it occurred in a much larger area of France. The function $f$ decreases very sharply close to 21.5 from 0.8 to 0 and is thus almost a threshold function ($\kappa$ estimated at 0.056, see Figure~\ref{fig.f_of_T.wrtkappa}). This may be caused by the poor survival of \textit{H.~fraxineus} at temperature above 35°C \citep{hauptman_temperature_2013}. Leaf temperature in the crown may reach value up to 5°C above the air temperature \citep{granier_evidence_2007}. Another cause of this strong threshold is probably the observation process. Ash dieback is reported by the forest health survey system when it reaches a level that impacts stand management. When \textit{H.~fraxineus} induced little shoot mortality because summer temperature are unfavorable, the trees are able to compensate by producing new foliage and do not show significant dieback. Ash dieback is then not reported as a forest health problem. Moreover, in southeast France, the species of ash trees \textit{F.~excelsior} is replaced by \textit{F.~angustifolia} and becomes less economically important in the forests, leading to less thorough observation by the DSF and fewer reports of ash dieback.

A feature taken into account in the model is the local ash density (average local basal area of ash per ha). Host density is recognized as a key factor in explaining the spread of pathogens, particularly invasive parasites that affect forests, either microorganisms or insects \citep{hudgins_predicting_2017, keesing_impacts_2010}. In the model, it scales the production of inoculum throughout the production of infected rachises that fall on the forest floor each autumn, and thus disease transmission. It was demonstrated that population of ash with a high basal area are more severely impacted by \textit{H.~fraxineus} \citep{grosdidier_landscape_2020, chumanova_predicting_2019}. The fast colonization of central France (Massif central) as well as the spread of the disease through a corridor between the massif central and the Pyrenées could be partly driven by high ash density in the area. However, we did not observe a slower spread toward the west of France after 2015, despite a very low ash density in the area. 
The ash density that we used is based on forest inventory data, which only accounts for ash trees in forests. It could inadequately represent ash density in western France where the acreage of forest is low while hedges are very frequent. Ash trees in the hedges are not taken into account by the forest inventory data, resulting in a likely underestimation of the overall ash density in western France.

By contrast, some factors known to affect ash dieback could be neglected in modeling the spread of the disease. It was shown that a component Allee effect significantly affects the population dynamics of \textit{H.~fraxineus} and is induced by limited encounters of sexual partners (i.e. the mating types) at low density of colonized ash rachises in the forest litter \citep{laubray_evidence_2023}. \cite{hamelin_mate_2016} showed that an Allee effect induced by a mating limitation significantly reduces the spread rate of \textit{Mycosphaerella fijiensis}. We did not find any evidence that this is the case for ash dieback. The Allee effect observed for \textit{H.~fraxineus} is a weak Allee effect, the production of ascospores remaining positive even at very low density of the colonized rachises. This may explain the limited impact of Allee effect on the spread of ash dieback. Another factor that could be neglected is the impact of rainfall.
\textit{H.~fraxineus} production of apothecia and ascopores has been shown to be positively related to spring and summer rainfall \citep{havrdova_environmental_2017, chumanova_predicting_2019, dvorak_vertical_2023} and we therefore expected that it would significantly affect the disease spread. This is not the case. A possibility is that rainfall in France during spring seldom hampers the pathogen sporulation completely. 
In fact, Figures~\ref{fig.spores.dynamics} and \ref{fig.saturatedspores.dynamics} suggest that, due to the saturation effect, the impact of the rainfall heterogeneity would have played a role only on the propagation front.
\cite{marcais_ability_2023} showed that rainfall enables the fulfillment of \textit{H.~fraxineus} cycle in all French regions, even in the southeast where the disease is not observed. It remain possible that spread of \textit{H.~fraxineus} will be reduced in mediteranean areas drier than southeast France. However, in these areas the period of apothecia production is likely to start earlier, in late winter or early spring, during a rainy enough period for the pathogen's cycle to be fulfilled. In the Pyrénées, apothecia were produced as early as May 1st \citep{marcais_ability_2023}. Also, the pathogen may be able to produced enough apothecia for efficient dispersal in favorable wet sites even during dry springs.  

In conclusion, we have proposed a model that well describes the dynamics of dieback symptoms caused by \textit{H.~fraxineus} in France. Our numerical analysis suggests that the density of ash trees and high temperatures, particularly the number of days in summer with temperatures exceeding 26-28°C, play significant roles in the spread of the pathogen and the development of dieback symptoms.
Additionally, the statistical estimation of the model parameters has allowed us to quantify the impact of these local environmental parameters, as well as the global propagation parameters such as the diffusion parameter $D$ and the reproduction parameter $\beta_0$.
In future research, we plan to further investigate the local speed of propagation in the reaction-diffusion model, which is dependent on the local ash tree density.

\appendix

\section{First rough estimation and model reduction}
\label{sec:appendix.first.estimation}

We present in this appendix the first estimation of the parameters leading to the model reduction \eqref{eq:model_reduction} as well as the initialization of Algorithm~\ref{algo.AMIS}. 
This step have been repeated for several temperature indices, however we show here the numerical results only for the temperature index T28 (\textit{i.e.} $T_a^i$ is the number of days of July and August of the year $a$ for which the temperature exceeds 28$^{\circ}$C in the quadrat $\omega_i$), which gives the best qualitative and quantitative behaviors (see Section~\ref{sec:comparison_temp} for a comparison of the model behavior for the different temperature indices).

\subsection{A Metropolis-Hasting algorithm with iteration dependent proposal distribution}
\label{sec:MH}

In order to obtain a first rough estimation of the parameters, we realized ten runs of a Metropolis-Hasting algorithm with 
a proposal distribution depending on the iteration step 
%iteration dependent proposal distribution 
(see Algorithm~\ref{algo.MH.separated_parameters}) initialized with different initial conditions.
At each iteration only one of the parameters $D,\beta_0,\beta_1,r,\gamma,\kappa,S,\Cinit,\Cpers$ is changed (we remind that $r_S$ is set to 1000, see Section~\ref{sec.identifiability}) according to a Gamma distribution with a \textit{not too small} variance.  
Testing parameters one by one allows to not reject a bloc of new parameters because only one parameter is not adequate and, combined with the large variances of the proposal distribution, allows users to capture a fast convergence of parameters which are not strongly correlated to the others ones towards the `main support' of the posterior distribution.
More precisely, for $i\in\llbracket 1, 9 \rrbracket$, at the $k=9\ell+i$-th iteration, and setting 
$\theta=(D,\beta_0,\beta_1,r,\gamma,\kappa,S,\Cinit,r_S,\Cpers) = (\theta_1,\dots,\theta_8,r_S,\theta_{9})$
then  
\begin{align}
\label{proposal.dist.MH}
\tilde \theta = (\tilde \theta_1,\dots,\tilde \theta_8,r_S,\tilde \theta_{9}) \sim Q^k(.|\theta) \Longleftrightarrow
\begin{cases}
	\tilde \theta_i \sim \Gamma\left(\lambda, \frac{\theta_i}{\lambda}\right) \\
	\tilde \theta_j = \theta_j, \,  \forall j \neq i
\end{cases}
\end{align}
and $q^k(\tilde \theta|\theta)$ is the density of the gamma distribution
 $\Gamma\left(\lambda, \frac{\theta_i}{\lambda}\right)$ with scale parameter $\lambda$ and shape parameter $\theta_i/\lambda$,  that is with mean $\theta_i$ and variance $\theta_i^2/\lambda$.
Moreover, without prior information,
we choose for $\tilde \pi$ a uniform distribution on $[0,\bar D]\times [0,\bar \beta_0]\times [0,\bar \beta_1]  \times [0,\bar r] \times [0,\bar \gamma]
	\times [0,\bar \kappa] \times [0,\bar S] \times [0,\widebar \Cinit]\times \{1000\} \times [0, 1]$
 with $\bar D$, $\bar \beta_0$, $\bar \beta_1$, $\bar r$, $\bar \gamma$, $\bar \kappa$, $\bar S$ and $\widebar \Cinit$ sufficiently large such that parameters don't reach the upper bounds of there domains.

\medskip

%--------------------------
\begin{algorithm}
\begin{center}
\begin{minipage}{9cm}
\small
\begin{algorithmic}
\STATE initialization of $\theta^0$% = (D^0,\beta_0^0,\beta_1^0, r^0, \gamma^0, \kappa^0, S^0, \Cinit^0, r_{symp},\Cpers)$
\STATE $k\leftarrow 0$
\WHILE {$k\leq N$}
  \STATE $\hat \theta \sim Q^k(. \, | \, \theta^k)$ 
  \COMMENT{new parameter to be tested}
   \STATE $\delta \leftarrow 
  		\frac{\L(\hat \theta)\,\tilde\pi(\hat \theta)\, q^k(\theta^k\, | \, \hat \theta)}
  		{\L(\theta^k)\,\tilde\pi(\theta^k)\, q^k(\hat \theta \, | \, \theta^k)}$ 
  \STATE $u\sim U[0,1]$ 
  \IF {$u\leq \delta$}
      \STATE $\theta^{k+1} \leftarrow \hat \theta$
      \COMMENT{acceptation of the new parameter}
  \ELSE 
  	  \STATE $\theta^{k+1} \leftarrow \theta^k$
     \COMMENT{rejection of the new parameter}
  \ENDIF
  \STATE $k\leftarrow k+1$
\ENDWHILE
\end{algorithmic}
\end{minipage}
\end{center}
\caption{Metropolis-Hastings Algorithm with a proposal distribution depending on the iteration step.}
\label{algo.MH.separated_parameters}
\end{algorithm}
%--------------------------

\begin{figure}
\captionsetup[subfigure]{justification=centering}
\begin{center}
\begin{subfigure}{0.32\textwidth}
\includegraphics[height=3.6cm, trim = 0cm 0cm 0cm 0cm, clip=true]{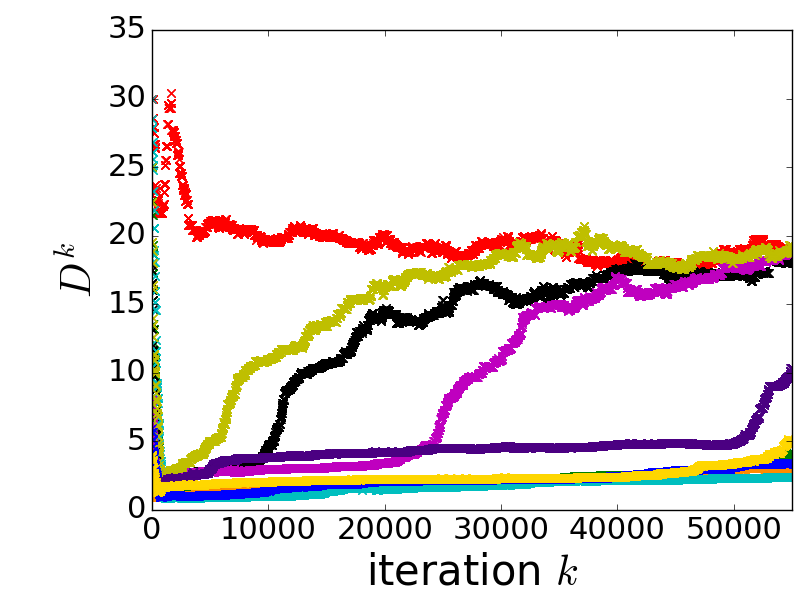}
\caption{\label{fig.MH.D}Evolution of $D$.\\ \strut}
\end{subfigure}
\begin{subfigure}{0.32\textwidth}
\includegraphics[height=3.6cm, trim = 0cm 0cm 0cm 0cm, clip=true]{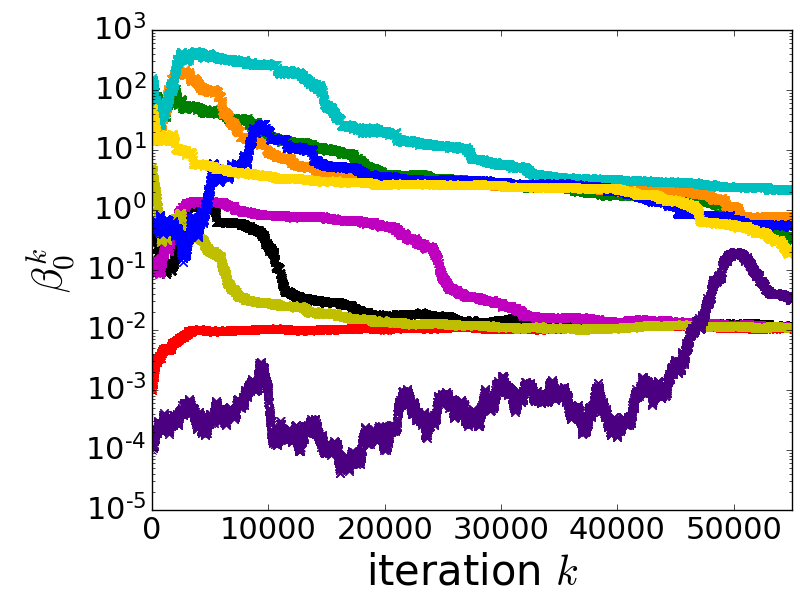}
\caption{\label{fig.MH.beta0}Evolution of $\beta_0$.\\ \strut}
\end{subfigure}
\begin{subfigure}{0.32\textwidth}
\includegraphics[height=3.6cm, trim = 0cm 0cm 0cm 0cm, clip=true]{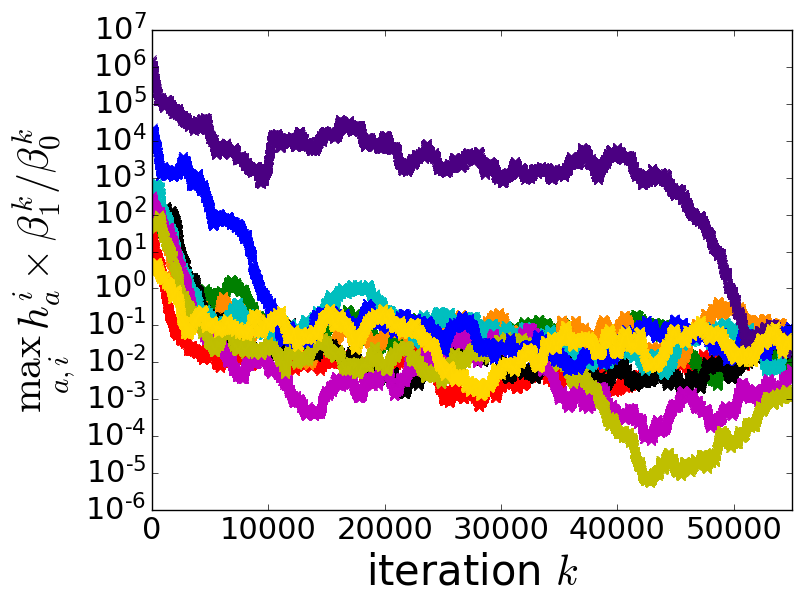}
\caption{\label{fig.MH.beta1_hmax_beta0}Evolution of $\max_{a,i} h_a^i\,{\beta_1}/{\beta_0}$.}
\end{subfigure}
\begin{subfigure}{0.32\textwidth}
\includegraphics[height=3.6cm, trim = 0cm 0cm 0cm 0cm, clip=true]{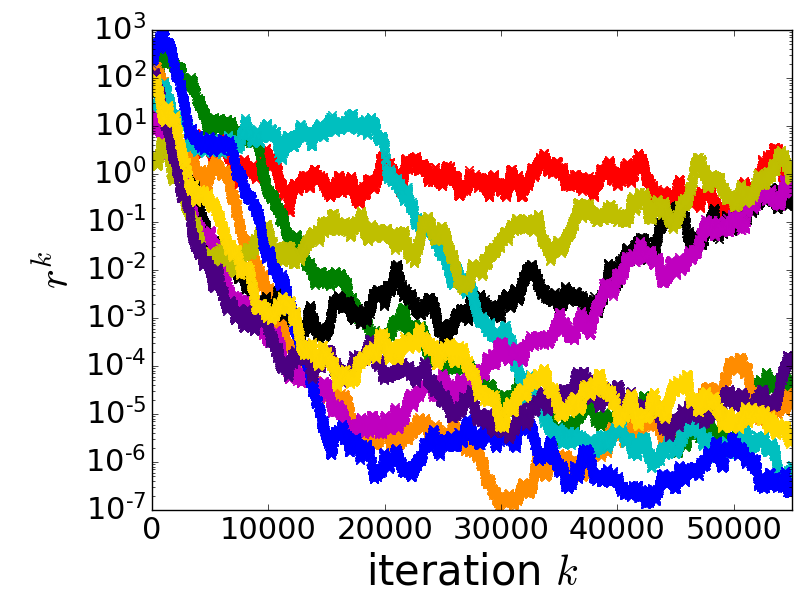}
\caption{\label{fig.MH.r}Evolution of $r$.}
\end{subfigure}
\begin{subfigure}{0.32\textwidth}
\includegraphics[height=3.6cm, trim = 0cm 0cm 0cm 0cm, clip=true]{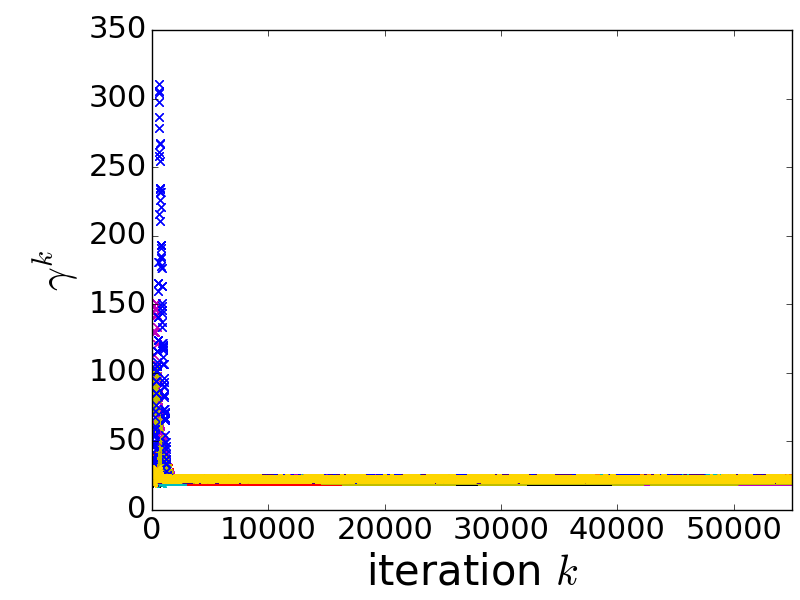}
\caption{\label{fig.MH.gamma}Evolution of $\gamma$.}
\end{subfigure}
\begin{subfigure}{0.32\textwidth}
\includegraphics[height=3.6cm, trim = 0cm 0cm 0cm 0cm, clip=true]{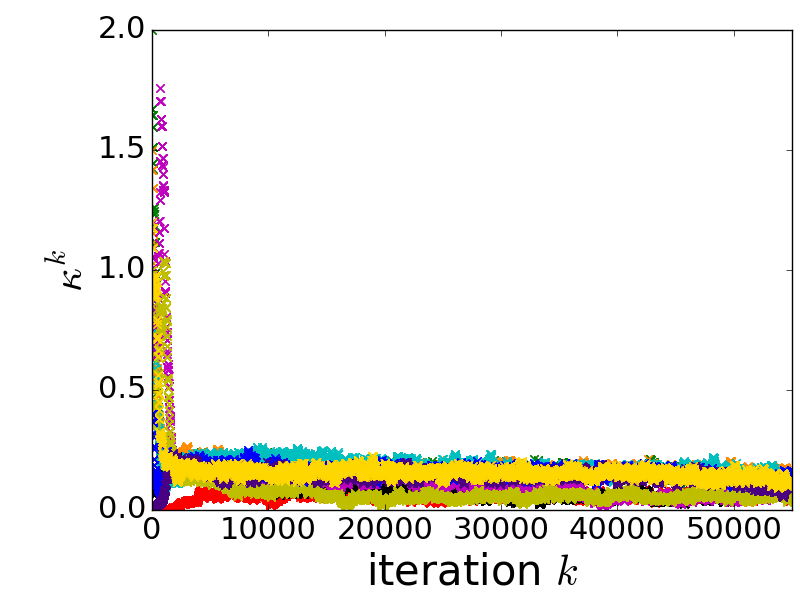}
\caption{\label{fig.MH.kappa}Evolution of $\kappa$.}
\end{subfigure}
\begin{subfigure}{0.32\textwidth}
\includegraphics[height=3.6cm, trim = 0cm 0cm 0cm 0cm, clip=true]{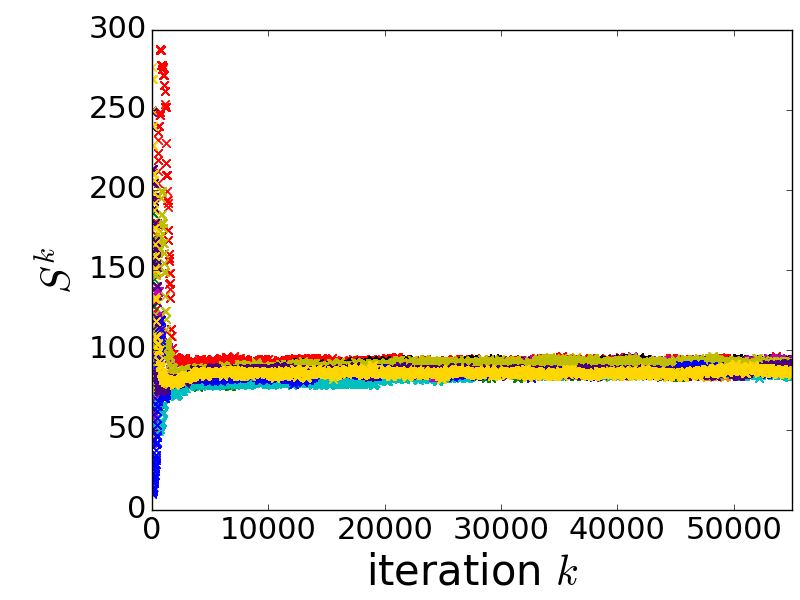}
\caption{\label{fig.MH.S}Evolution of $S$.}
\end{subfigure}
\begin{subfigure}{0.32\textwidth}
\includegraphics[height=3.6cm, trim = 0cm 0cm 0cm 0cm, clip=true]{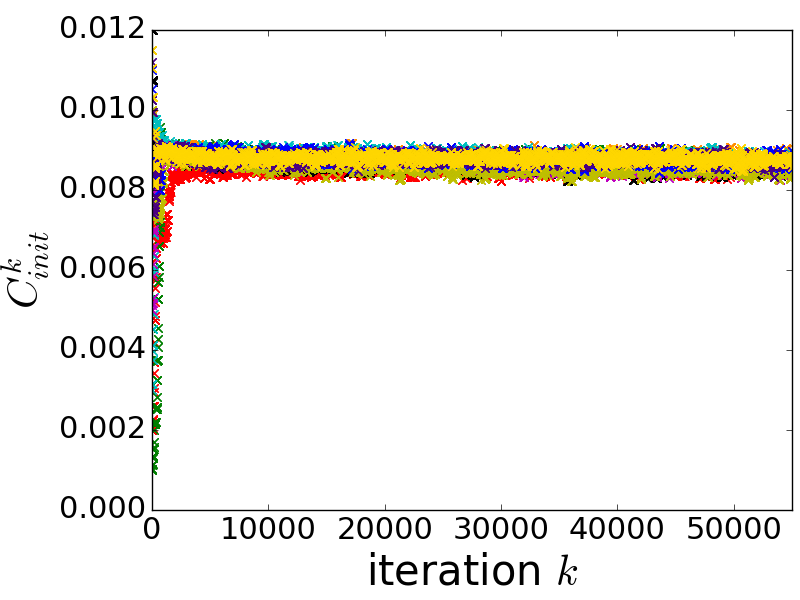}
\caption{\label{fig.MH.Cinit}Evolution of $\Cinit$.}
\end{subfigure}
\begin{subfigure}{0.32\textwidth}
\includegraphics[height=3.6cm, trim = 0cm 0cm 0cm 0cm, clip=true]{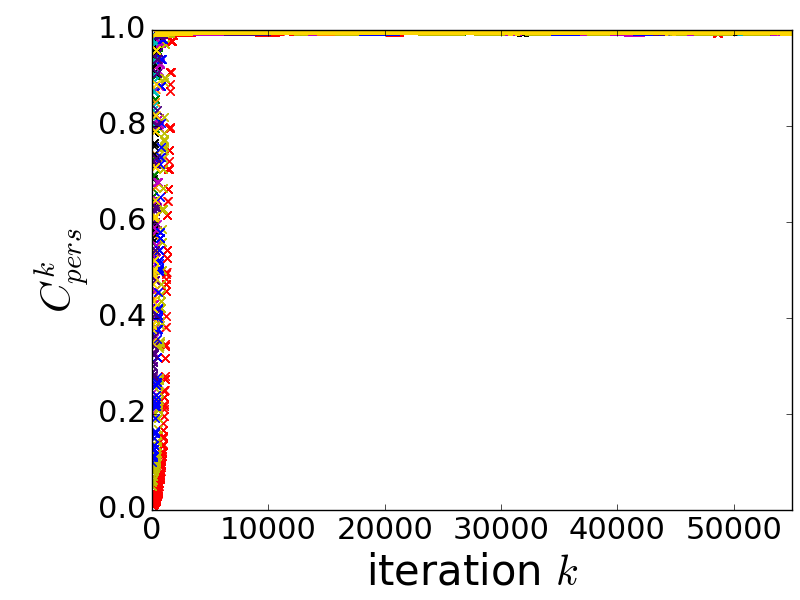}
\caption{\label{fig.MH.Cpers}Evolution of $\Cpers$.}
\end{subfigure}
\begin{subfigure}{0.32\textwidth}
\includegraphics[height=3.6cm, trim = 0cm 0cm 0cm 0cm, clip=true]{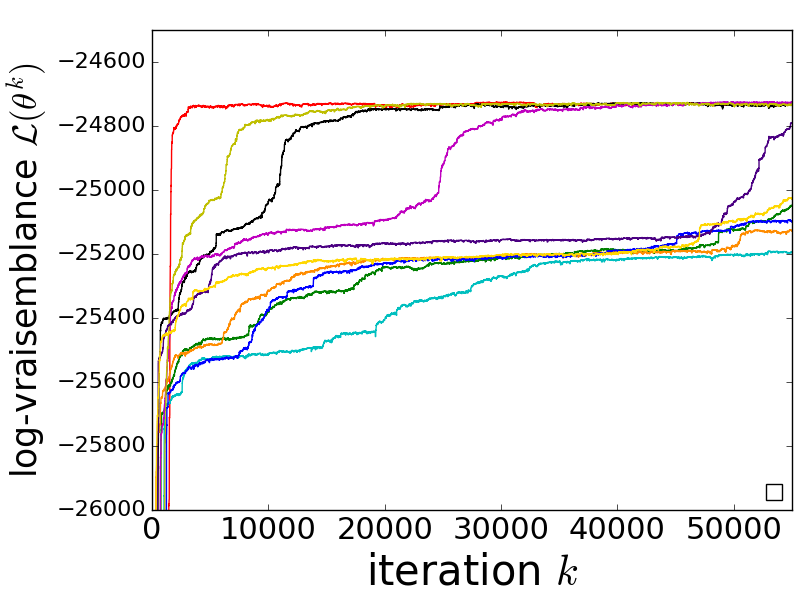}
\caption{Evolution of the log-likelihood.}
\end{subfigure}
\end{center}
\caption{\label{fig.parameters.convergence} Dynamics of the parameters and the log-likelihood for the Metropolis-Hastings Algorithm \ref{algo.MH.separated_parameters},  with the iteration dependent proposal distribution \eqref{proposal.dist.MH} with $\lambda=200$, starting from 10 different initial parameters $\theta^0$. Note that Figures~\ref{fig.MH.beta0}, \ref{fig.MH.beta1_hmax_beta0} and \ref{fig.MH.r} are represented on a log-scale.}
\end{figure}

Figure~\ref{fig.parameters.convergence} shows the evolution of the parameters computed by Algorithm~\ref{algo.MH.separated_parameters},  with the iteration dependent proposal distribution \eqref{proposal.dist.MH} with $\lambda=200$, starting from ten different initial parameters.
Notice that the initial prevalence shape given by $\Psi_{2008}$ and $\Psi_{2009}$ constraints some parameters. In fact, the conditions $q_{2008}<1$ and $q_{2009}<1$ combined to Eq.~\eqref{eq.initialisation} imply that $\Cinit<0.0125$. In addition, the initially non infected quadrats (i.e. such that $\Psi^i_{2009}=0$) with positive dieback observation in 2010 (i.e. $p^i_{2010}>0$) have to satisfy $q^i_{2010}=\tilde q^i_{2010}>0$ (see Eq.~\eqref{eq.q}) which implies, for the temperature index T28, that $\gamma>19.5$.

For five parameters, the convergence is very quick and seems independent from the four other parameters. In fact, we observe a fast convergence of the probability of the symptom persistence  $\Cpers$ towards 1 (Figure~\ref{fig.MH.Cpers}). 
In addition, the parameters $\gamma$,  $\kappa$, $S$ and $\Cinit$ quickly converge around fixed values or in restricted ranges of values (Figures~\ref{fig.MH.gamma}, \ref{fig.MH.kappa}, \ref{fig.MH.S} and \ref{fig.MH.Cinit}).

The parameter $\beta_1$ reaches values such that the contribution of the term $\beta_1\,h_a$ in $H_a$ defined by \eqref{eq.H_a} is small with respect to $\beta_0$. In fact, Figure~\ref{fig.MH.beta1_hmax_beta0} represents the evolution of $\max_{a,i}h_a^i\frac{\beta_1}{\beta_0}$, where $\max_{a,i}h_a^i=322.2$ is the maximum of the local rainfall in June in France between 2008 and 2019 (i.e. the period of data used for the estimation). This extreme value is reached in the South East of France in 2010 (see Figure~\ref{fig.pluv.2010}). In Figure~\ref{fig.MH.beta1_hmax_beta0}, all runs reach values below 0.06 suggesting that $\beta_1\,h_a$ is neglectable with respect to $\beta_0$.

Even if a range of values of parameters $D$, $\beta_0$ and $r$ seems to emerge for the runs with the best likelihoods, the convergence of these parameters is not clear. A rough estimation of these parameters is done in the next session.  

\medskip

Finally this step allows to reduce the model setting $\Cpers=1$ and $\beta_1=0$ and to identify suitable zones for the parameters $\gamma$,  $\kappa$, $S$ and $\Cinit$.

\subsection{Log-likelihood on a coarse grid}
\label{sec:coarse_grid}

Following the model reduction given by the previous section, we then compute the log-likelihood on a coarse grid in order to catch a suitable zone for the parameter initialization of the AMIS algorithm~\ref{algo.AMIS}. 
Without idea of the scale of the parameters (particulary for $\beta_0$ and $r$), we used a large coarse grid that we refined toward the identification of the best zone of parameters. 

\begin{figure}
\captionsetup[subfigure]{justification=centering}
\begin{center}
\begin{subfigure}{0.4\textwidth}
\includegraphics[height=4.2cm]{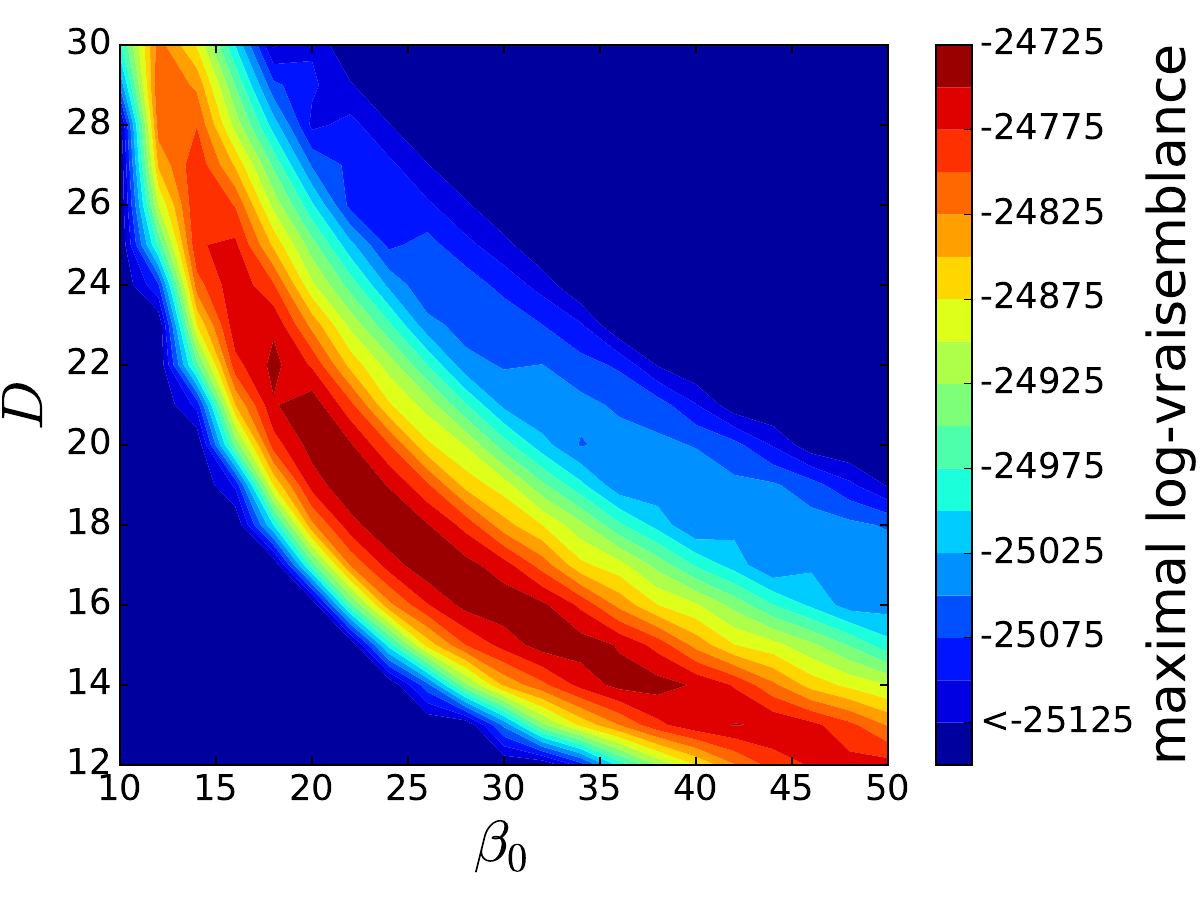}
\caption{\label{fig.grid.D.beta0.likeli}Maximal log-likelihood \eqref{eq.max.loglikeli}.}
\end{subfigure}
\begin{subfigure}{0.4\textwidth}
\includegraphics[height=4.2cm]{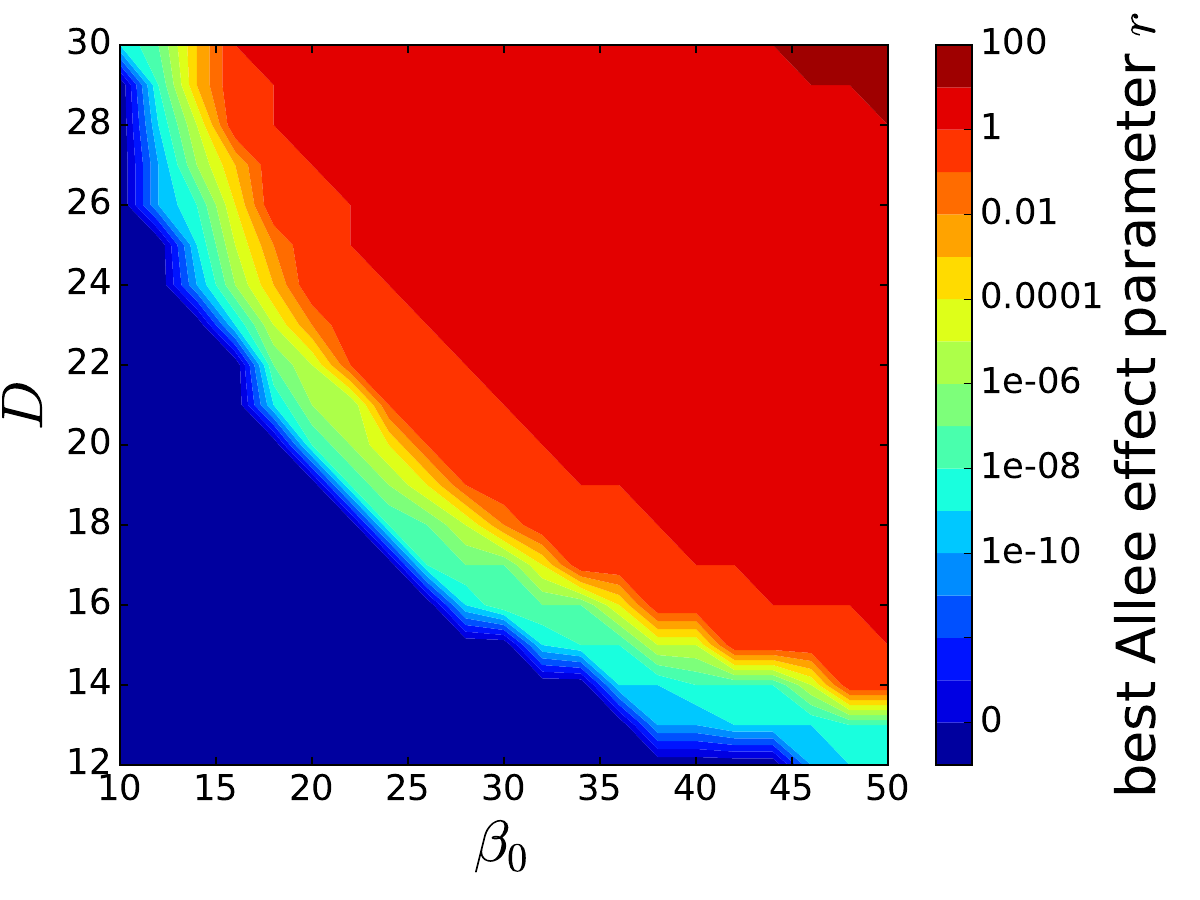}
\caption{\label{fig.grid.D.beta0.best_r}Best Allee effect parameter $r$.}
\end{subfigure}
\end{center}
\caption{\label{fig.grid.D.beta0} Maximal log-likelihood \eqref{eq.max.loglikeli} (left) and value of the Allee effect parameter $r$ leading to this maximal log-likelihood (right) for the coarse grid \eqref{eq:choise_coarse_grid}, $\beta \in \{10, 12,\dots, 50\}$, 
$D \in \{12, 13, \dots, 30\}$ and with the model reduction $\beta_1=0$, $\Cpers = 1$ and $r_S=1000$.}
\end{figure}

Figure~\ref{fig.grid.D.beta0.likeli} represents, for each fixed parameters $\beta_0$ and $D$ on a (refined) coarse grid, the maximal log-likelihood computed for different values of $r$, $\gamma$, $\kappa$, $S$ and $\Cinit$, that is
\begin{equation}
\label{eq.max.loglikeli}
	\max_{r\in \mathcal{I}_r; \gamma \in \mathcal{I}_\gamma; \kappa\in \mathcal{I}_\kappa; S \in \mathcal{I}_S; \Cinit \in \mathcal{I}_{\Cinit}}\{\mathcal{L}(D,\beta_0,\beta_1=0,r,\gamma,\kappa,S,\Cinit,r_S=1000,\Cpers=1)\}
\end{equation}
where 
\begin{align}
\nonumber
\mathcal{I}_r&=\{0, 10^{-10}, 10^{-9} \dots, 10^{4}, 10^{5}\},&
\mathcal{I}_\gamma&=\{21.001, 21.501, \dots, 23.001\},\\
\nonumber
\mathcal{I}_\kappa&=\{0.025, 0.05, \dots,0.25\},&
\mathcal{I}_S&=\{80, 90, 100\},\\
\label{eq:choise_coarse_grid}
\mathcal{I}_{\Cinit}&=\{0.008, 0.0085, 0.009\}
\end{align}
are the discretisation points on the coarse grid for the $r$, $\gamma$, $\kappa$, $S$ and $\Cinit$ components respectively. 
Note that the chose of the set $\mathcal{I}_\gamma$ is due to some irregularities of the log-likelihood in the variable $\gamma$ (see Section~\ref{sec:degenerated_dist_gamma}). 

The maximal log-likelihood on the coarse grid equals -24727.26 and is obtained for the set of parameters
\[
\begin{matrix}
\label{para.max.like2}
D=18 ; & \beta_0 = 24  ;  & \kappa =  0.05 ; & \gamma = 21.501\\
S = 90 ; & \Cinit=0.0085 ; & r=10^{-8}.
\end{matrix}
\]
Moreover, we identify a banana shape zone for the choice of the parameters $\beta_0$ and $D$ (see Figure~\ref{fig.grid.D.beta0.likeli}).

Figure~\ref{fig.grid.D.beta0.best_r} represents the value of the Allee effect parameter $r$ leading to this maximal log-likelihood, that is the argument of the maximum in $\mathcal{I}_r$ of the maximal log-likelihood~\eqref{eq.max.loglikeli}. Note that this figure is represented on a kind of log-scale.
We observe that the previous banana shape zone is included in the region where the Allee effect parameter $r$ is less than 0.0001 and negligeable. In fact
although the maximal value of the log-likelihood is obtained for $r=10^{-8}$, setting $r=0$ insteed of $10^{-8}$ makes the log-likelihood goes from -24727.26054 to -24727.26067.

\begin{figure}
\captionsetup[subfigure]{justification=centering}
\begin{center}
\begin{subfigure}{0.19\textwidth}
\includegraphics[height=2.8cm, trim = 0cm 0cm 4.9cm 0cm, clip=true]{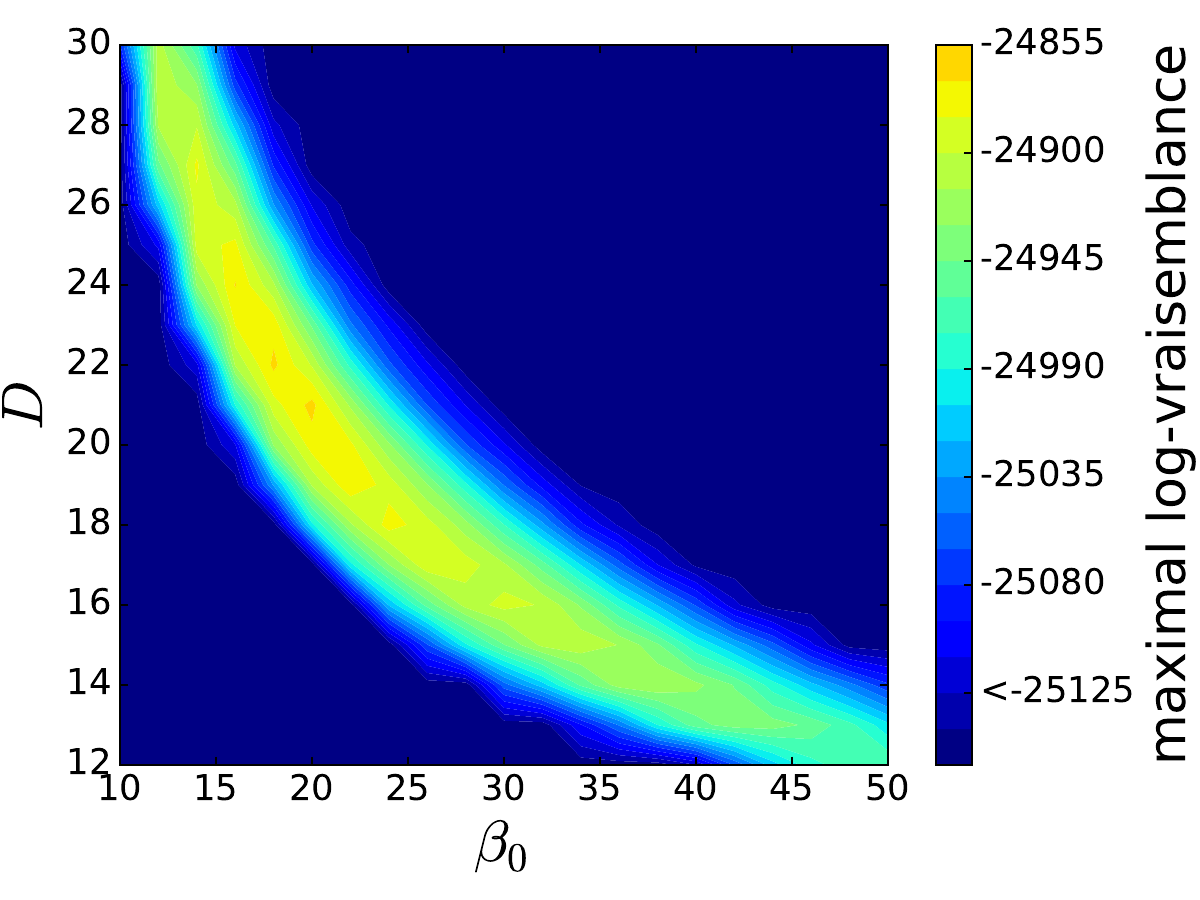}
\caption{$v_\gamma=21.001$.}
\end{subfigure}
\begin{subfigure}{0.19\textwidth}
\includegraphics[height=2.8cm, trim = 0cm 0cm 4.9cm 0cm, clip=true]{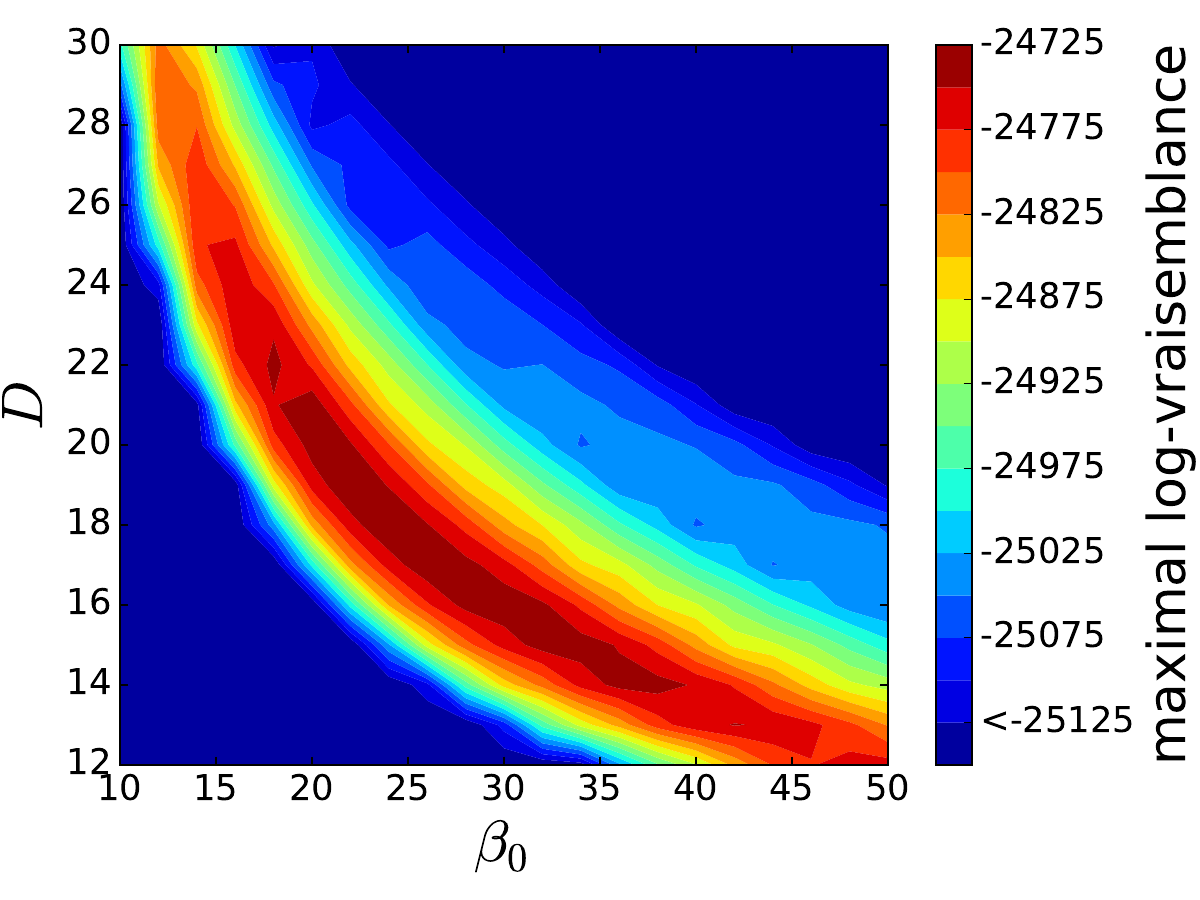}
\caption{$v_\gamma=21.501$.}
\end{subfigure}
\begin{subfigure}{0.19\textwidth}
\includegraphics[height=2.8cm, trim = 0cm 0cm 4.9cm 0cm, clip=true]{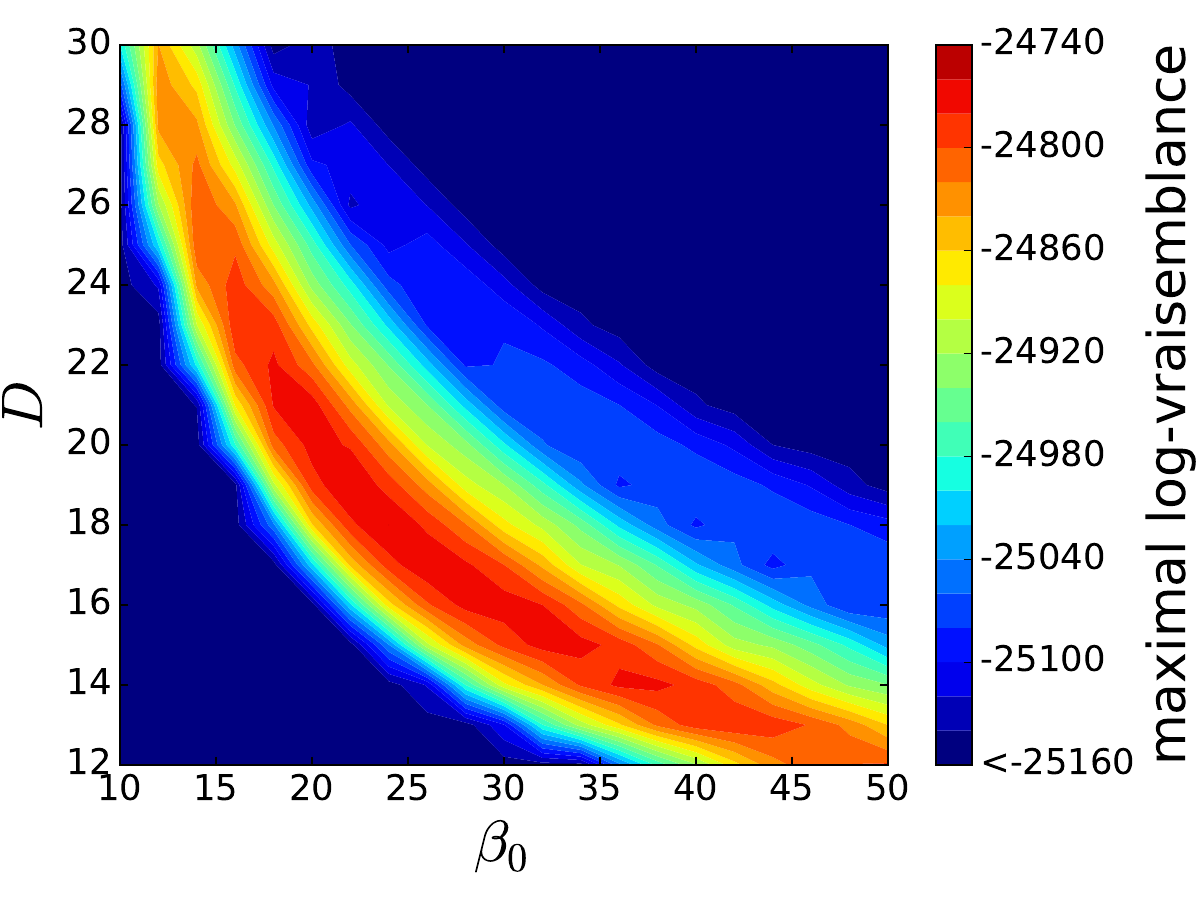}
\caption{$v_\gamma=22.001$.}
\end{subfigure}
\begin{subfigure}{0.19\textwidth}
\includegraphics[height=2.8cm, trim = 0cm 0cm 4.9cm 0cm, clip=true]{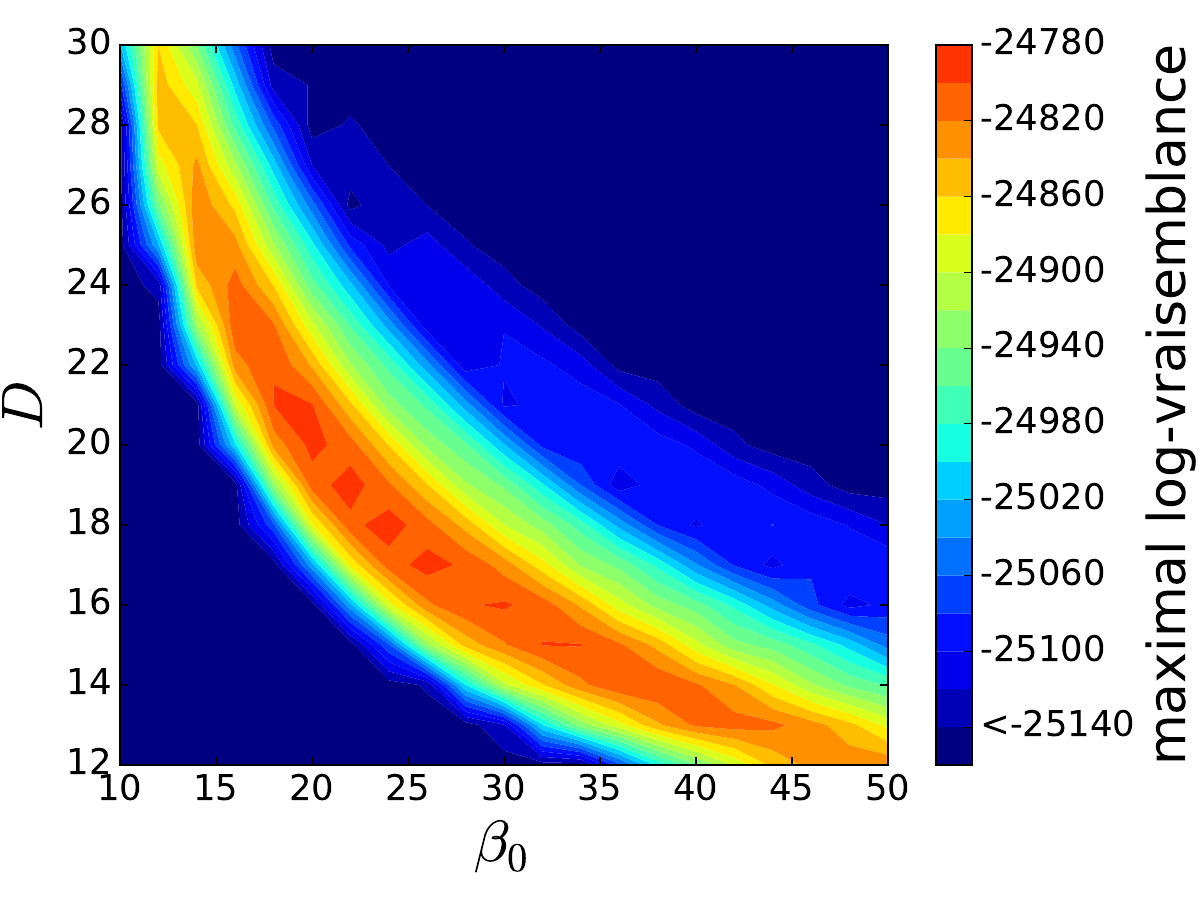}
\caption{$v_\gamma=22.501$.}
\end{subfigure}
\begin{subfigure}{0.2\textwidth}
\includegraphics[height=2.8cm]{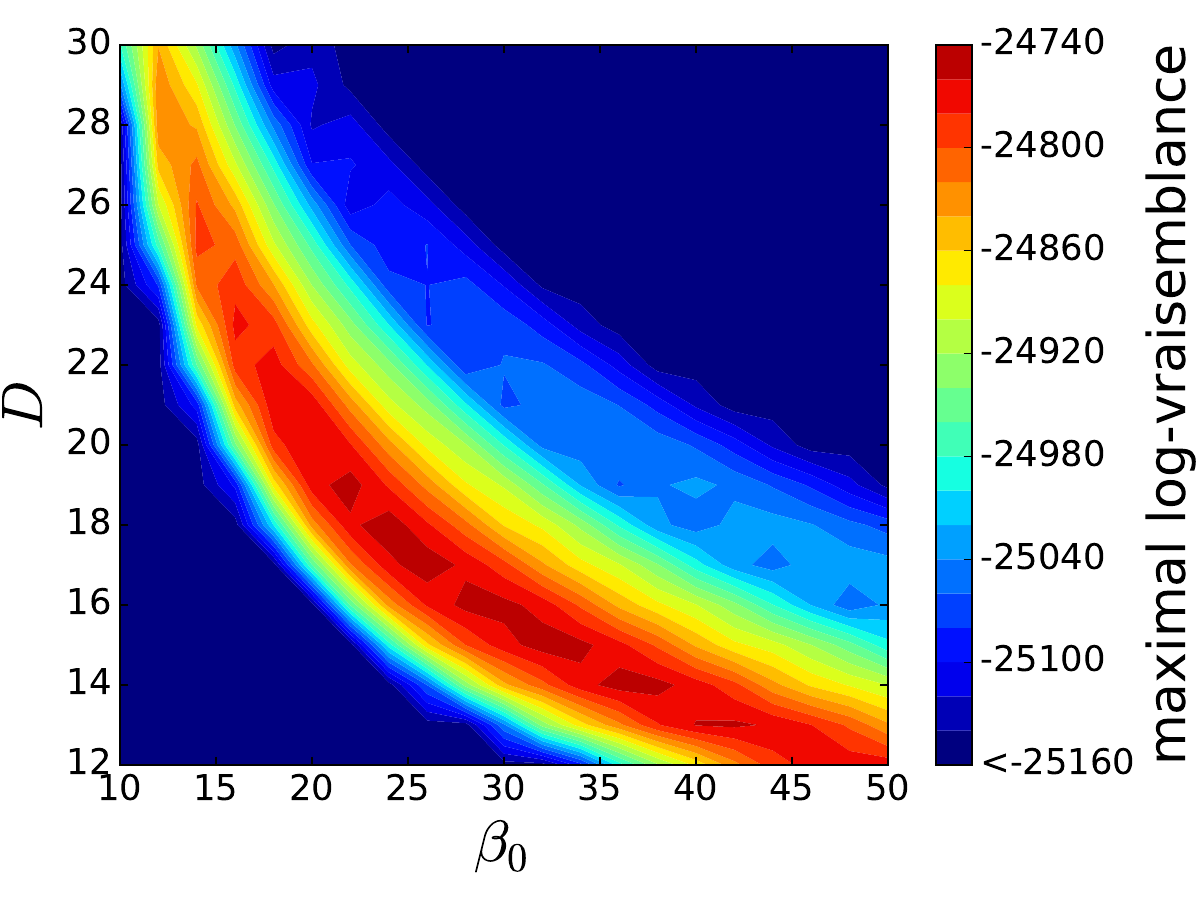}
\caption{$v_\gamma=23.001$.}
\end{subfigure}
\end{center}
\caption{\label{fig.loglikeli_gamma_fixed}Maximal log-likelihood \eqref{eq.max.loglikeli} for the coarse grid \eqref{eq:choise_coarse_grid} except for the parameter $\gamma$ which is fixed to the value $v_\gamma$ and with the model reduction $\beta_1=0$, $\Cpers = 1$ and $r_S=1000$.}
\end{figure}

Figure~\ref{fig.loglikeli_gamma_fixed} represents the maximal log-likelihood with respect to $\beta_0$ and $D$, as in Figure~\ref{fig.grid.D.beta0.likeli}, but for different fixed values of $\gamma$. We can notice that the maximal likelihood is non-monotonic in $\gamma$.
However, we identify that the best likelihood is for $\gamma \approx_{>} 21.5$.

Finally, this step leads  to the model reduction $r=0$. Moreover, we identified an initialisation zone for the parameters $D$ and $\beta_0$ as well as refined zone for the initialisation of $\gamma$.

\subsection{An almost degenerated posterior distribution for $\gamma$}
\label{sec:degenerated_dist_gamma}

Further to the reduction model and the first rough estimations of parameters induced by the steps of Sections~\ref{sec:MH} and \ref{sec:coarse_grid}, we run an AMIS algorithm for the parameters estimations (see \cite{cornuet2012a}).
We do not give details here on the parameters of the algorithm, however we refer to Section~\ref{sec:AMIS} where the algorithm and the parameters are given for the final estimation step (after the reduction model given in this section).

Even if we observe a fast convergence of the support of the one-dimensional posterior distributions as well as their rough shapes, the stabilisation of the distributions was not observed after 150 runs of the AMIS algorithm. The slow convergence is due, in particular, to the almost degenerated and non-symmetric form of the  posterior distribution for the parameters $\gamma$ which weights values very close to (but larger than) $21.5$. % (see Figure \ref{fig.posterior.gamma.Njs28}).
In fact sampled parameters, generated by normal proposal distributions by the AMIS algorithm, contain a large number of ``non appropriated'' parameters.

%\begin{figure}
%\begin{center}
%\includegraphics[height=4.5cm]{fig/AMIS_NjS28_230209094431/loi_gamma.pdf}
%\caption{\label{fig.posterior.gamma.Njs28}Posterior distribution of the parameter $\gamma$ estimated by an AMIS algorithm.}
%\end{center}
%\end{figure}

\medskip

\begin{figure}
\captionsetup[subfigure]{justification=centering}
\begin{center}
\begin{subfigure}{0.4\textwidth}
\includegraphics[height=4.5cm]{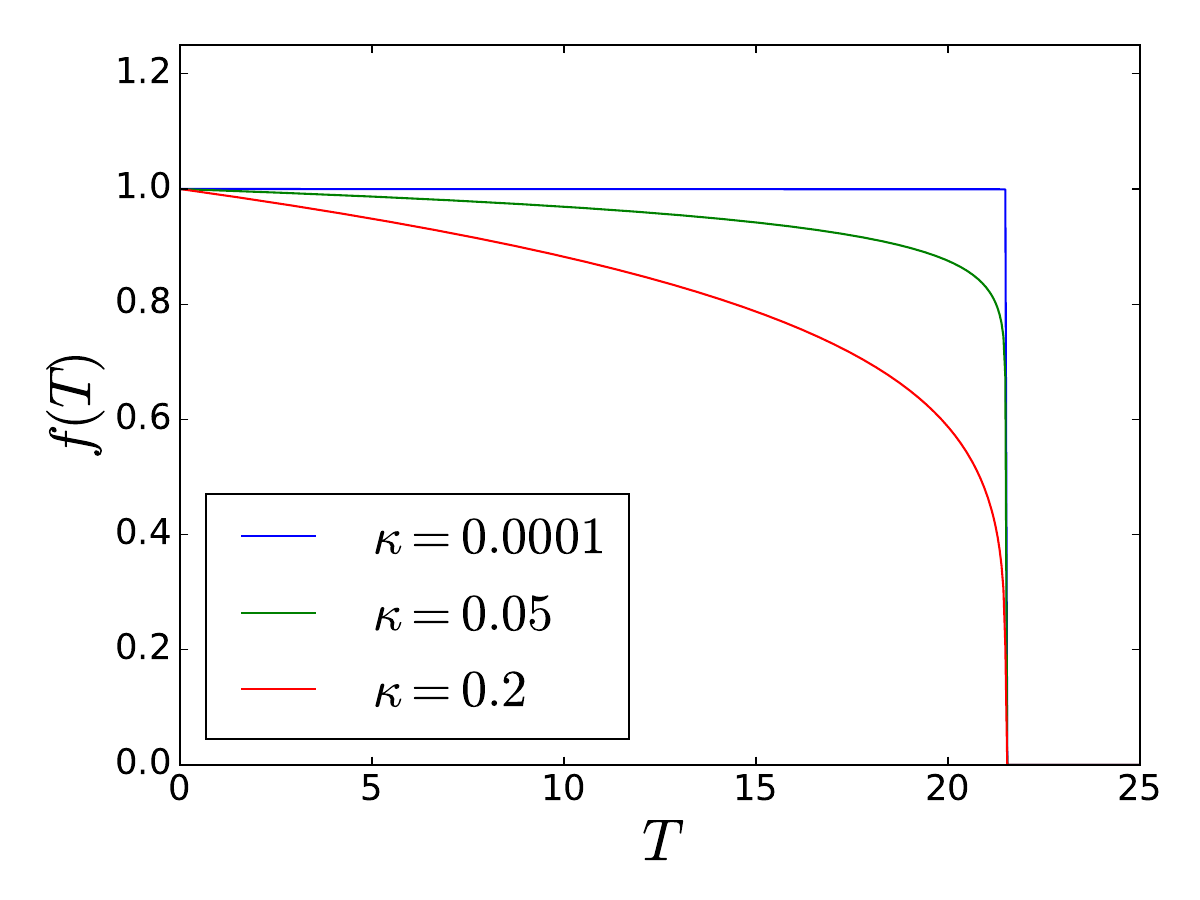}
\caption{\label{fig.f_of_T.wrtkappa}Variation of $f$ with respect to $\kappa$ ($\gamma=21.501$ fixed).}
\end{subfigure}
\begin{subfigure}{0.4\textwidth}
\includegraphics[height=4.5cm]{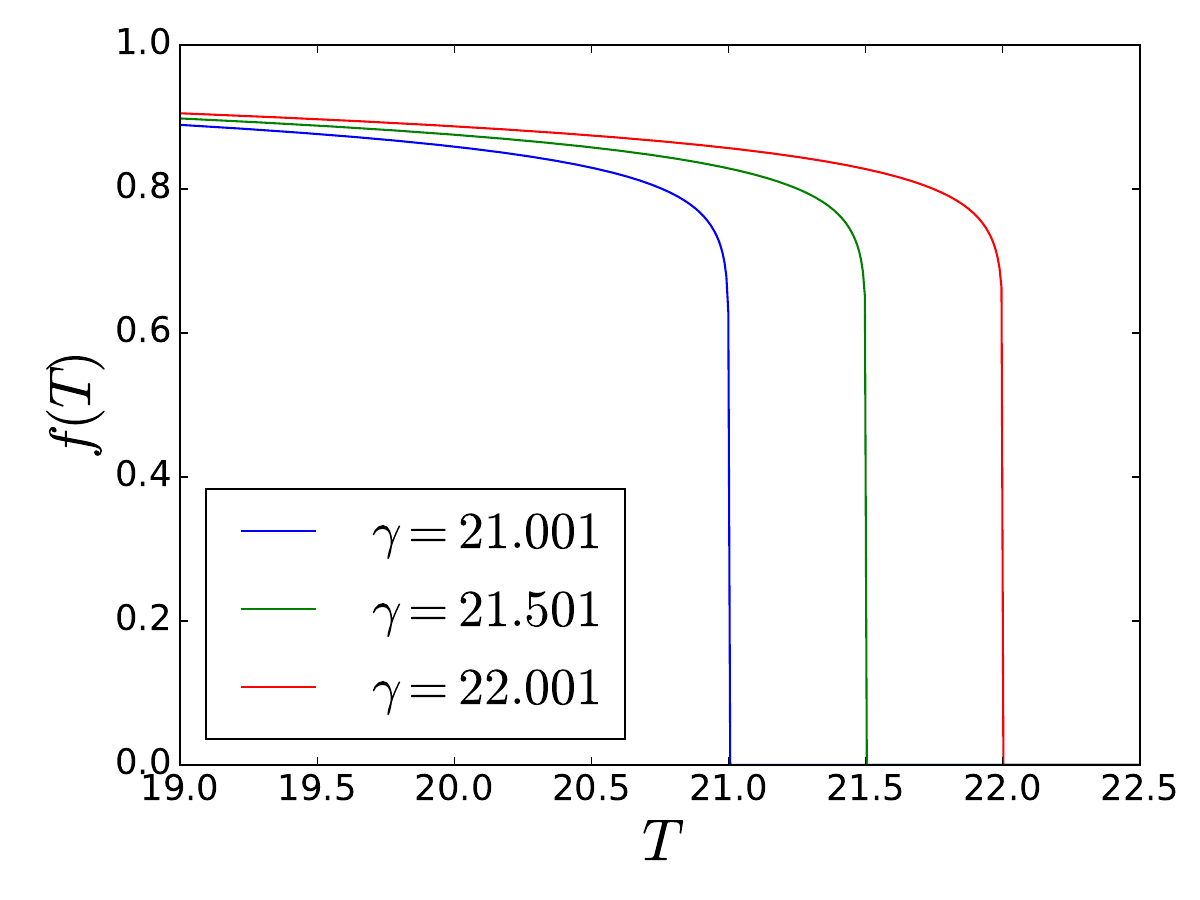}
\caption{Variation of $f$ with respect to $\gamma$ ($\kappa=0.05$ fixed).}
\end{subfigure}
\end{center}
\caption{\label{fig.f_of_T}Function of the temperature impact $f$ defined by \eqref{eq.fct_temp} for different values of $\gamma$ and $\kappa$.}
\end{figure}

The non regularity of the posterior distribution of the parameter $\gamma$ in 21.5 is due to the form of the function of the temperature impact $f$ given by \eqref{eq.fct_temp}. 
In Figure~\ref{fig.MH.kappa}, we observe that the parameter $\kappa$ converges toward small values (less than 0.2), and in fact around 0.05 after convergence in the best zone (see Section~\ref{sec:coarse_grid} and runs of Figure~\ref{fig.MH.kappa} with the best log-likelihoods). Then, for the best zone of parameters, the function of the temperature impact $T\mapsto f(T)$ is abrupt close to $T=\gamma$ (see Figure~\ref{fig.f_of_T}).

Moreover, due to the discrete nature of the temperature data, which are a mean of four entire values (see Section~\ref{append:data}), the log-likelihood contains some irregularities in these discrete values, in particular in $\gamma=21.5$.

\medskip

By the previous arguments (almost degenerated posterior distribution and best log-likelihood for $\gamma$ very close to, but strictly larger than $21.5$), we reduce the model setting $\gamma=21.50001$.

\section{Convergence of the AMIS algorithm}
\label{sec:converenge_AMIS}

We discuss, in this section, the convergence of the AMIS Algorithm~\ref{algo.AMIS} which led to the numerical estimation of $(D,\beta_0,\kappa,S,\Cinit)$ presented in Section~\ref{sec.results}.
Following \cite{abboud2019dating}, we compute the deviation measure between two consecutive iterations $k-1$ and $k$ of the algorithm, defined for a partition $\mathcal{P}$ of the space of parameters $\mathbb{R}_+^5$ by
\begin{equation}
\label{eq:dev_measure}
    \mathcal{D}(\mathcal{P},k-1,k)=\max_{c\in \mathcal{P}}|\rho_k(c)-\rho_{k-1}(c)|
\end{equation}
with $\rho_k(c)$ the estimated probability of the element $c$ of the partition $\mathcal{P}$ given by
\[
\rho_k(c) = \sum_{i=1}^k\sum_{\ell=1}^N w_{\ell i}
1_{\theta_{\ell i} \in c}
\]
where $\theta_{\ell i}$ and $w_{\ell i}$ are the parameters sampled by the algorithm and their weights respectively.

\begin{figure}
\begin{center}
\includegraphics[height=3.7cm, trim = 0cm 0cm 0.5cm 0cm, clip=true]{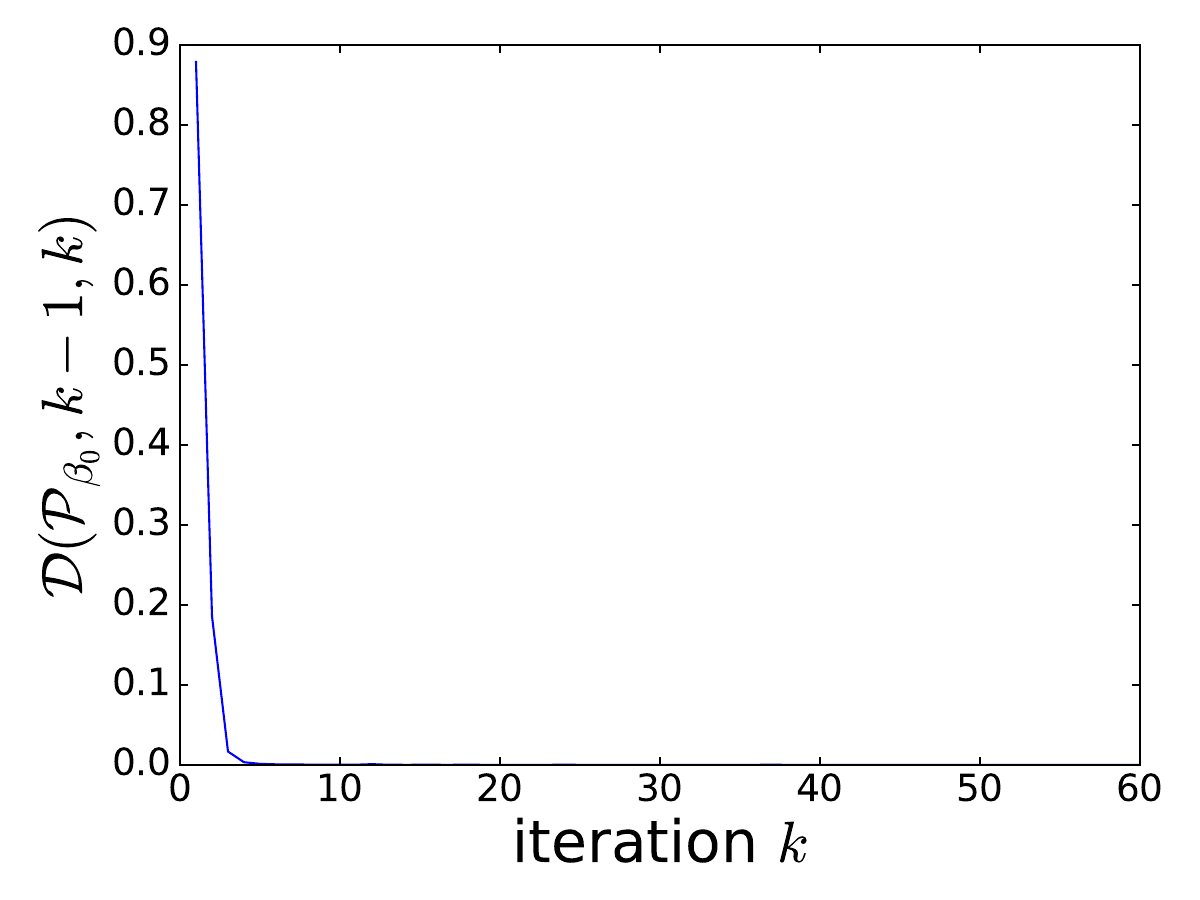}
\includegraphics[height=3.7cm, trim = 0cm 0cm 0.5cm 0cm, clip=true]{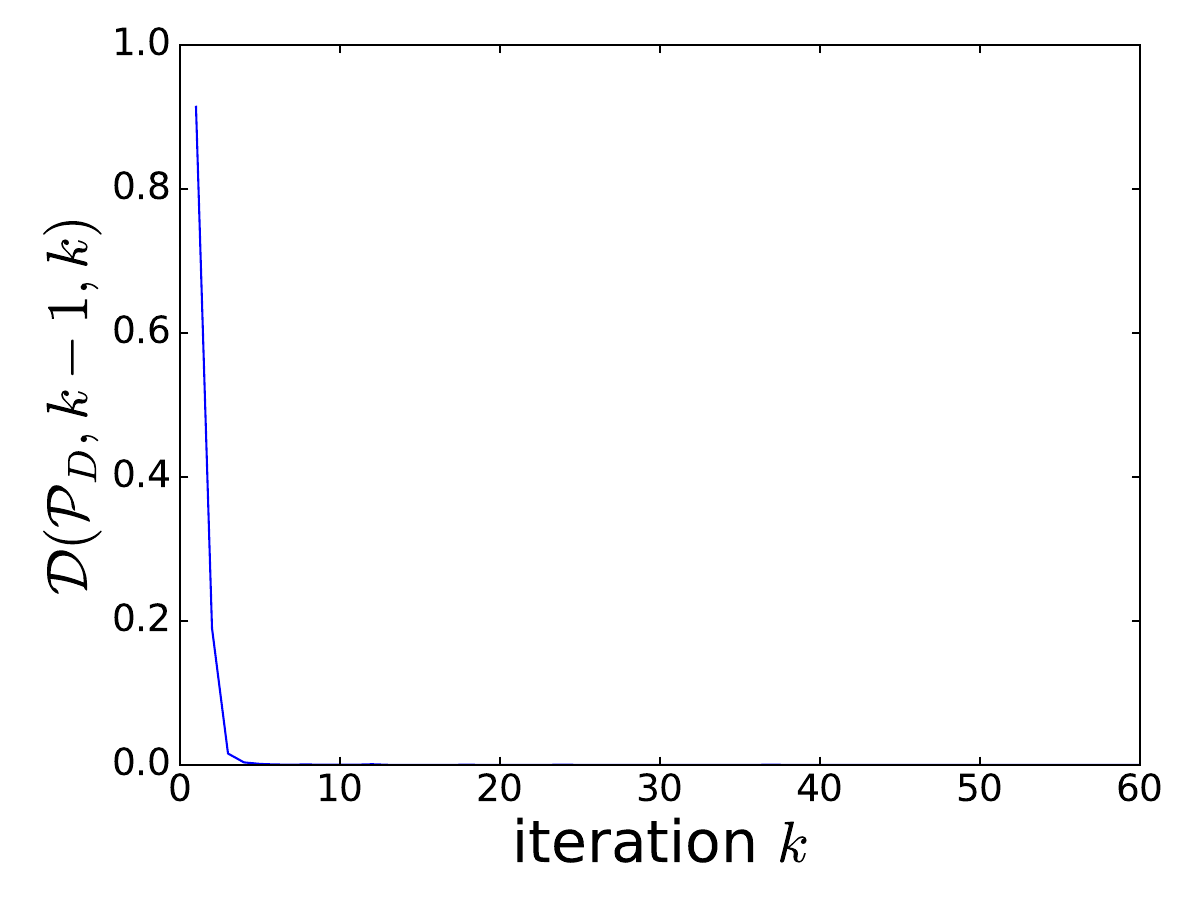}
\includegraphics[height=3.7cm, trim = 0cm 0cm 0.5cm 0cm, clip=true]{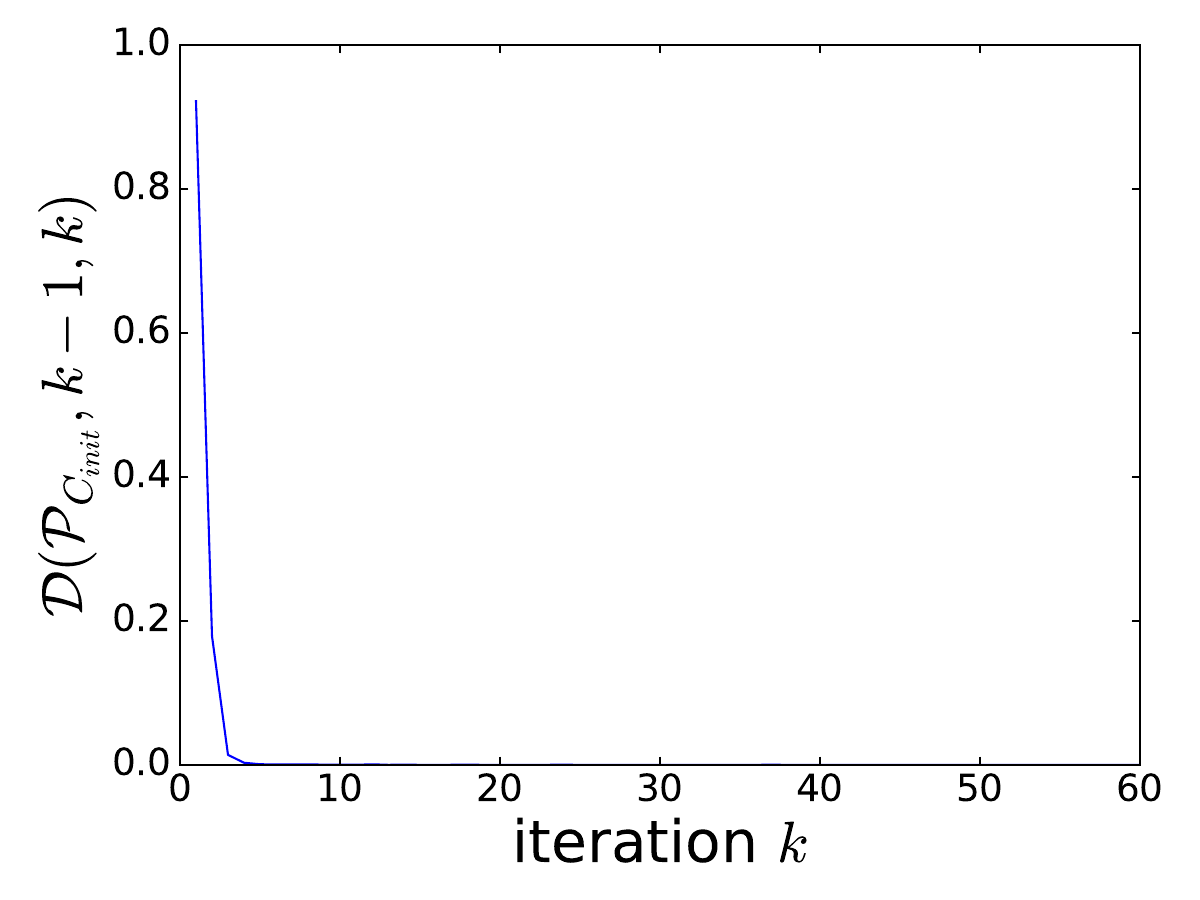}
\\
\includegraphics[height=3.7cm, trim = 0cm 0cm 0.5cm 0cm, clip=true]{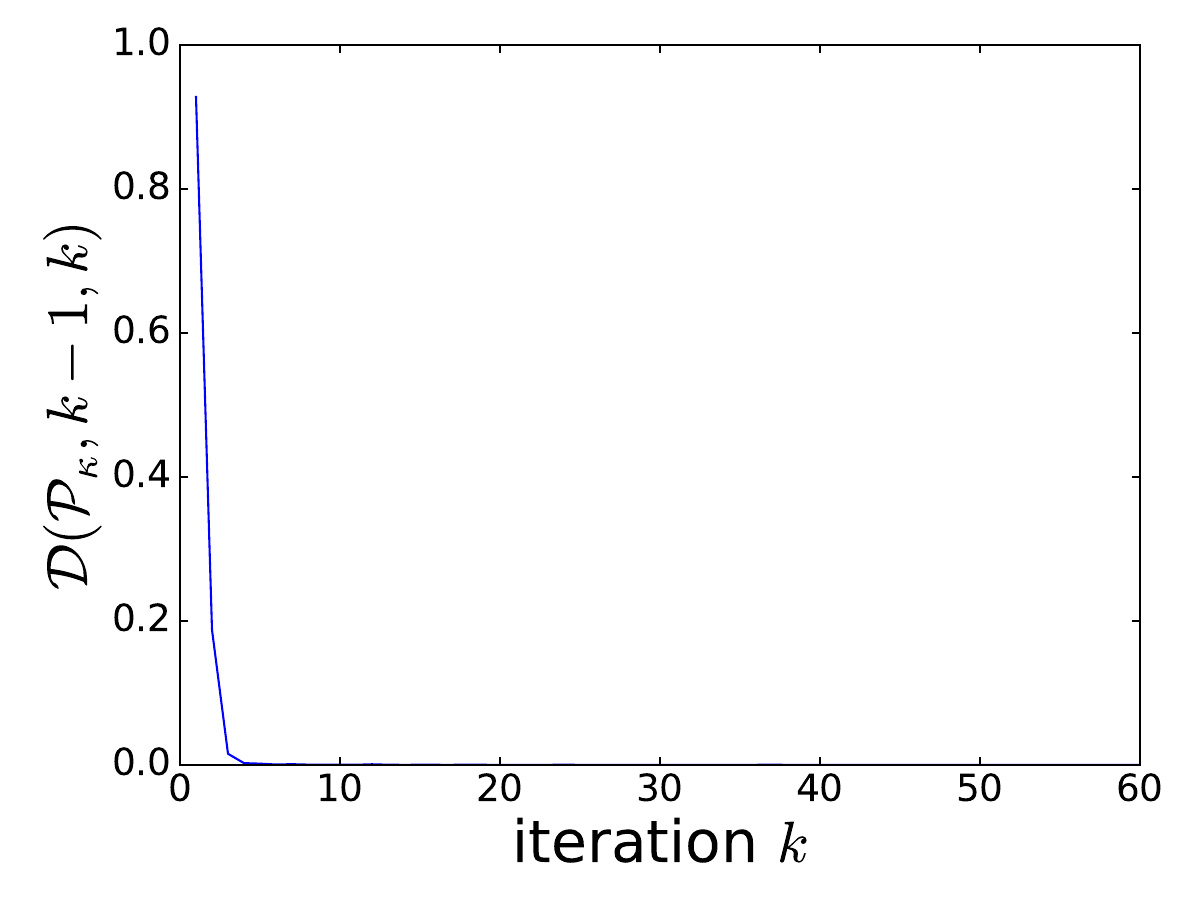}
\includegraphics[height=3.7cm, trim = 0cm 0cm 0.5cm 0cm, clip=true]{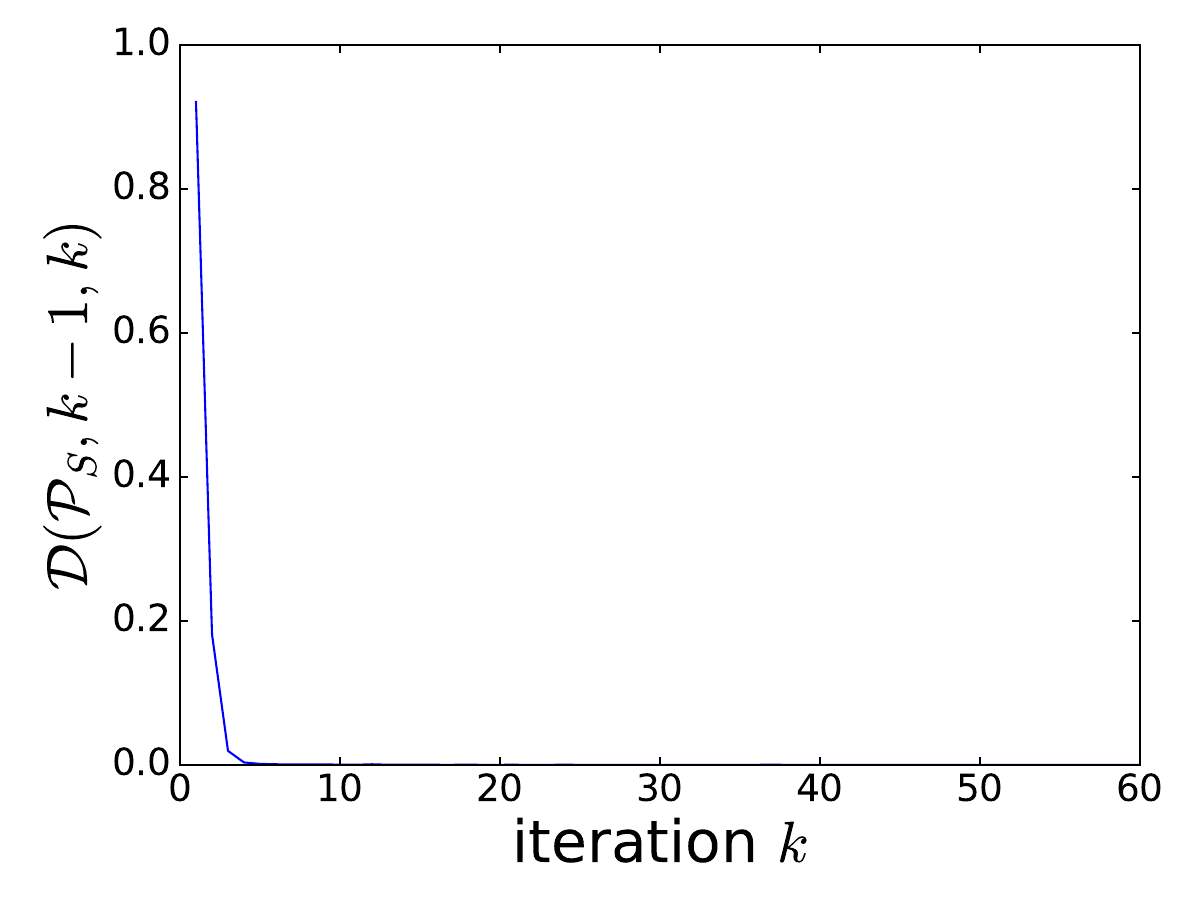}
\end{center}
\caption{\label{fig:dev_measure_1D} Deviation measure~\eqref{eq:dev_measure} for five partitions $\mathcal{P}_{\beta_0}$, $\mathcal{P}_{D}$, $\mathcal{P}_{\Cinit}$, $\mathcal{P}_{\kappa}$, $\mathcal{P}_{S}$
illustrating the convergence of the five one-dimensional distributions.}
\end{figure}

Figure~\ref{fig:dev_measure_1D} represents the deviation measure for five partitions, illustrating the convergence of the five one-dimensional posterior distributions. For example, the first picture on Figure~\ref{fig:dev_measure_1D} is the deviation measure for a partition $\mathcal{P}_{\beta_0}= \left([n\,c_{\beta},(n+1)c_{\beta}[\right)_{n\in\mathbb{N}} \times \mathbb{R}^4$ with $c_{\beta}$ small and illustrates the convergence of the marginal posterior distribution of the parameter $\beta_0$. Although there are not presented here, we also checked the convergence of the 2-dimensional posterior distributions.
We observe that the convergence of the AMIS algorithm occurs in about 10 iterations.

Importantly, notice that the convergence of the AMIS algorithm toward the posterior distribution is not proved. In fact, although we observe a convergence of our algorithm, it is not excluded that it converges toward parameters for which the log-likelihood are around a local maximum. 
To restrict this effect, the computation of the log-likelihood on a coarse grid described in Section~\ref{sec:coarse_grid} have been done first for a large scale of parameters values, before to be refined on the coarse grid presented in Section~\ref{sec:coarse_grid}.

\section{Model behavior for different temperature indices}
\label{sec:comparison_temp}

% Les différentes variables de température sont décrites dans la section "DATA" - à voir si on remonte tout ici ?

%We tested five temperature indices (T28, T30, T35, TX78 and STs28, explained below) for the temperature variable $(T^i_a)_{a,i}$ impacting the symptoms development (see Eq.~\eqref{eq.rachis_infection}). The threshold of 30 and 35$^{\circ}$C were chosen because \textcite{hauptman_temperature_2013} showed that \textit{H.~fraxineus} survive only few hour in infected ash shoots at temperature above 35$^{\circ}$C. the temperature of 28$^{\circ}$C is the maximal temperature at which the fungus still grow \textit{in vitro}. 
%Each data $T^i_a$ is a mean of the data (depending on the chosen index) of four (occasionally one, two or three, on the border of France) meteorological stations located on the quadrat $\omega_i$.
%The temperature indices T28, T30 and T35 denote the mean number of days, between July 1 and August 31, for which the daily maximal temperature is above 28$^{\circ}$C, 30$^{\circ}$C, 35$^{\circ}$C respectively.
%The temperature indices TX78 denotes the mean of the daily maximal temperatures for July and August.
%The temperature indices STs28 denotes the mean of the sums of daily temperatures over 28$^{\circ}$C for July and August.

We tested five different temperature indices for the choice of the temperature variable $(T^i_a)_{a,i}$ impacting the symptoms development (see Eq.~\eqref{eq.rachis_infection}): T24, T26, T28, T30 and T35, denoting the mean number of days, between July 1 and August 31, for which the daily maximal temperature exceeds 24$^{\circ}$C, 26$^{\circ}$C, 28$^{\circ}$C, 30$^{\circ}$C and 35$^{\circ}$C respectively.
The threshold of 30$^{\circ}$C and 35$^{\circ}$C were chosen because \cite{hauptman_temperature_2013} showed that \textit{H.~fraxineus} survives only few hour in infected ash shoots at temperature above 35$^{\circ}$C. The temperature of 28$^{\circ}$C is the maximal temperature where the fungus grows \textit{in vitro} \citep{hauptman_temperature_2013}. However, leaves temperature often exceeds air temperature by several degrees \citep{granier_evidence_2007} and we thus tried lower temperature thresholds (24$^{\circ}$C and 26$^{\circ}$C).

\medskip

%Figure~\ref{compare.temp} compares the results obtained with the different tested temperature indices. Quantitatively, all parameters indices give comparative log-likelihoods. We observed however that the best log-likelihood is obtained for the temperature index T28, for the period 2008-2019 (period of the data used for the estimation) as well as for the period 2008-2023 (where 2020-2023 is the predictive period). Both parameters TX78 and T30 give very similar results. The temperature index T35 gives the worst log-likelihood.
%Qualitatively, contrary to other temperature indices, the index T35 allows to describe the limited expansion neither in the southeast of France nor in the Garonne valley. The limited expansion in the Garonne valley seems less strong for temperature indexes STs28 ans T30.

\newcommand{\tabelm}[1]{\begin{tabular}{c}#1 \end{tabular}}

\begin{figure}
\begin{tabular}{|c|c|}
\hline
\begin{minipage}{7.4cm}\bigskip
\center T24\\
\begin{minipage}{3.4cm} \center
\includegraphics[height=3.5cm, trim = 3.2cm 1.8cm 5.4cm 0.5cm, clip=true]{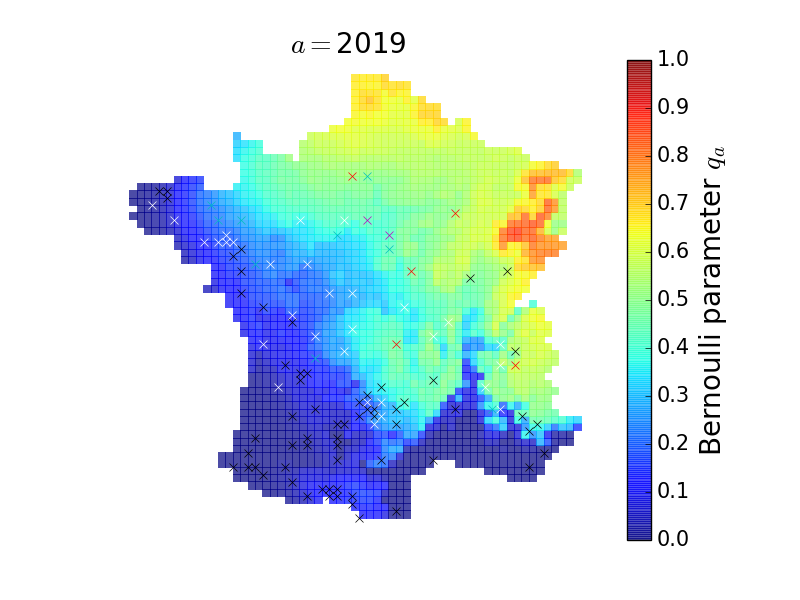}
log-likelihood : -24733 
\end{minipage}
\begin{minipage}{3.4cm} \center
\includegraphics[height=3.5cm, trim = 3.2cm 1.8cm 5.4cm 0.5cm, clip=true]{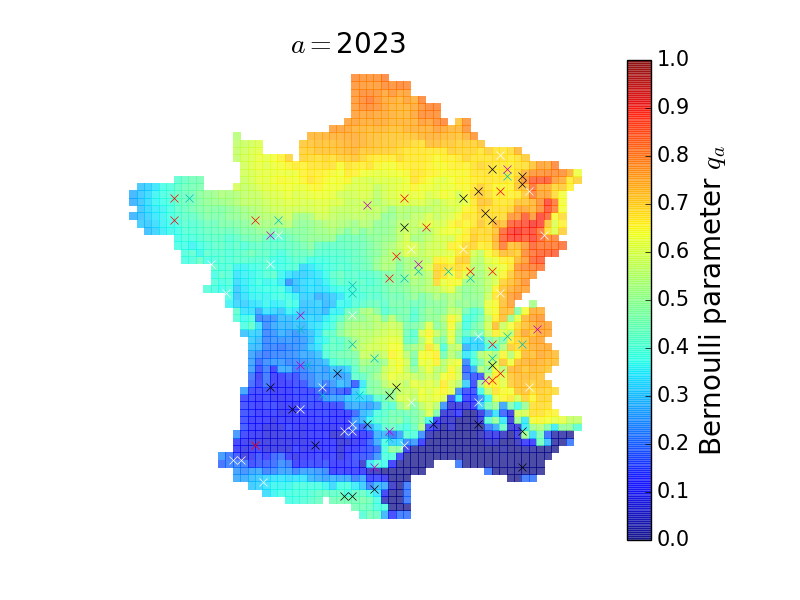}
log-likelihood : -32013
\end{minipage}\bigskip
\end{minipage}
&
\begin{minipage}{7.4cm}\bigskip
\center T26\\
\begin{minipage}{3.4cm} \center
\includegraphics[height=3.5cm, trim = 3.2cm 1.8cm 5.4cm 0.5cm, clip=true]{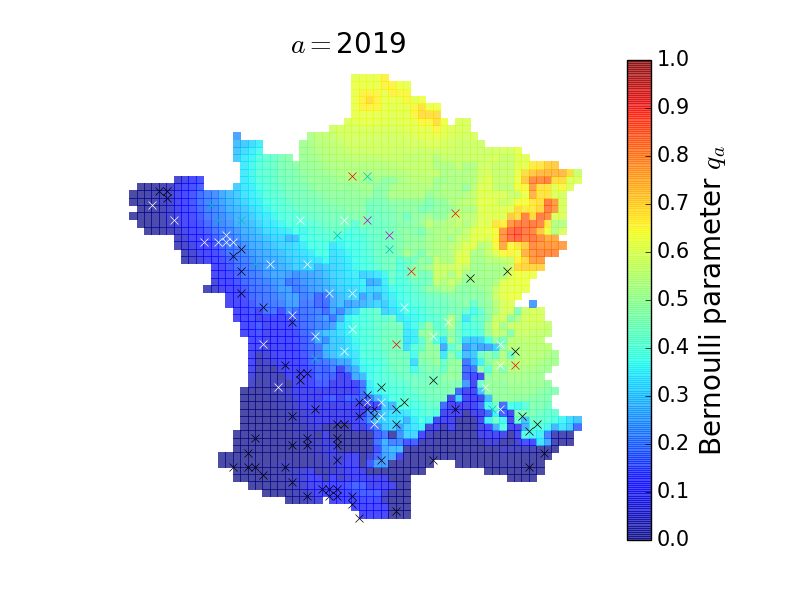}
log-likelihood : -24686 
\end{minipage}
\begin{minipage}{3.4cm} \center
\includegraphics[height=3.5cm, trim = 3.2cm 1.8cm 5.4cm 0.5cm, clip=true]{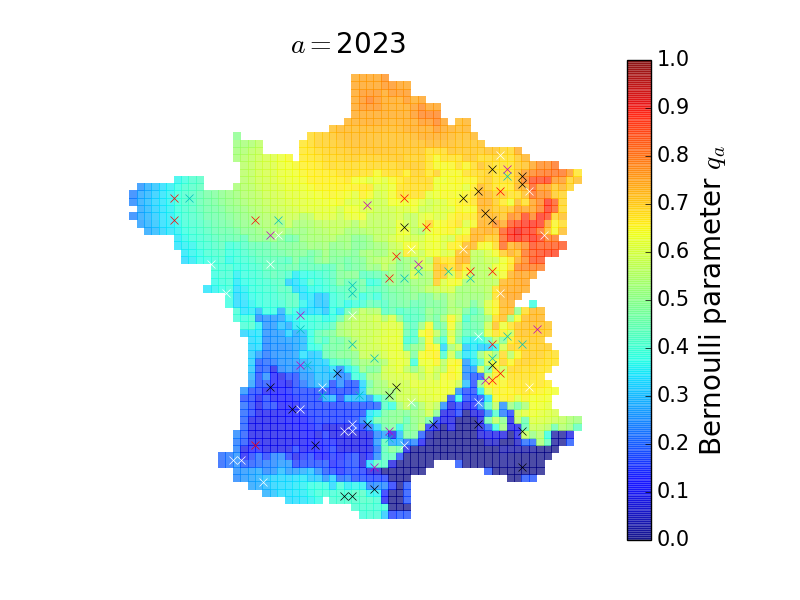}
log-likelihood : -31967
\end{minipage}\bigskip
\end{minipage}\\
\hline
\begin{minipage}{7.4cm}\bigskip
\center T28\\
\begin{minipage}{3.4cm} \center
\includegraphics[height=3.5cm, trim = 3.2cm 1.8cm 5.4cm 0.5cm, clip=true]{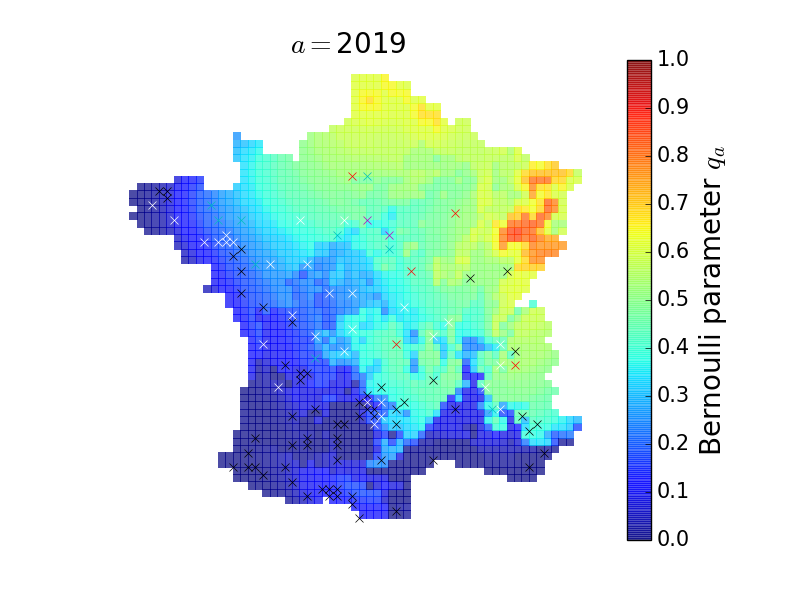}
log-likelihood : -24717 
\end{minipage}
\begin{minipage}{3.4cm} \center
\includegraphics[height=3.5cm, trim = 3.2cm 1.8cm 5.4cm 0.5cm, clip=true]{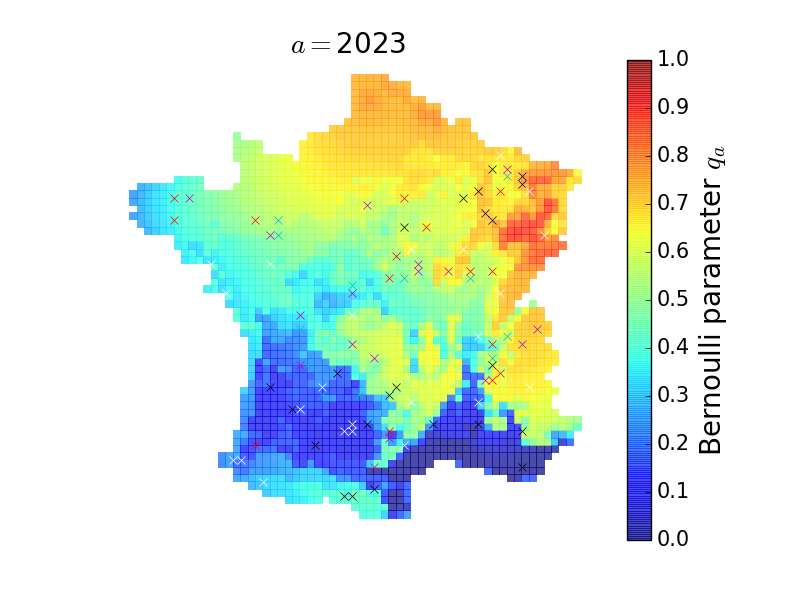}
log-likelihood : -31872
\end{minipage}\bigskip
\end{minipage}
&
\begin{minipage}{7.4cm}\bigskip
\center T30\\
    \begin{minipage}{3.4cm} \center
\includegraphics[height=3.5cm, trim = 3.2cm 1.8cm 5.4cm 0.5cm, clip=true]{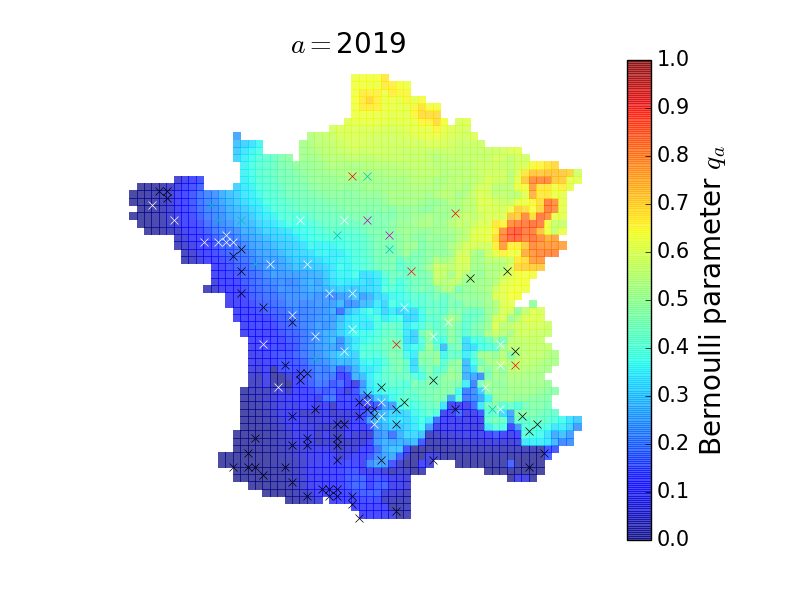}
log-likelihood : -24850 
\end{minipage}
\begin{minipage}{3.4cm} \center
\includegraphics[height=3.5cm, trim = 3.2cm 1.8cm 5.4cm 0.5cm, clip=true]{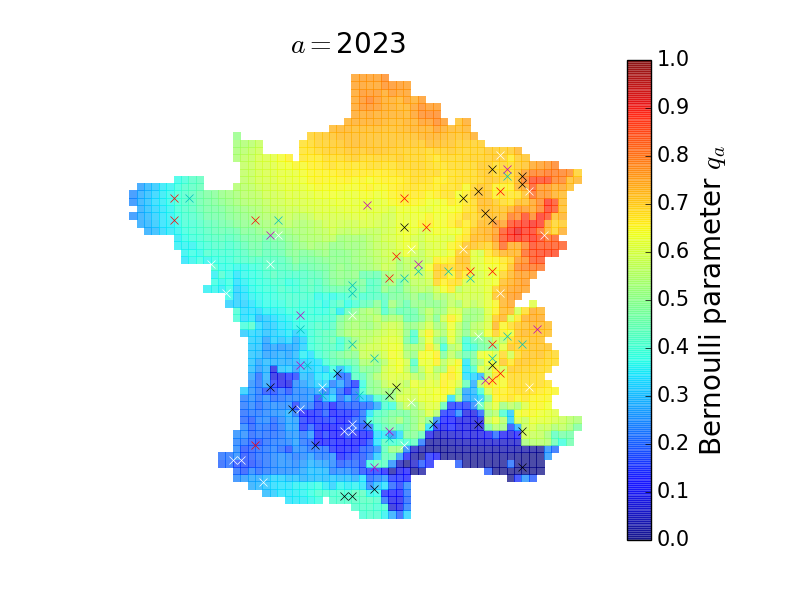}
log-likelihood : -32292
\end{minipage}\bigskip
\end{minipage}\\
\hline
\multicolumn{2}{|c|}{\begin{minipage}{7.4cm}\bigskip
\center T35\\
    \begin{minipage}{3.4cm} \center
\includegraphics[height=3.5cm, trim = 3.2cm 1.8cm 5.4cm 0.5cm, clip=true]{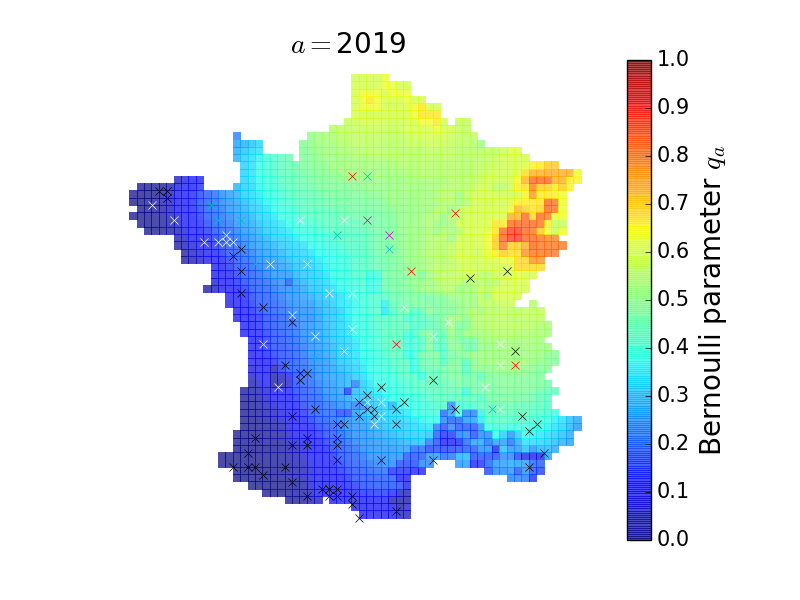}
log-likelihood : -25240 
\end{minipage}
\begin{minipage}{3.4cm} \center
\includegraphics[height=3.5cm, trim = 3.2cm 1.8cm 5.4cm 0.5cm, clip=true]{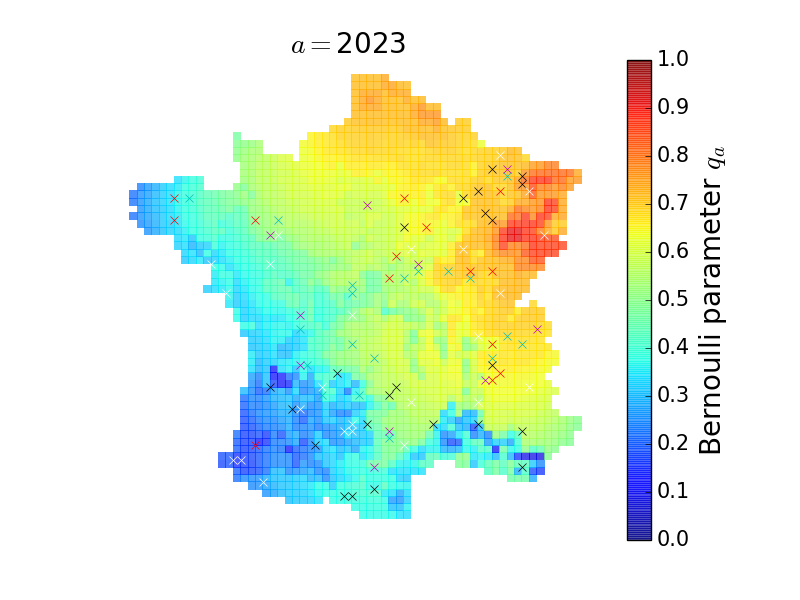}
log-likelihood : -33333
\end{minipage}\bigskip
\end{minipage}}\\
\hline
\end{tabular}
\caption{\label{compare.temp} Bernoulli parameters \eqref{eq.q} (heat map) and log-likelihood at the years 2019 and 2023 for the set of parameters which maximizes the likelihood and for the temperature indices T24, T26, T28, T30 and T35, as well as the observation data (crosses) of both years (see Figure~\ref{fig.chalarose.dynamics} for the color coding).}
\end{figure}

Figure~\ref{compare.temp} shows the results obtained with the different temperature indices tested. Quantitatively, all parameters indices give comparative log-likelihoods. We observed however that the best log-likelihood is obtained for the temperature index T26, for the period 2008-2019 (period of the data used for the estimation) but that T28 gives a better likelihood for the period 2008-2023 (where 2020-2023 is the predictive period). The temperature index T35 gives the worst log-likelihood.
Qualitatively, contrary to other temperature indices, the index T35 allows to describe the limited expansion neither in the southeast of France nor in the Garonne valley. The limited expansion in the Garonne valley seems less strong for the temperature index T30 than T24, T26 and T28.

\medskip

\begin{figure}[H]
\begin{center}
\includegraphics[height=4.cm, trim = 1cm 0cm 1.cm 0cm, clip=true]{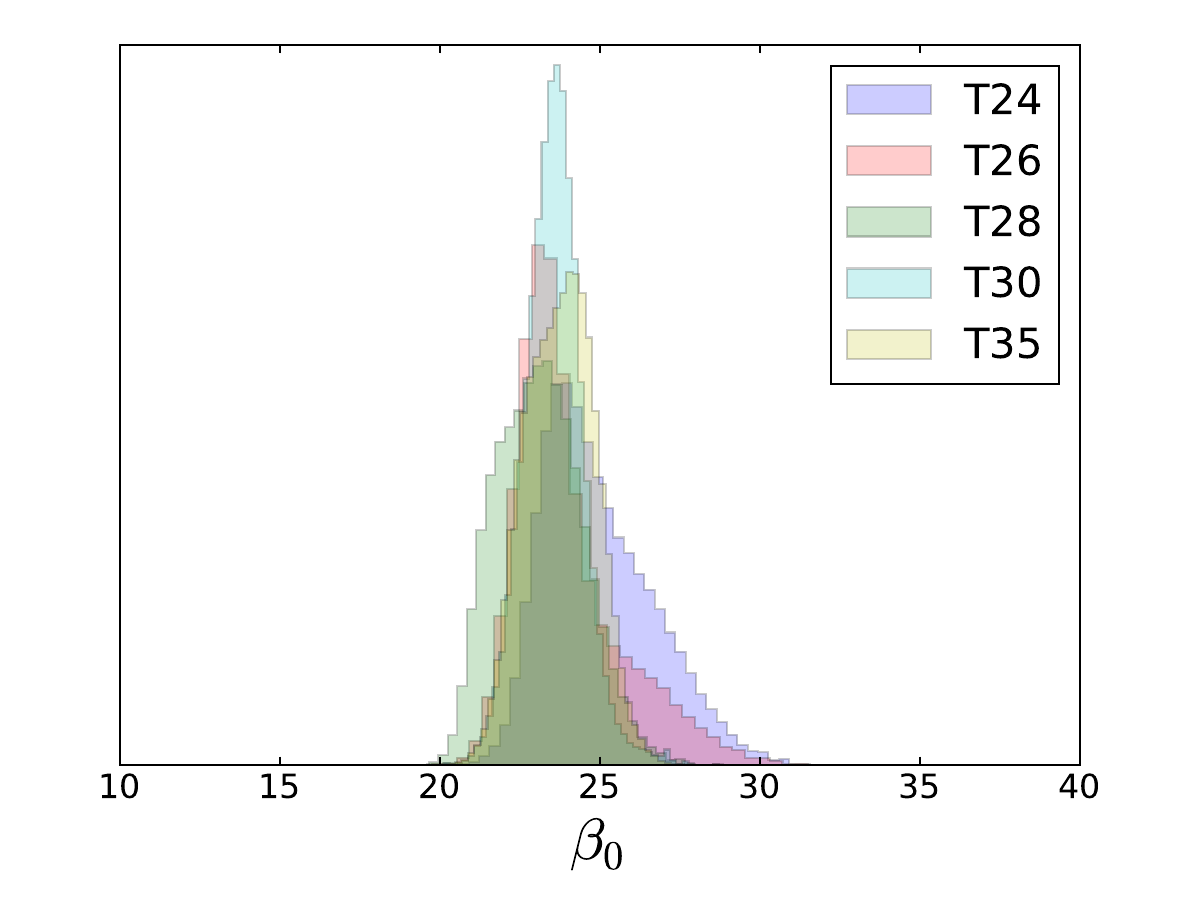}
\includegraphics[height=4.cm, trim = 1cm 0cm 1.cm 0cm, clip=true]{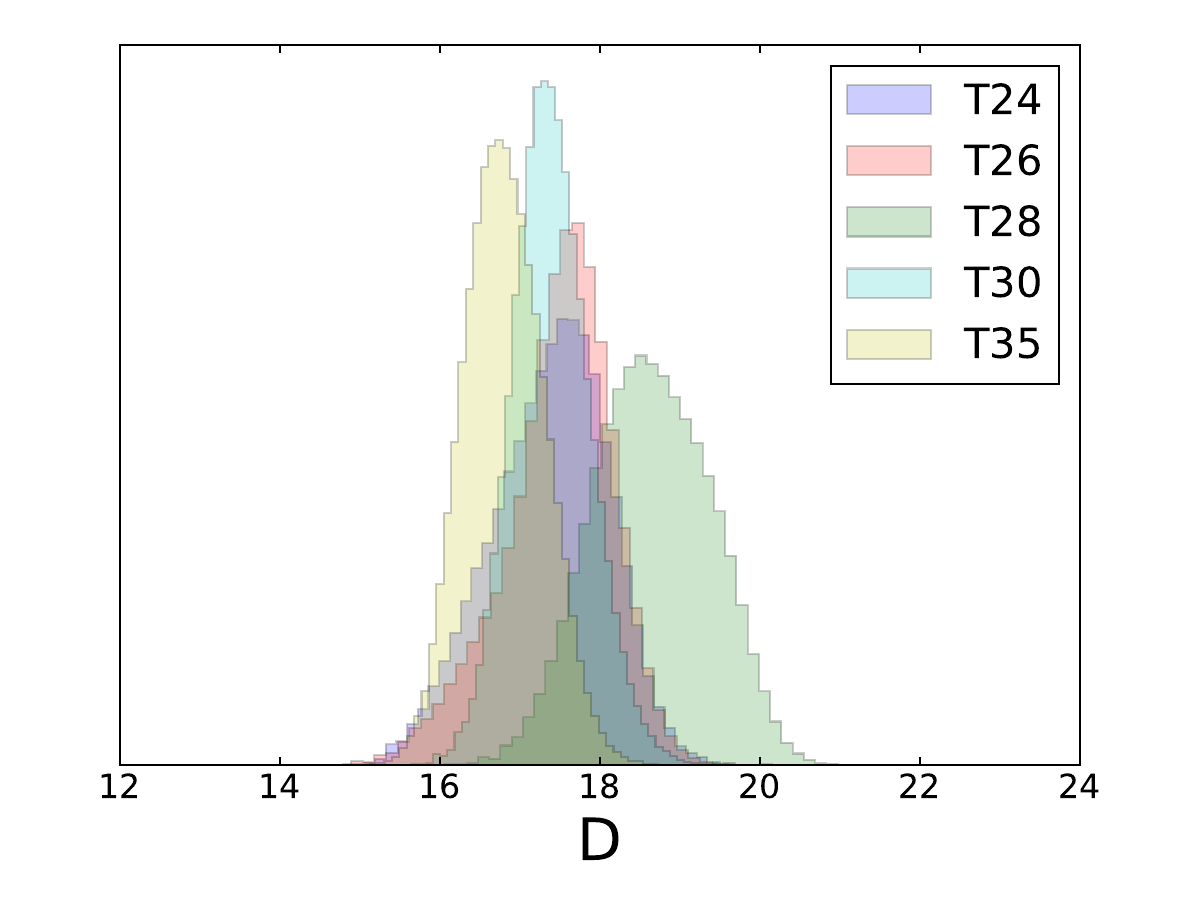}
\includegraphics[height=4.cm, trim = 1cm 0cm 1.cm 0cm, clip=true]{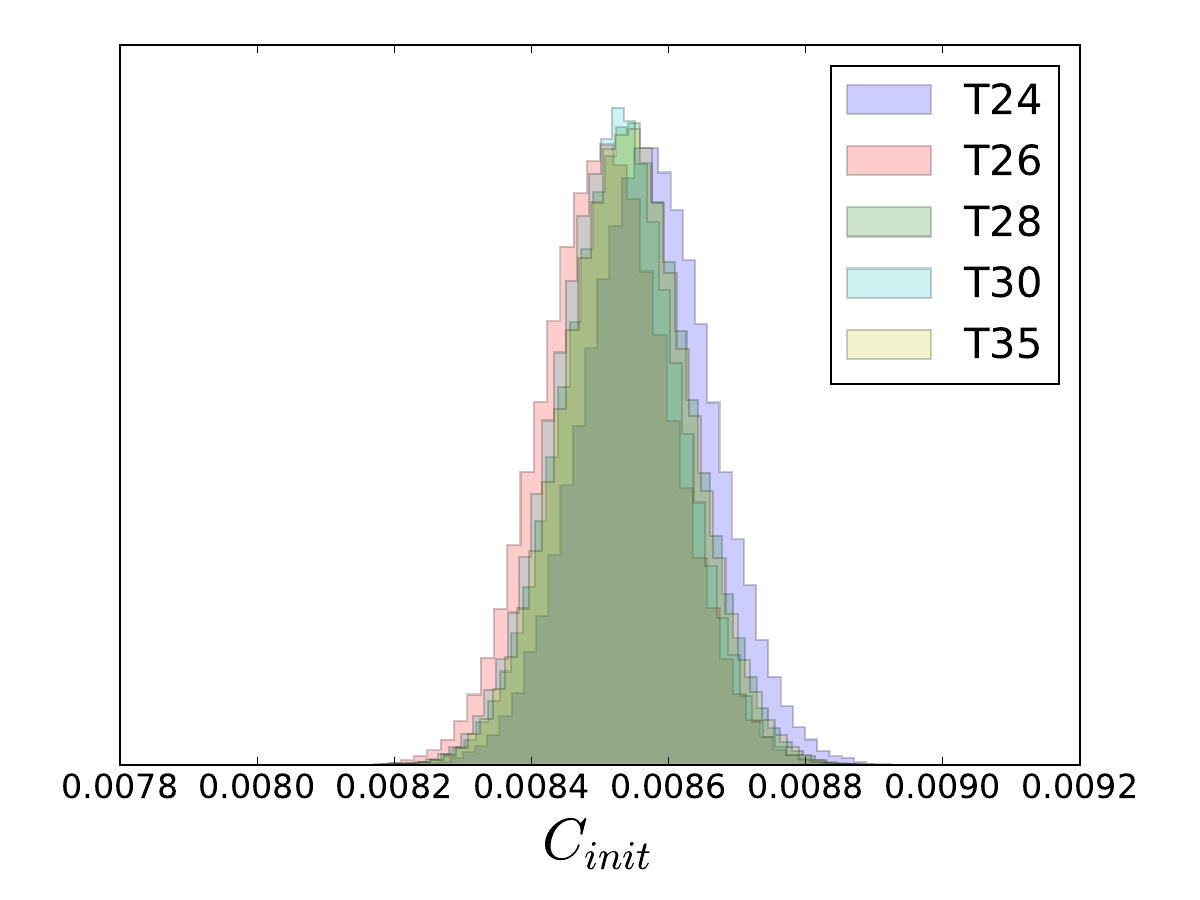}
\\
\includegraphics[height=4cm, trim = 1cm 0cm 1.cm 0cm, clip=true]{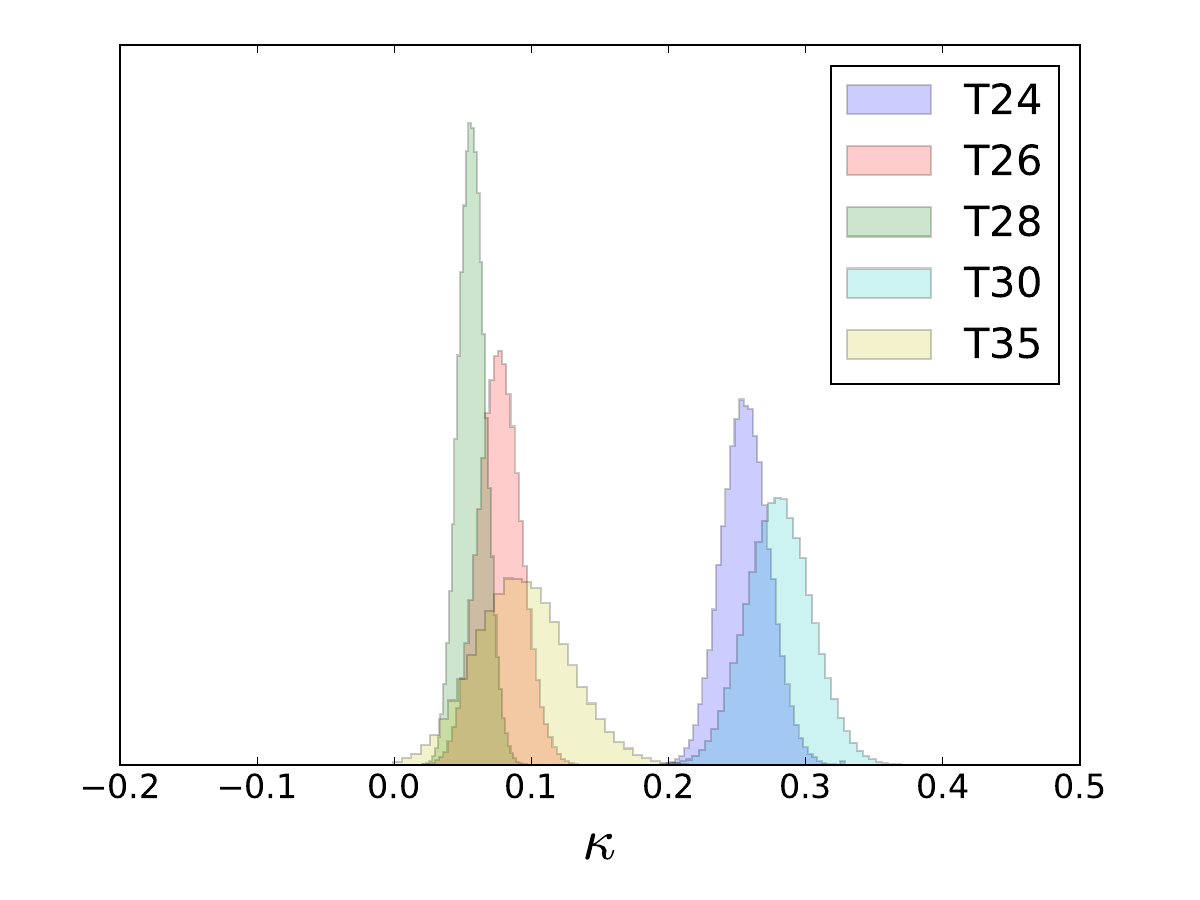}
\includegraphics[height=4cm, trim = 1cm 0cm 1.cm 0cm, clip=true]{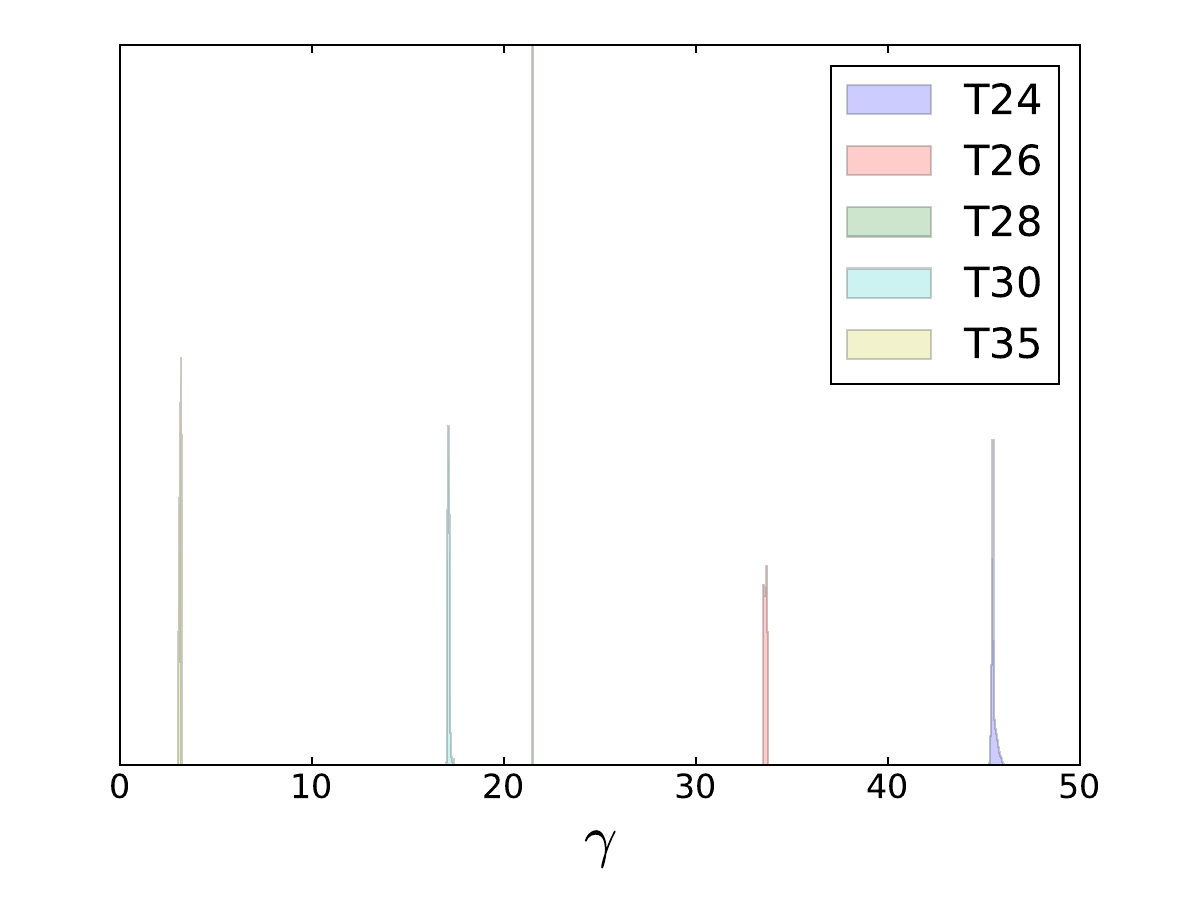}
\includegraphics[height=4cm, trim = 1cm 0cm 1.cm 0cm, clip=true]{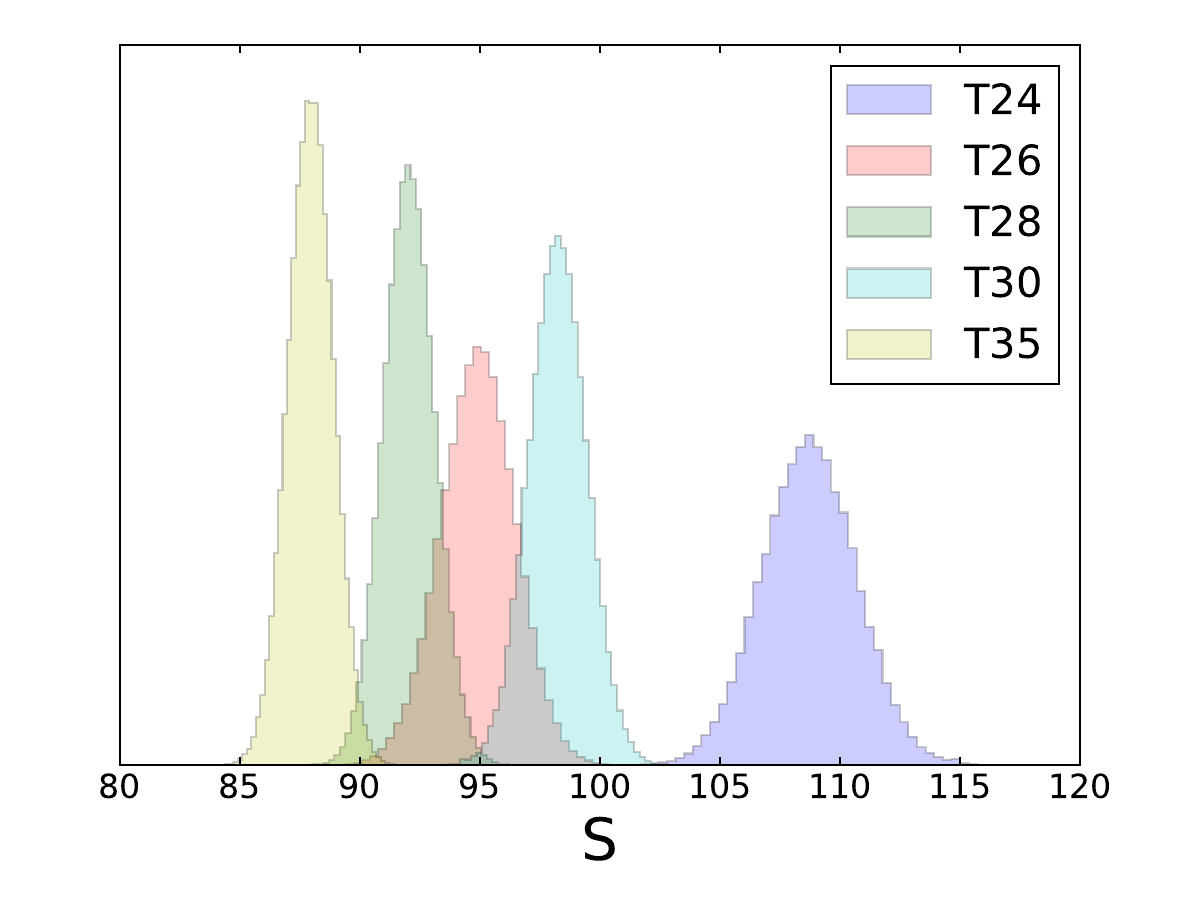}
\caption{\label{fig.compare.posterior} Marginal posterior distributions estimated by Algorithm~\ref{algo.AMIS} for the different temperature indices T24, T26, T28, T30 and T35.}
\end{center}
\end{figure}

Figure~\ref{fig.compare.posterior} shows the marginal posterior distributions estimated by Algorithm~\ref{algo.AMIS} for the different temperature indexes. We first observe that the posterior distribution of the parameter $\Cinit$ is almost the same for the five temperature indices. This parameter, in fact, describes the intensity of the prevalence of the dieback symptoms for the years 2008 and 2009, given by \eqref{eq.initialisation}, and does not directly depend on the temperature index. 
The posterior distributions of parameters $\beta_0$ and $D$ are with relatively similar supports. Therefore, although the distributions are quite different, all temperature indexes lead to the same order for the intensity of spore production $\beta_0$ as well as the diffusion coefficient $D$.
The posterior distributions of the parameter $\gamma$ can obviously not been compared as they represent different quantities. However, contrary to the temperature index T28 which led to an almost degenerated distribution to $\gamma$ (see Section~\ref{sec:degenerated_dist_gamma}), we obtained a posterior distribution with density, although spiky, for other temperature indexes. 
%{\color{red}Les lois des saturations $S$ décalées : pas d'explication triviale pour le moment.}

\section{Data visualisation}
\label{append:data}
\begin{figure}
\begin{center}
\begin{tabular}{cc}
\makecell{\includegraphics[height=3.5cm, trim = 3.2cm 1.8cm 5.4cm 0.5cm, clip=true]{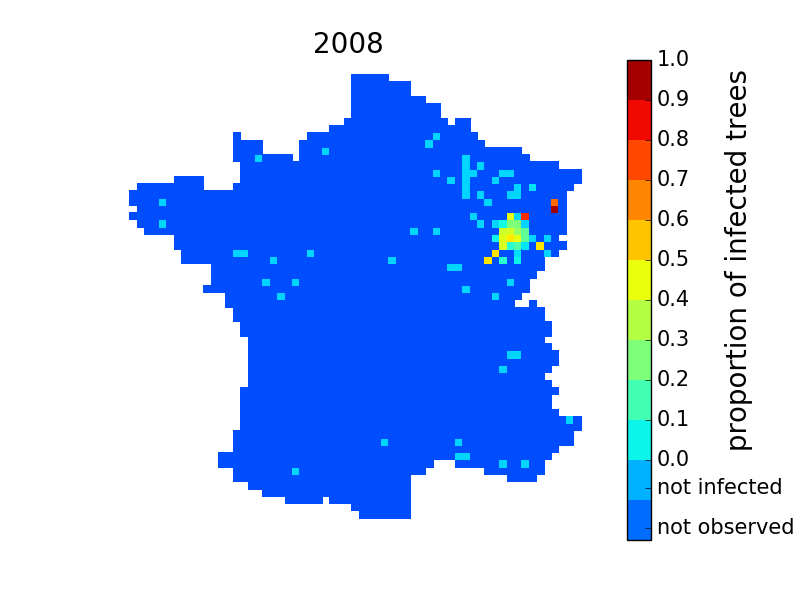}
\includegraphics[height=3.5cm, trim = 3.2cm 1.8cm 5.4cm 0.5cm, clip=true]{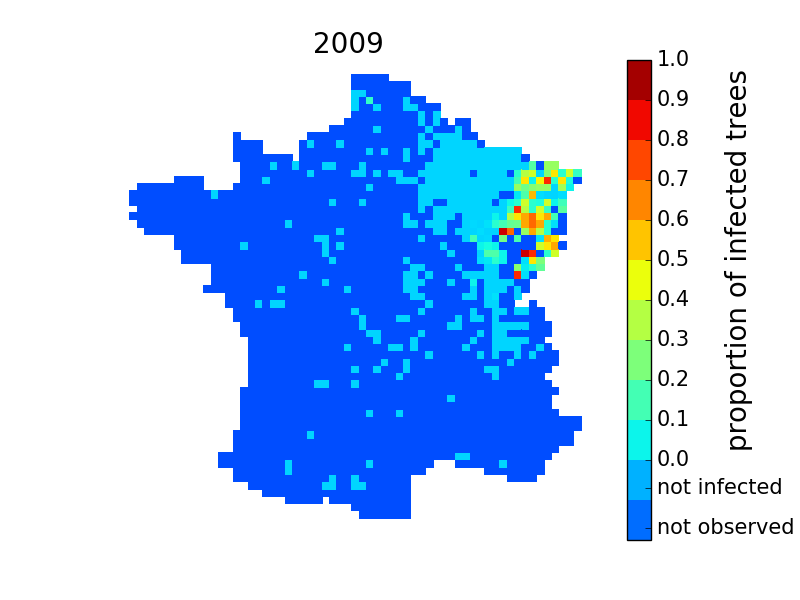}
\includegraphics[height=3.5cm, trim = 3.2cm 1.8cm 5.4cm 0.5cm, clip=true]{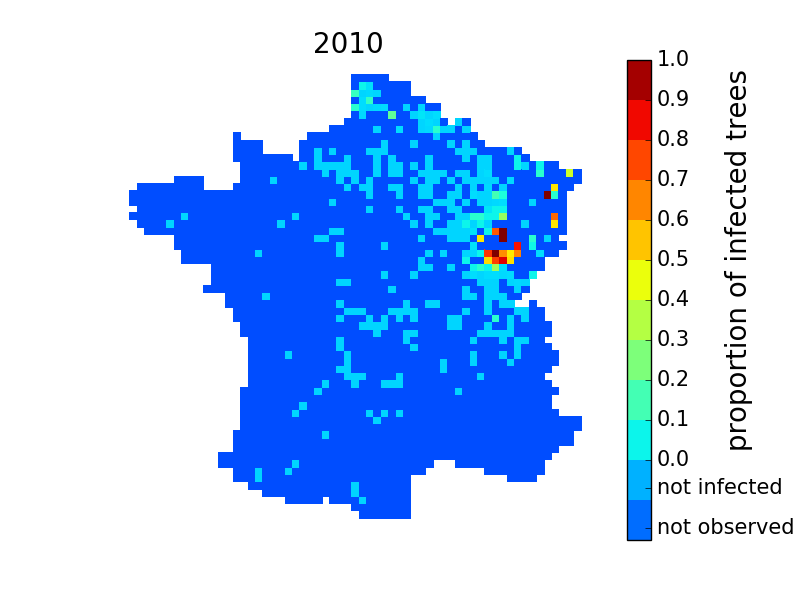}
\includegraphics[height=3.5cm, trim = 3.2cm 1.8cm 5.4cm 0.5cm, clip=true]{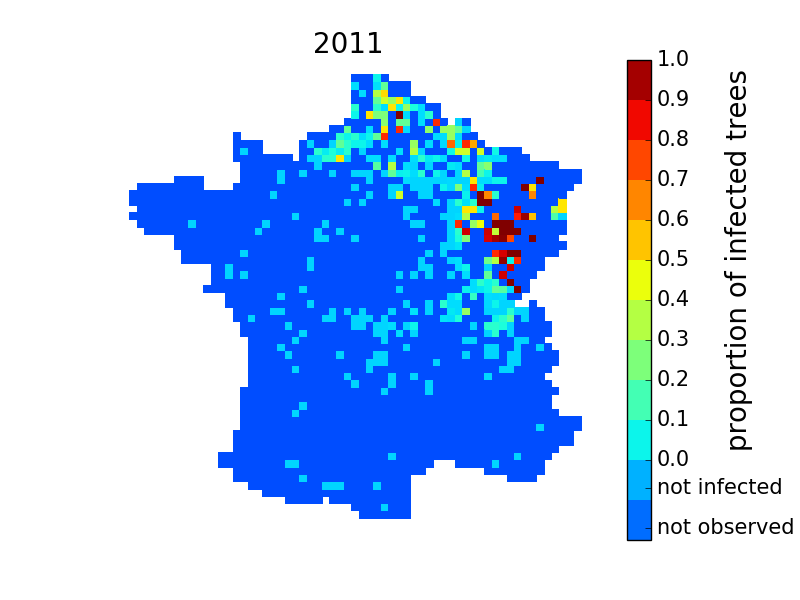}\\
\includegraphics[height=3.5cm, trim = 3.2cm 1.8cm 5.4cm 0.5cm, clip=true]{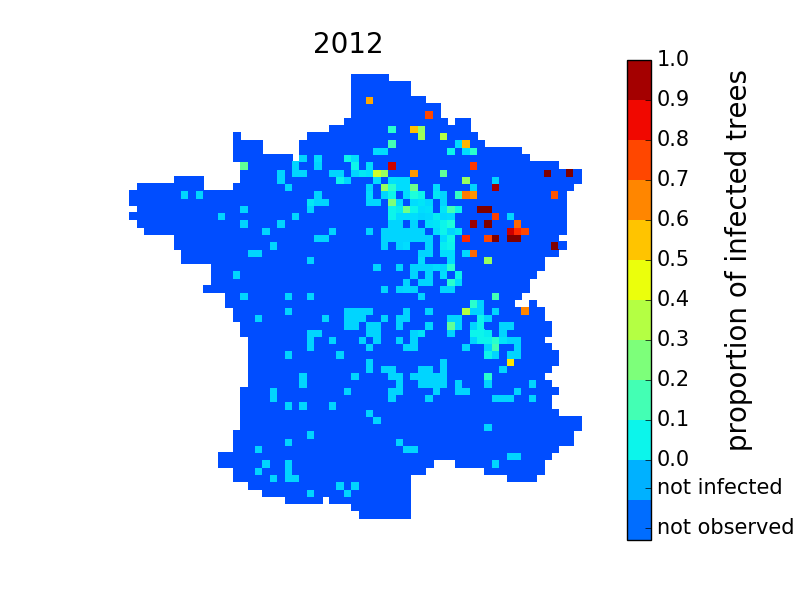}
\includegraphics[height=3.5cm, trim = 3.2cm 1.8cm 5.4cm 0.5cm, clip=true]{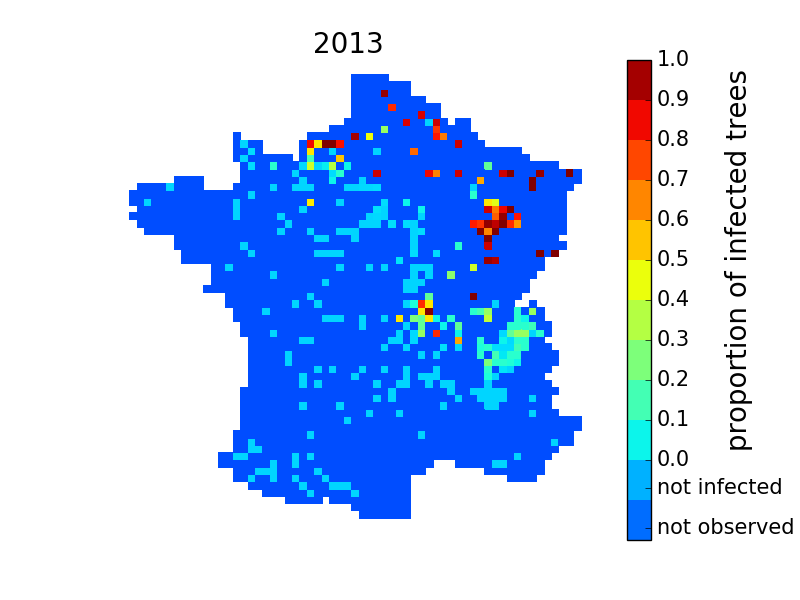}
\includegraphics[height=3.5cm, trim = 3.2cm 1.8cm 5.4cm 0.5cm, clip=true]{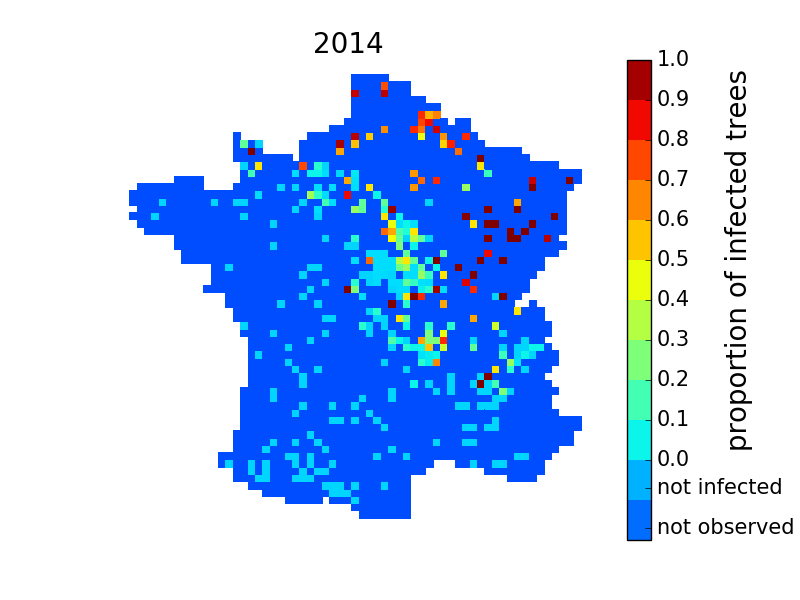}
\includegraphics[height=3.5cm, trim = 3.2cm 1.8cm 5.4cm 0.5cm, clip=true]{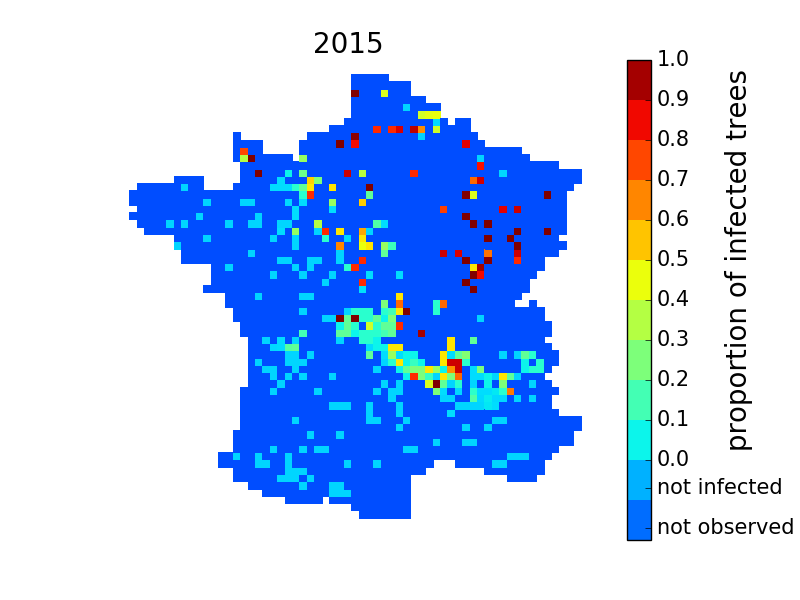}\\
\includegraphics[height=3.5cm, trim = 3.2cm 1.8cm 5.4cm 0.5cm, clip=true]{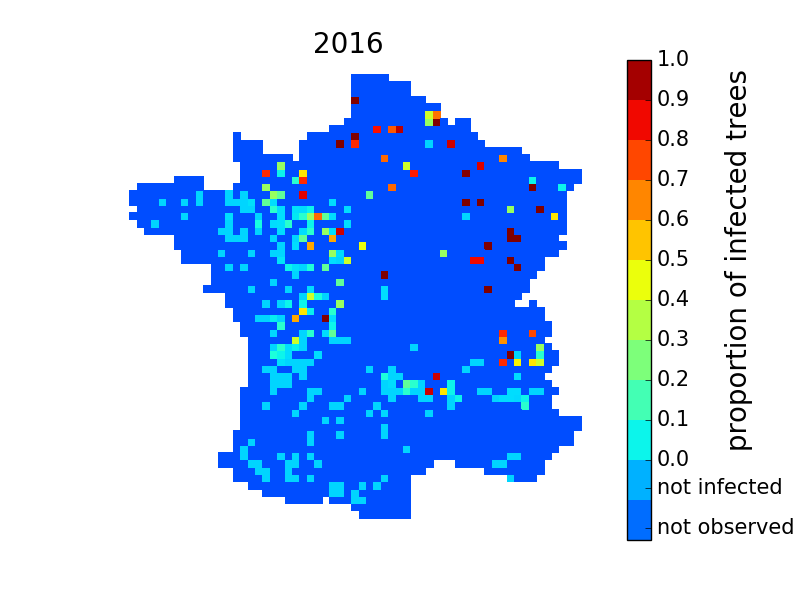}
\includegraphics[height=3.5cm, trim = 3.2cm 1.8cm 5.4cm 0.5cm, clip=true]{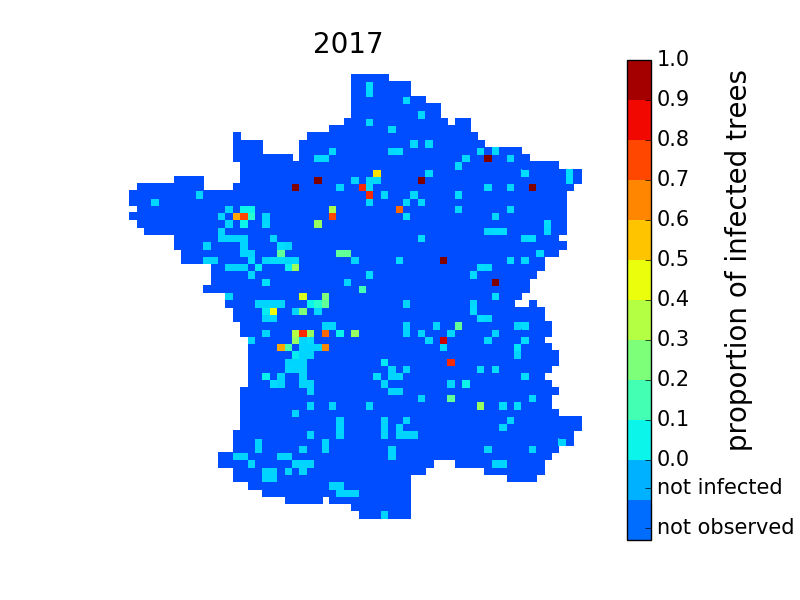}
\includegraphics[height=3.5cm, trim = 3.2cm 1.8cm 5.4cm 0.5cm, clip=true]{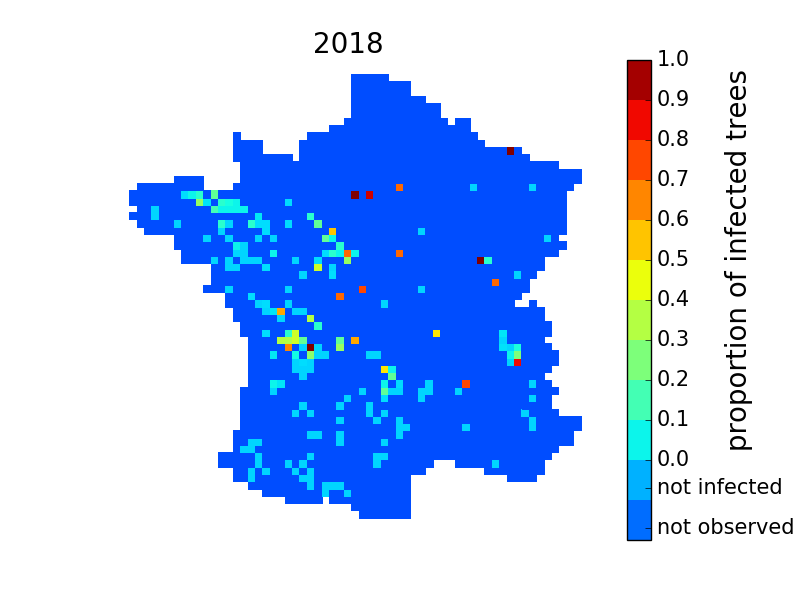}
\includegraphics[height=3.5cm, trim = 3.2cm 1.8cm 5.4cm 0.5cm, clip=true]{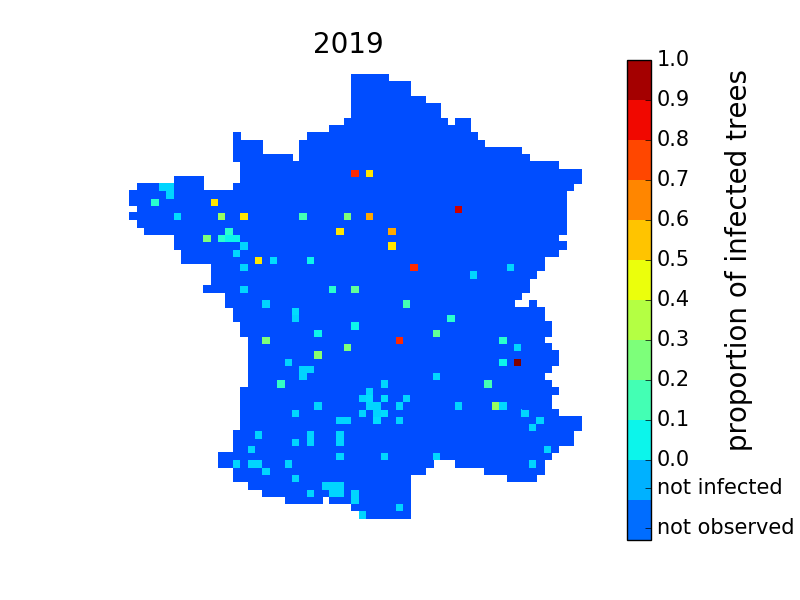}\\
\includegraphics[height=3.5cm, trim = 3.2cm 1.8cm 5.4cm 0.5cm, clip=true]{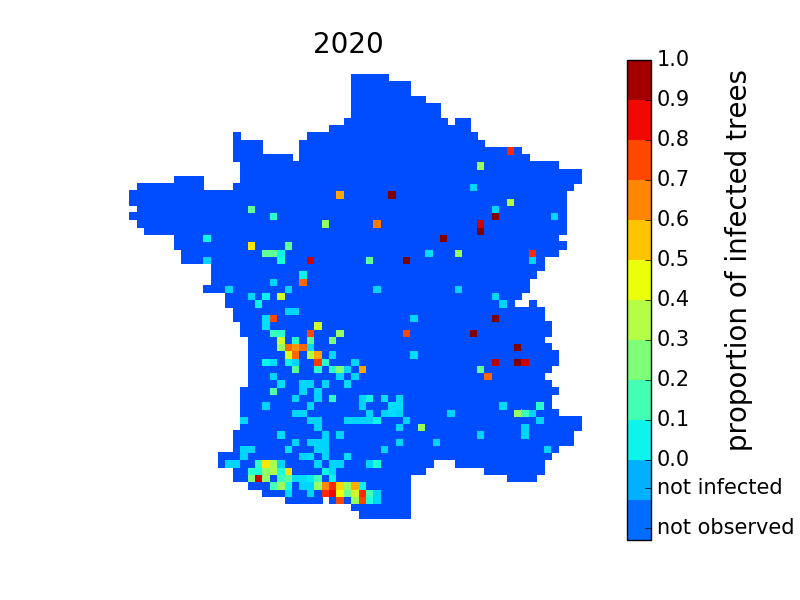}
\includegraphics[height=3.5cm, trim = 3.2cm 1.8cm 5.4cm 0.5cm, clip=true]{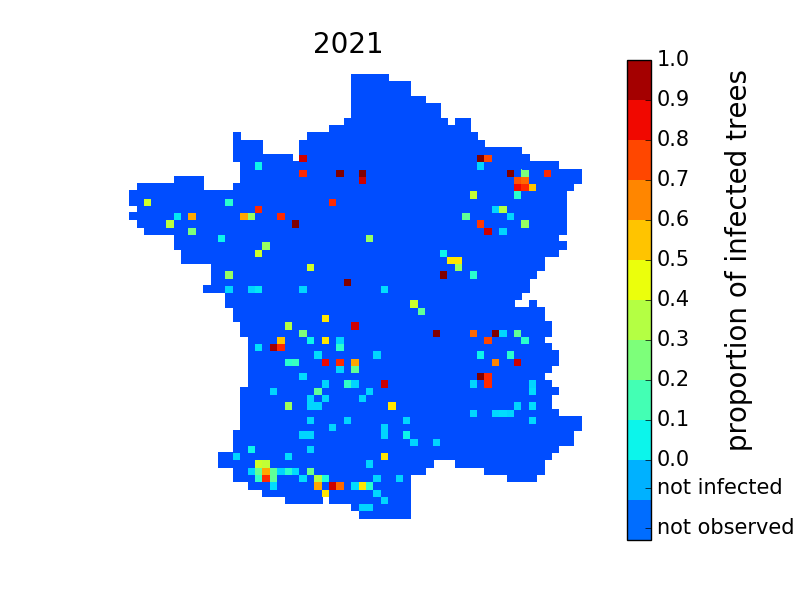}
\includegraphics[height=3.5cm, trim = 3.2cm 1.8cm 5.4cm 0.5cm, clip=true]{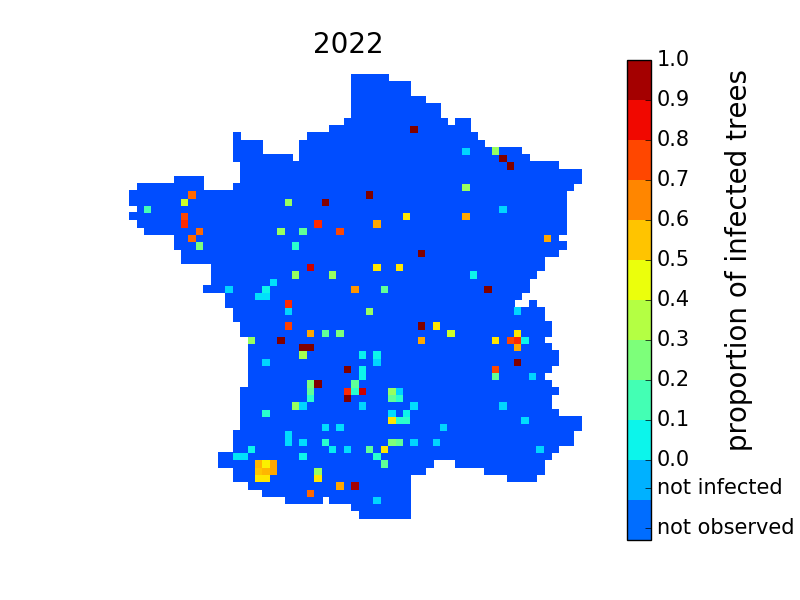}
\includegraphics[height=3.5cm, trim = 3.2cm 1.8cm 5.4cm 0.5cm, clip=true]{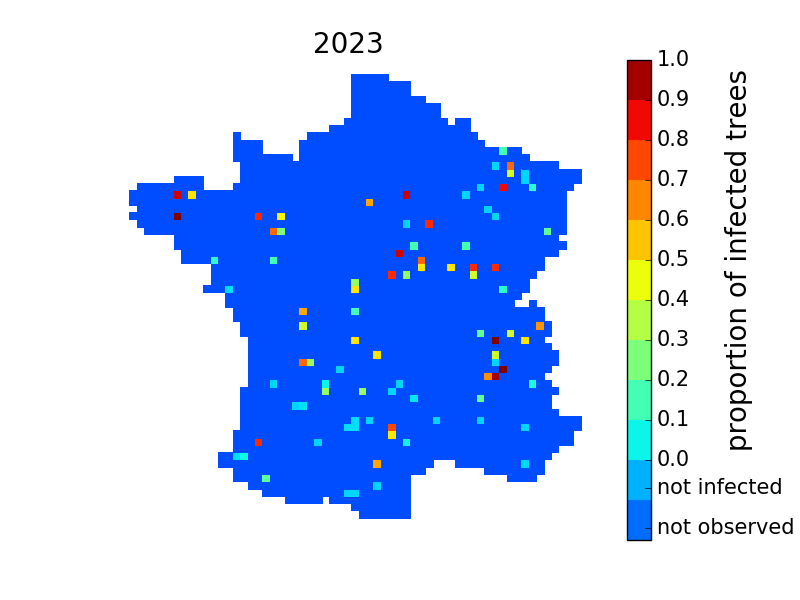}}
&
\makecell{
\includegraphics[height=6cm, trim = 15.5cm 1.2cm 0.1cm 0.5cm, clip=true]{fig_chalarose_observee_2023.png}}
\end{tabular}
\end{center}
\caption{\label{fig.observations} Proportion of infected trees among the observed trees from 2008 to 2023. Dark blue quadrats are non observed quadrats. Medium blue quadrats are non infected observed quadrats. From light blue to red quadrats: few infected to strongly infected observed quadrats.}
\end{figure}

\begin{figure}
\begin{center}
\includegraphics[width=7cm, trim = 2.5cm 1.2cm 0cm 0.5cm, clip=true]{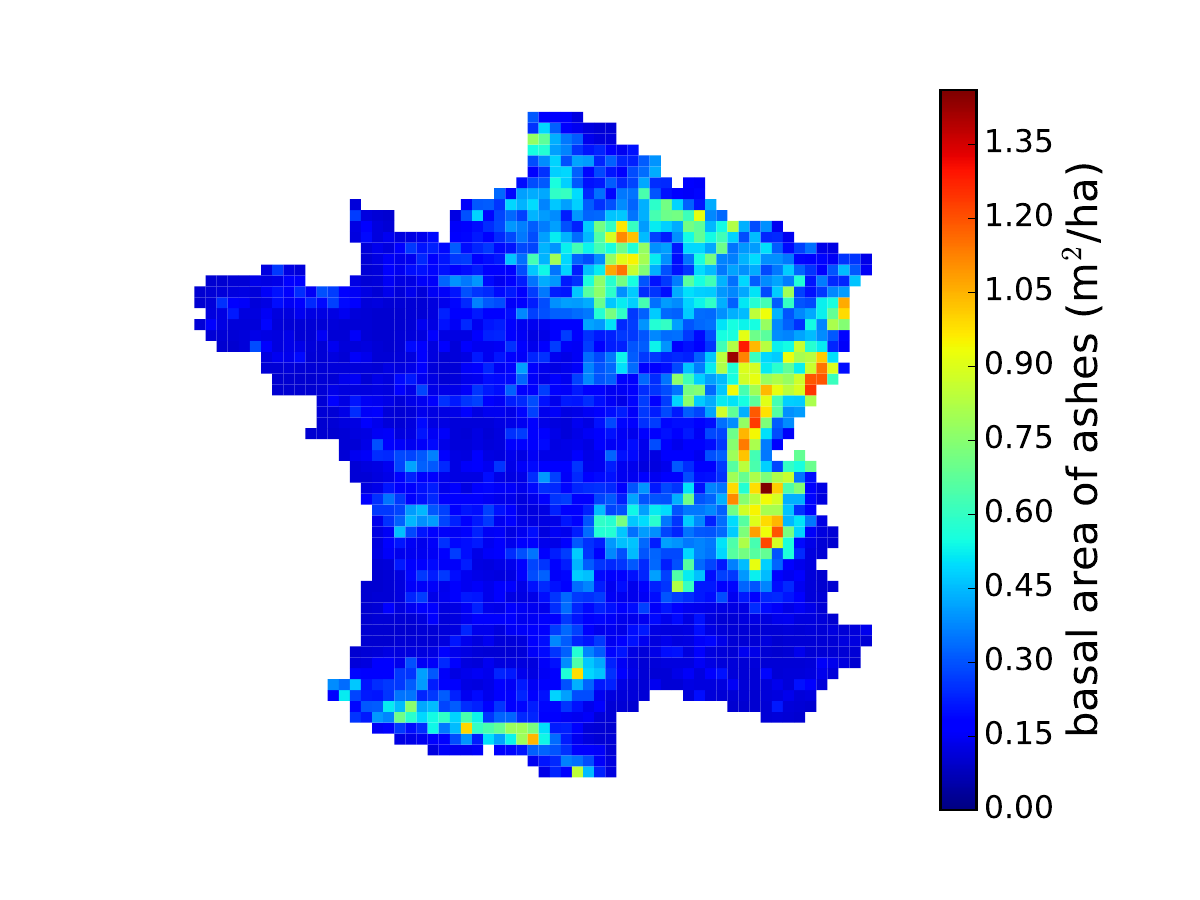}
\end{center}
\caption{\label{fig.ash.density} Basal area of ashes (\textit{F. excelsior} and \textit{F. angustifolia}) in France.}
\end{figure}

\begin{figure}
\begin{center}
\begin{tabular}{cc}
\makecell{
%---------------------
\begin{subfigure}{2.5cm}
\includegraphics[height=2.6cm, trim = 3.2cm 2cm 5.4cm 1.4cm, clip=true]{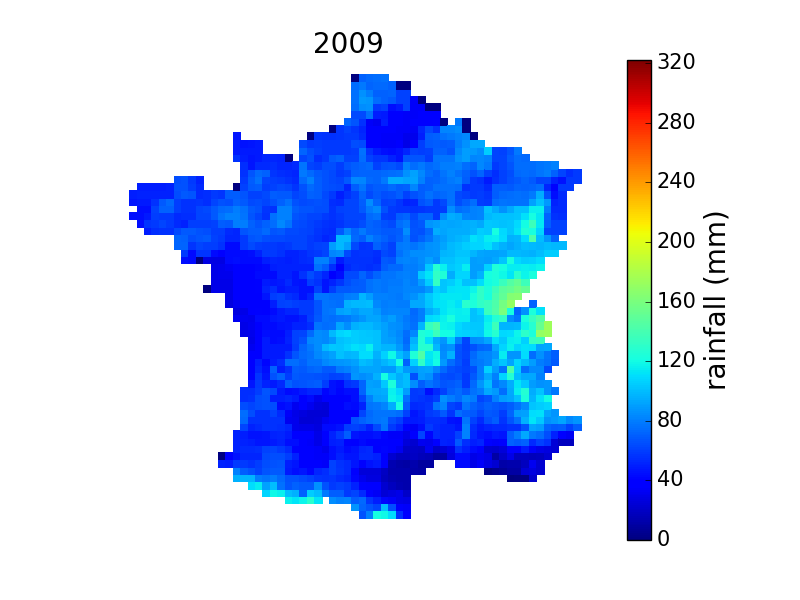}
\caption{2009}
\end{subfigure}
%---------------------
\begin{subfigure}{2.5cm}
\includegraphics[height=2.6cm, trim = 3.2cm 2cm 5.4cm 1.4cm, clip=true]{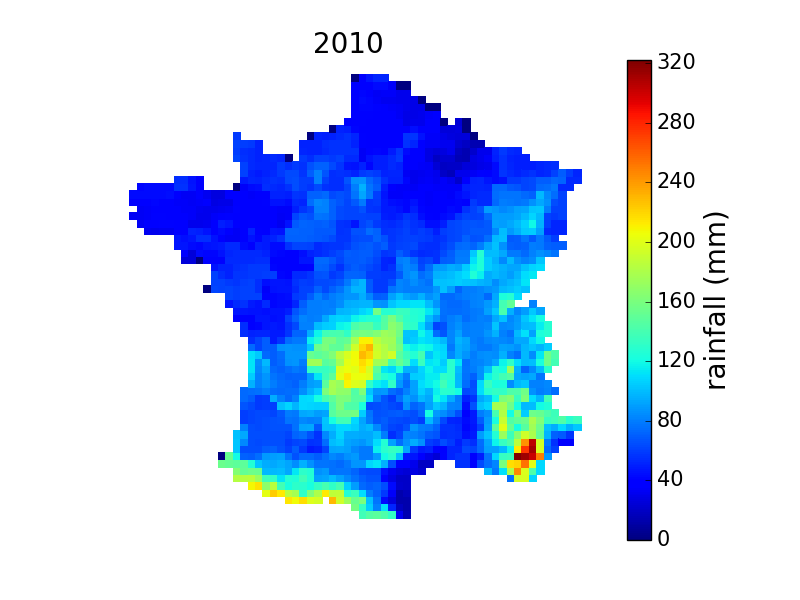}
\caption{\label{fig.pluv.2010}2010}
\end{subfigure}
%---------------------
\begin{subfigure}{2.5cm}
\includegraphics[height=2.6cm, trim = 3.2cm 2cm 5.4cm 1.4cm, clip=true]{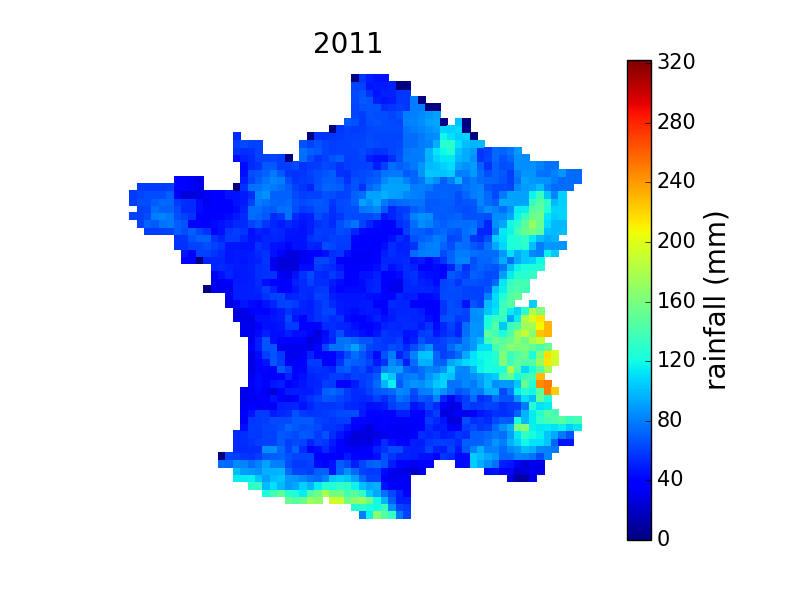}
\caption{2011}
\end{subfigure}
%---------------------
\begin{subfigure}{2.5cm}
\includegraphics[height=2.6cm, trim = 3.2cm 2cm 5.4cm 1.4cm, clip=true]{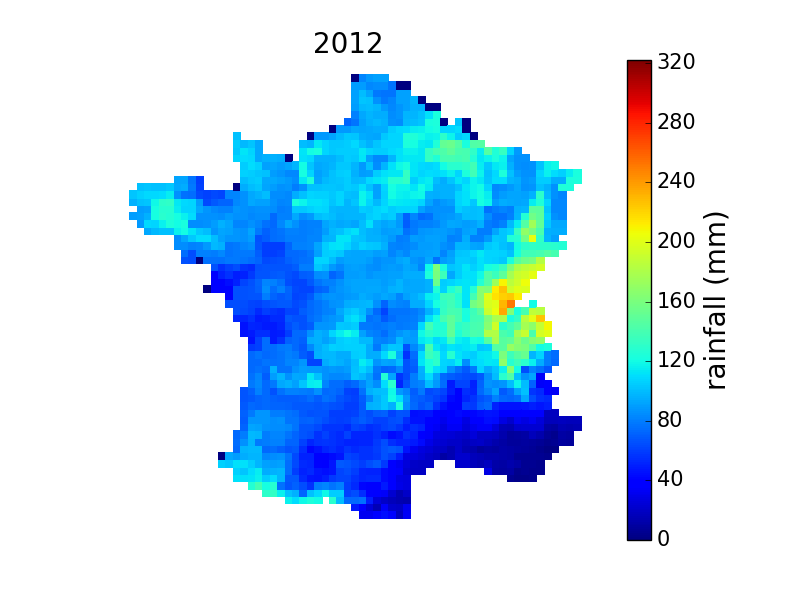}
\caption{2012}
\end{subfigure}
%---------------------
\begin{subfigure}{2.5cm}
\includegraphics[height=2.6cm, trim = 3.2cm 2cm 5.4cm 1.4cm, clip=true]{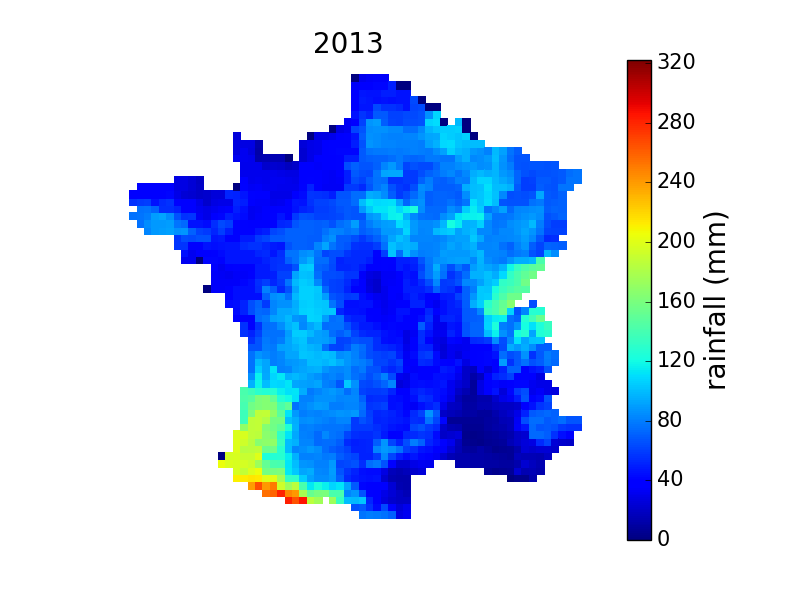}
\caption{2013}
\end{subfigure}
\\
%---------------------
\begin{subfigure}{2.5cm}
\includegraphics[height=2.6cm, trim = 3.2cm 2cm 5.4cm 1.4cm, clip=true]{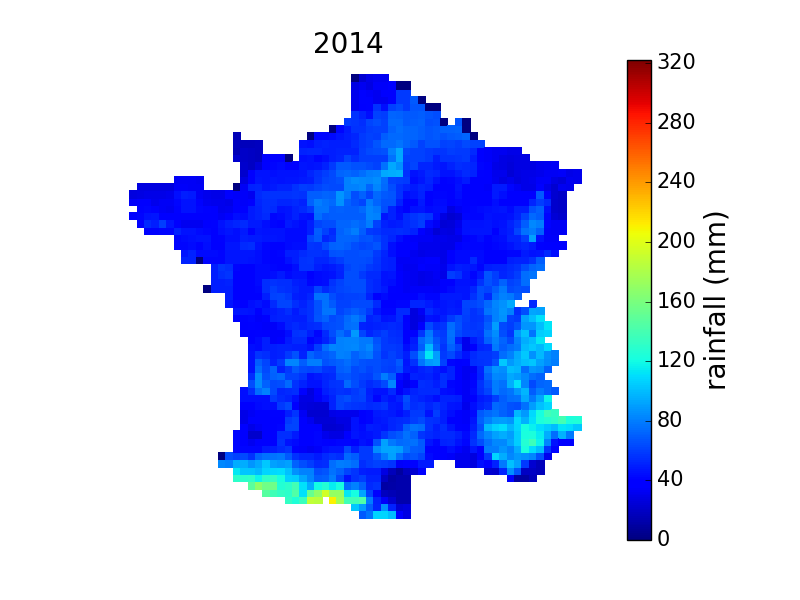}
\caption{2014}
\end{subfigure}
%---------------------
\begin{subfigure}{2.5cm}
\includegraphics[height=2.6cm, trim = 3.2cm 2cm 5.4cm 1.4cm, clip=true]{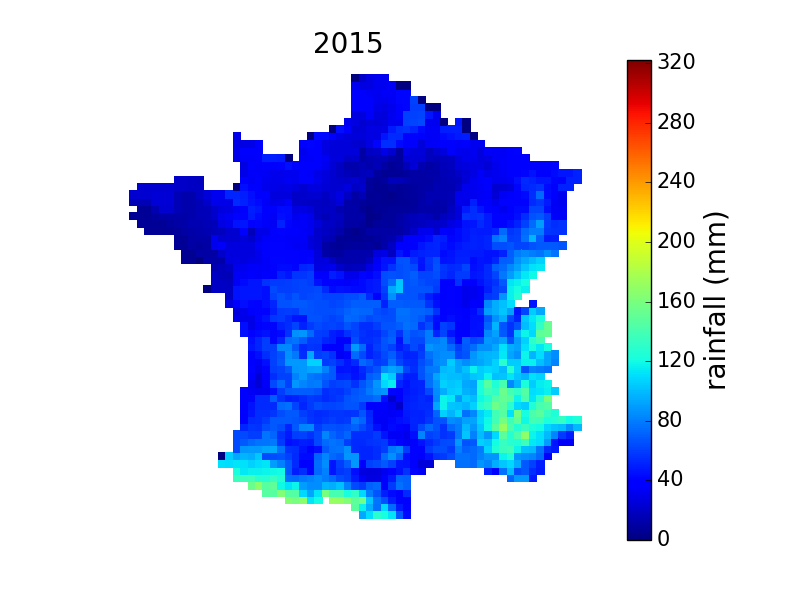}
\caption{2015}
\end{subfigure}
%---------------------
\begin{subfigure}{2.5cm}
\includegraphics[height=2.6cm, trim = 3.2cm 2cm 5.4cm 1.4cm, clip=true]{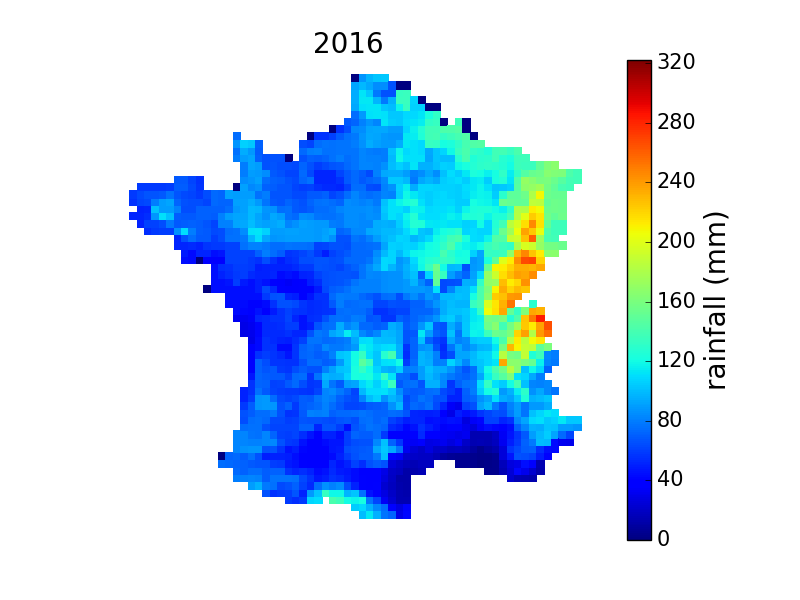}
\caption{2016}
\end{subfigure}
%---------------------
\begin{subfigure}{2.5cm}
\includegraphics[height=2.6cm, trim = 3.2cm 2cm 5.4cm 1.4cm, clip=true]{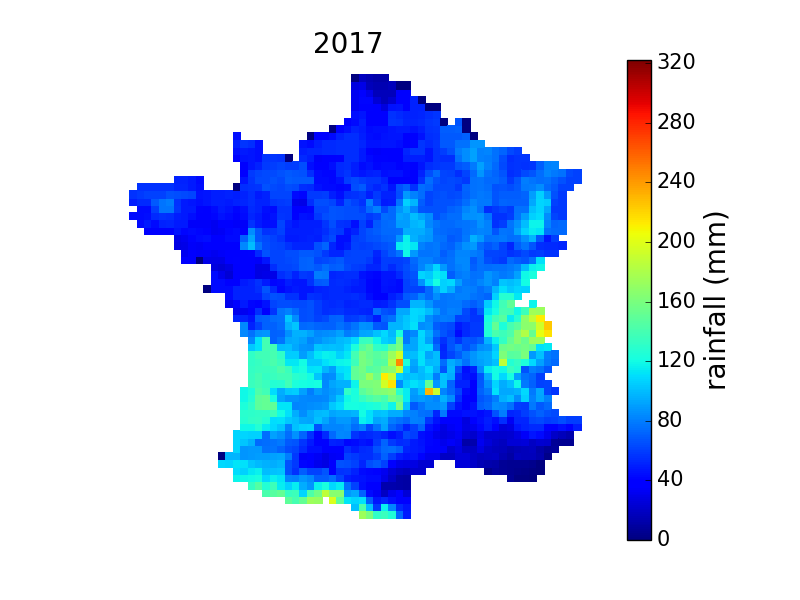}
\caption{2017}
\end{subfigure}
%---------------------
\begin{subfigure}{2.5cm}
\includegraphics[height=2.6cm, trim = 3.2cm 2cm 5.4cm 1.4cm, clip=true]{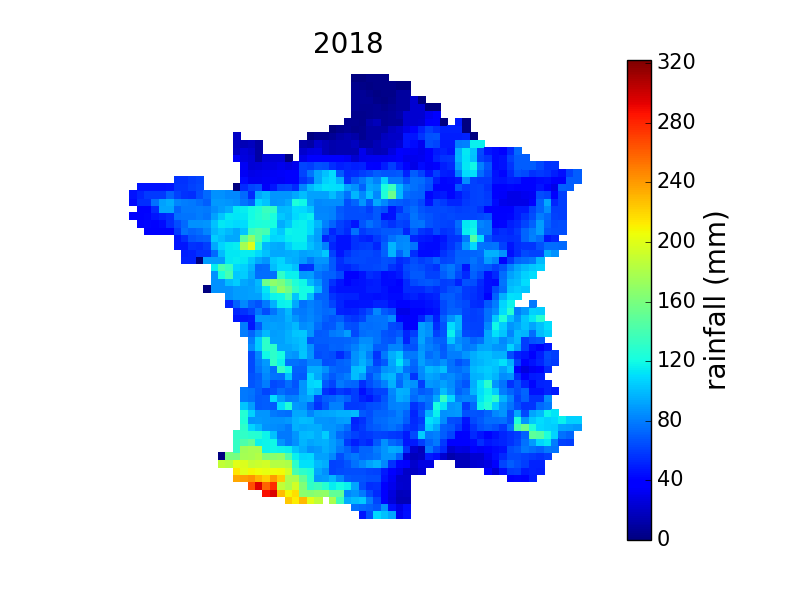}
\caption{2018}
\end{subfigure}
}
&
\makecell{\includegraphics[height=6cm, trim = 15.5cm 1.2cm 1.6cm 1.cm, clip=true, valign=m]{fig_pluviometry2018} \\ \medskip}
\end{tabular}
\end{center}
\caption{\label{fig.rainfall} Rainfall of June (sum of the solid and liquid precipitation) from 2009 to 2018. (Source: Safran data from Météo-France).}
\end{figure}

\begin{figure}
\begin{center}
\begin{tabular}{cc}
\makecell{
%---------------------
\begin{subfigure}{3cm}
\includegraphics[height=3cm, trim = 3.2cm 2cm 5.4cm 1.4cm, clip=true]{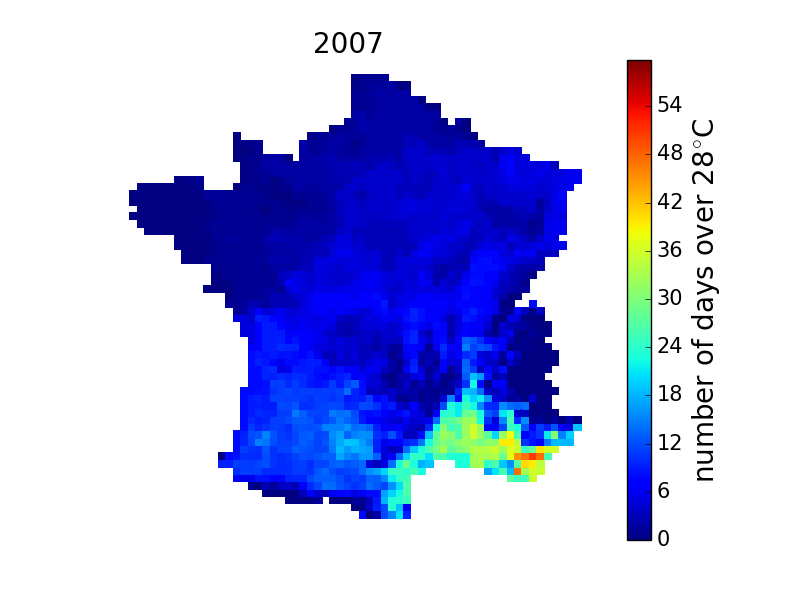}
\caption{2007}
\end{subfigure}
%---------------------
\begin{subfigure}{3cm}
\includegraphics[height=3cm, trim = 3.2cm 2cm 5.4cm 1.4cm, clip=true]{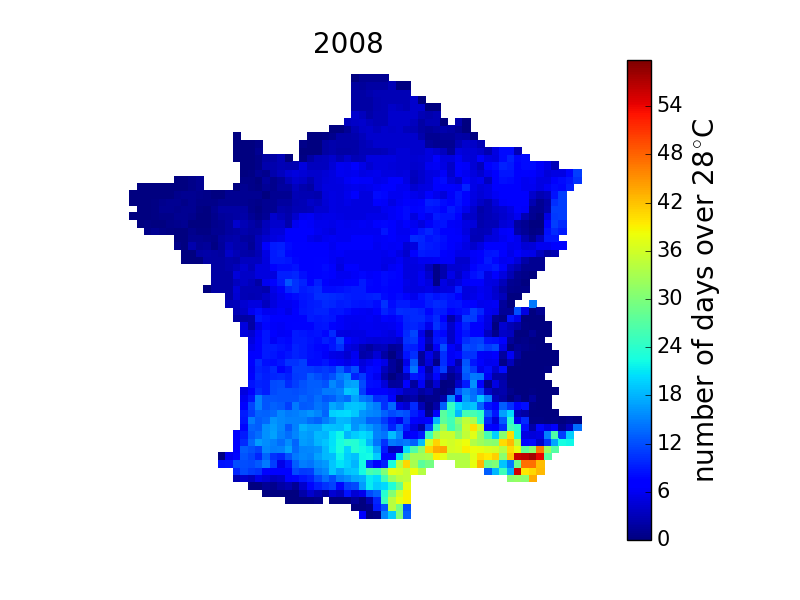}
\caption{2008}
\end{subfigure}
%---------------------
\begin{subfigure}{3cm}
\includegraphics[height=3cm, trim = 3.2cm 2cm 5.4cm 1.4cm, clip=true]{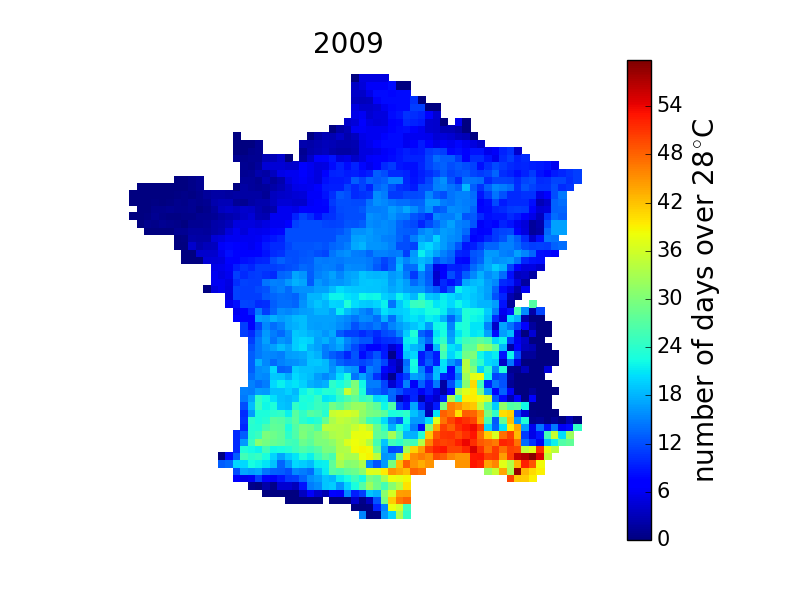}
\caption{2009}
\end{subfigure}
%---------------------
\begin{subfigure}{3cm}
\includegraphics[height=3cm, trim = 3.2cm 2cm 5.4cm 1.4cm, clip=true]{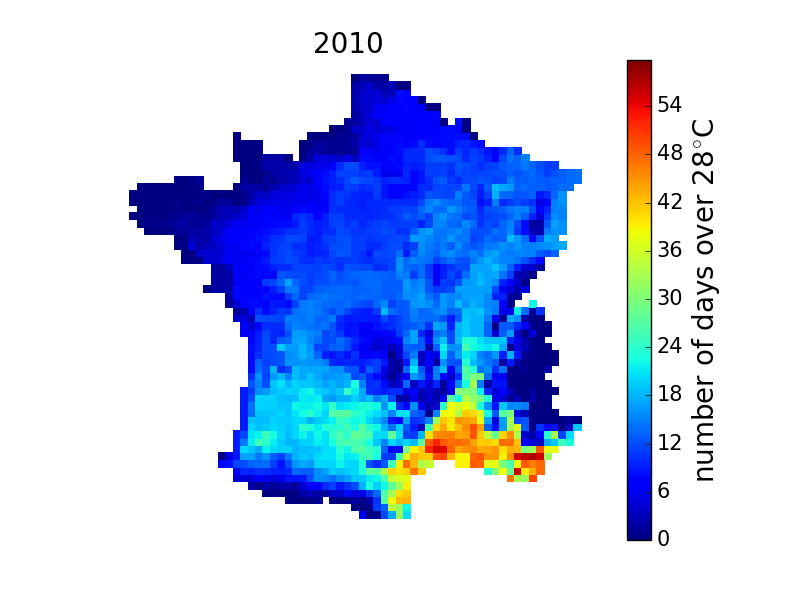}
\caption{2010}
\end{subfigure}
\\
%---------------------
\begin{subfigure}{3cm}
\includegraphics[height=3cm, trim = 3.2cm 2cm 5.4cm 1.4cm, clip=true]{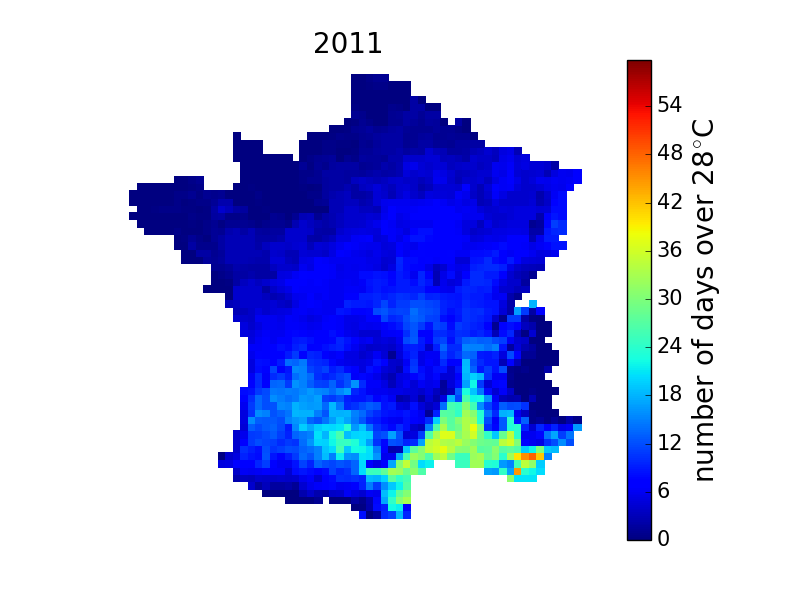}
\caption{2011}
\end{subfigure}
%---------------------
\begin{subfigure}{3cm}
\includegraphics[height=3cm, trim = 3.2cm 2cm 5.4cm 1.4cm, clip=true]{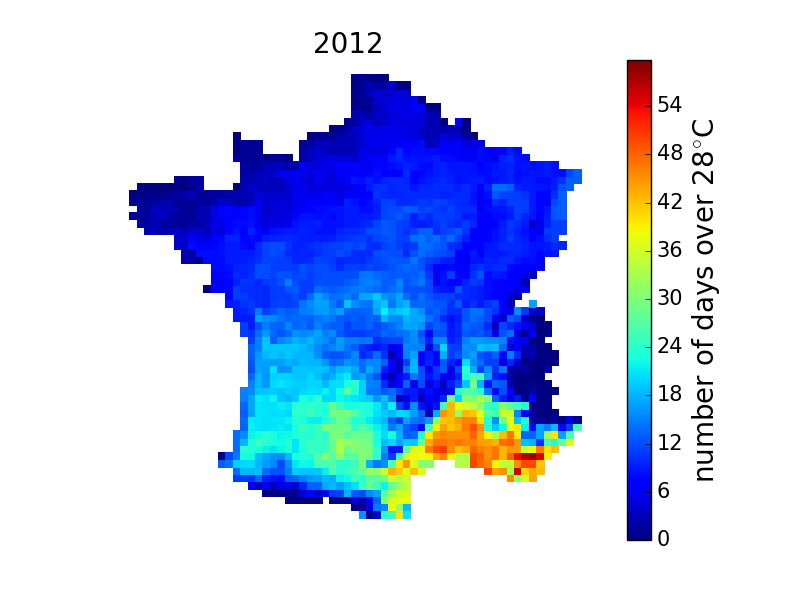}
\caption{2012}
\end{subfigure}
%---------------------
\begin{subfigure}{3cm}
\includegraphics[height=3cm, trim = 3.2cm 2cm 5.4cm 1.4cm, clip=true]{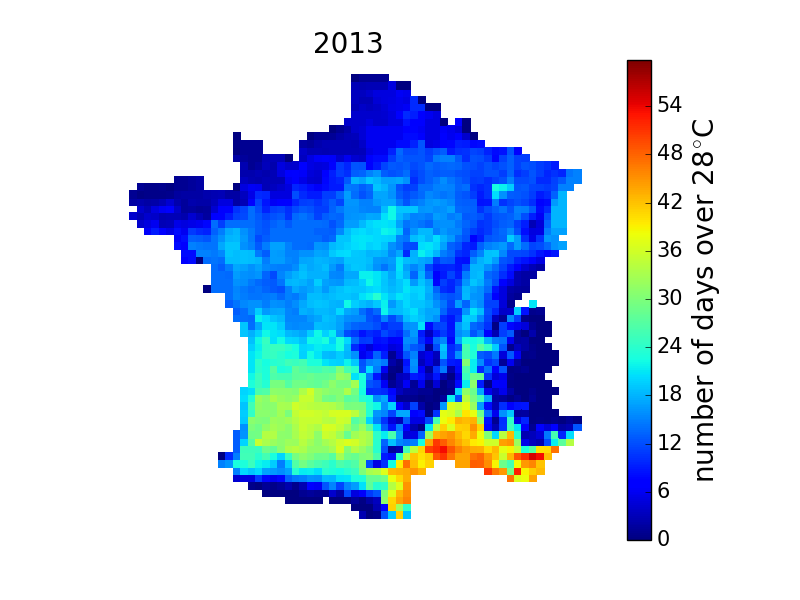}
\caption{2013}
\end{subfigure}
%---------------------
\begin{subfigure}{3cm}
\includegraphics[height=3cm, trim = 3.2cm 2cm 5.4cm 1.4cm, clip=true]{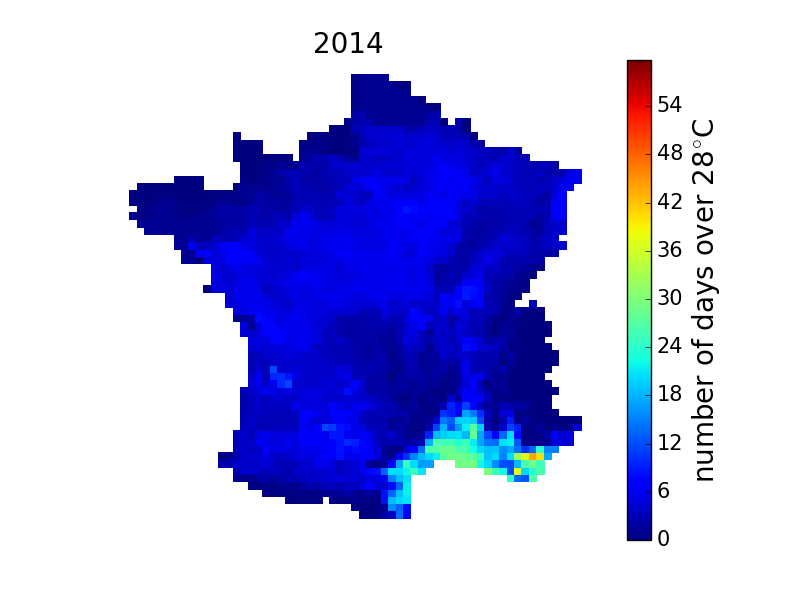}
\caption{2014}
\end{subfigure}
\\
%---------------------
\begin{subfigure}{3cm}
\includegraphics[height=3cm, trim = 3.2cm 2cm 5.4cm 1.4cm, clip=true]{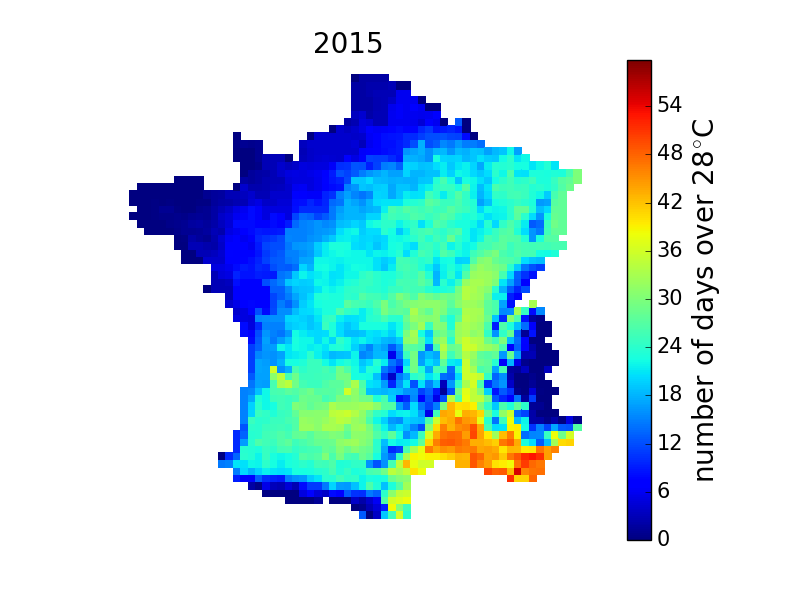}
\caption{2015}
\end{subfigure}
%---------------------
\begin{subfigure}{3cm}
\includegraphics[height=3cm, trim = 3.2cm 2cm 5.4cm 1.4cm, clip=true]{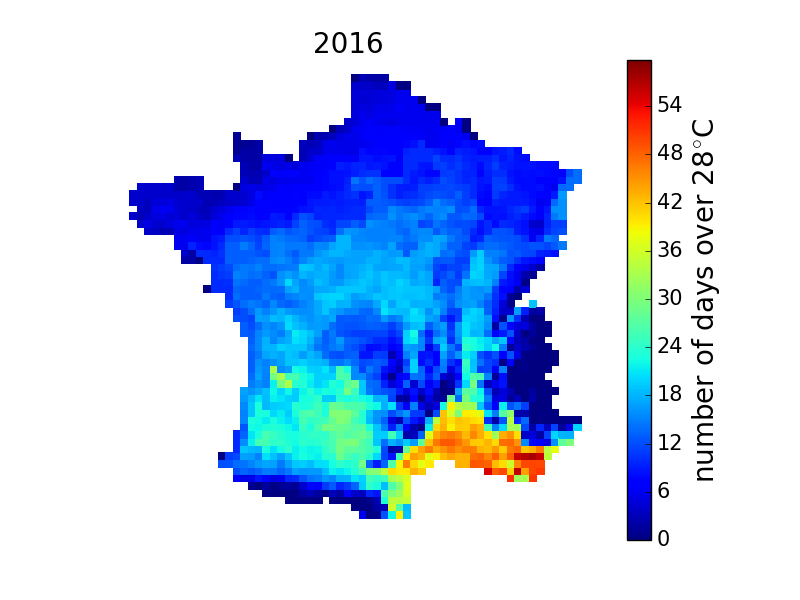}
\caption{2016}
\end{subfigure}
%---------------------
\begin{subfigure}{3cm}
\includegraphics[height=3cm, trim = 3.2cm 2cm 5.4cm 1.4cm, clip=true]{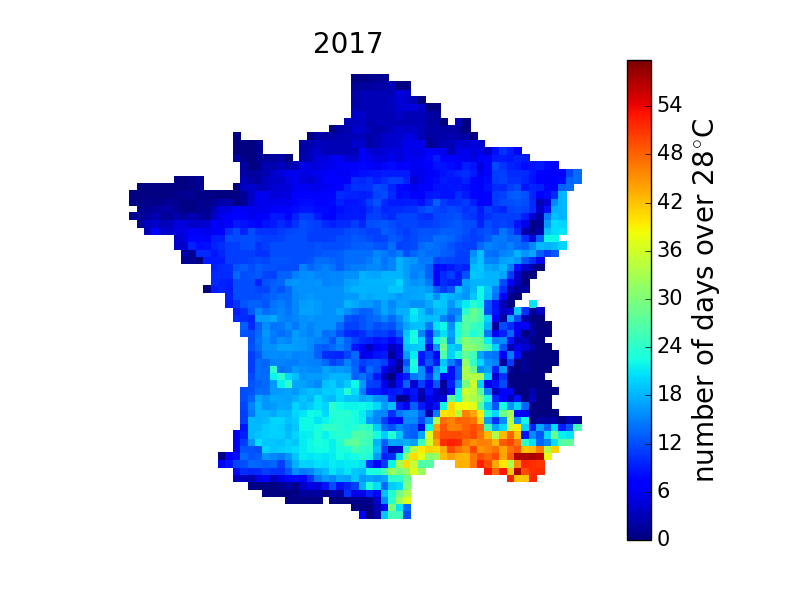}
\caption{2017}
\end{subfigure}
%---------------------
\begin{subfigure}{3cm}
\includegraphics[height=3cm, trim = 3.2cm 2cm 5.4cm 1.4cm, clip=true]{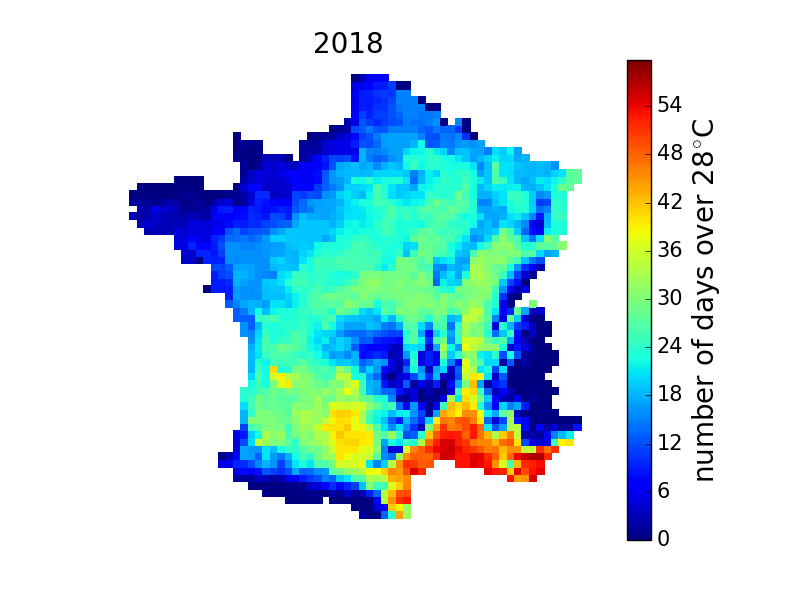}
\caption{2018}
\end{subfigure}
\\
%---------------------
\begin{subfigure}{3cm}
\includegraphics[height=3cm, trim = 3.2cm 2cm 5.4cm 1.4cm, clip=true]{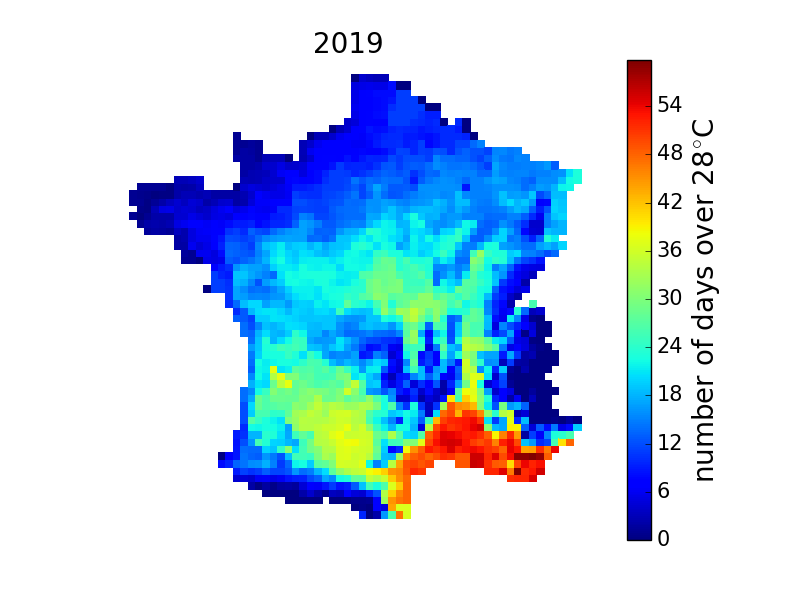}
\caption{2019}
\end{subfigure}
%---------------------
\begin{subfigure}{3cm}
\includegraphics[height=3cm, trim = 3.2cm 2cm 5.4cm 1.4cm, clip=true]{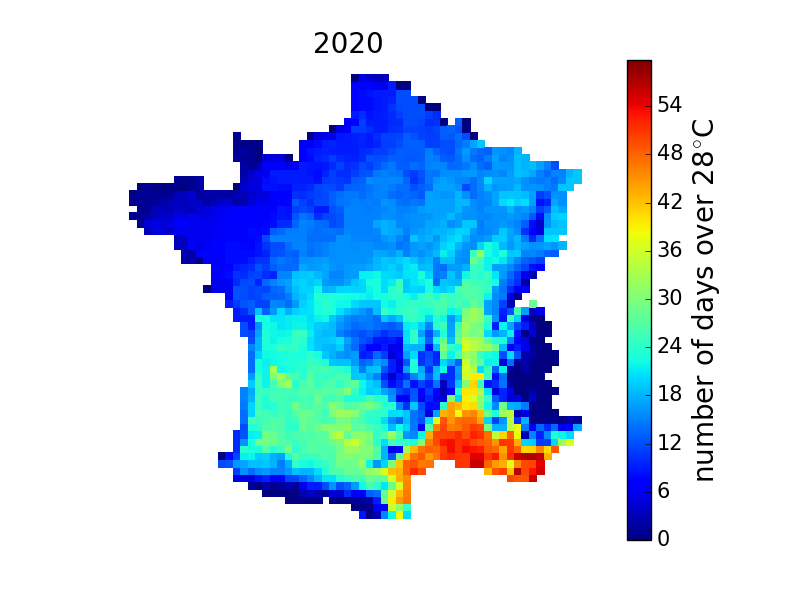}
\caption{2020}
\end{subfigure}
%---------------------
\begin{subfigure}{3cm}
\includegraphics[height=3cm, trim = 3.2cm 2cm 5.4cm 1.4cm, clip=true]{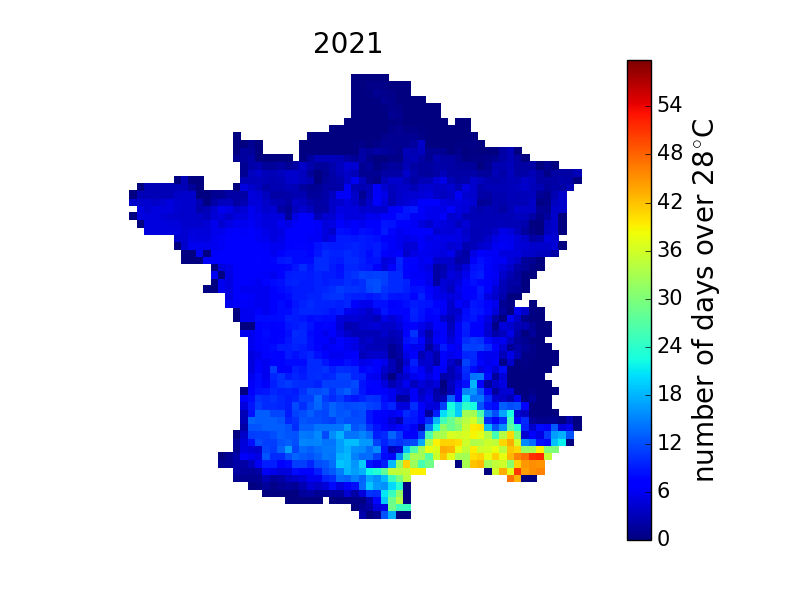}
\caption{2021}
\end{subfigure}
%---------------------
\begin{subfigure}{3cm}
\includegraphics[height=3cm, trim = 3.2cm 2cm 5.4cm 1.4cm, clip=true]{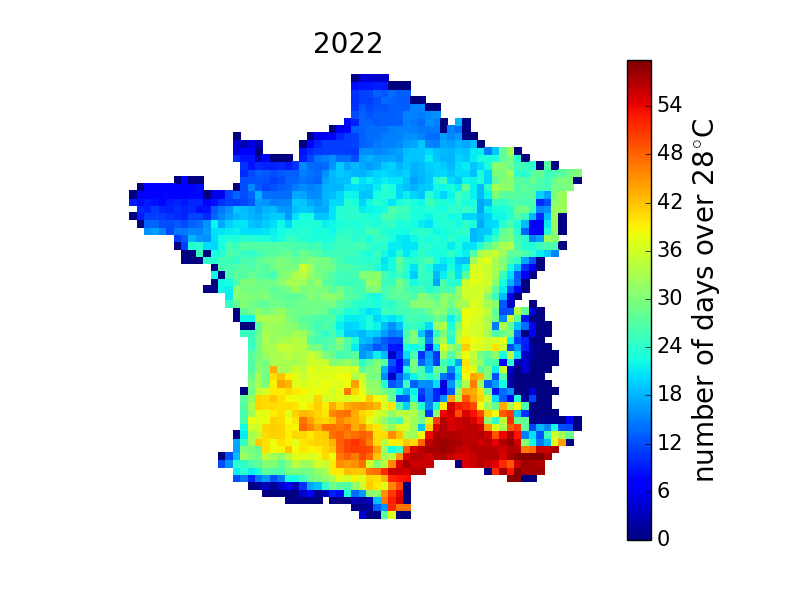}
\caption{2022}
\end{subfigure}
}
&
\includegraphics[height=6cm, trim = 15.5cm 1.2cm 1.6cm 1.cm, clip=true]{fig_temperature_NjS28_240126_2022}
\end{tabular}
\end{center}
\caption{\label{fig.NjS28} Number of days with maximal temperature over 28$^\circ$C during the summer (July and August) in France from 2009 to 2022. (Source: Safran data from Météo-France).}
\end{figure}

This section describes different data used for the inference. We give first the variable of interest that is the proportion of infected trees from 2008 to 2023 (Figure~\ref{fig.observations}). Density of ashes in France is given in Figure~\ref{fig.ash.density}. The density values are derived from the IGN data from 2006 to 2015 (https://inventaire-forestier.ign.fr/dataifn/).
The meteorological data used are Safran data from Météo-France. Safran data are computed on a 8 x 8 km grid over France \citep{quintana-segui_analysis_2008}. We computed relevant meteorological variables and then averaged them over the 16 x 16 km quadrat. Each data is then a mean of four (occasionally one, two or three, on the border of France) meteorological stations located on the quadrat.
Figures~\ref{fig.rainfall} and \ref{fig.NjS28} describe the June rainfall and the number of days with temperature above 28$^{\circ}$C in July and August, respectively.

\section*{Acknowledgments}

We acknowledge Matthieu Brachet for our discussions about the numerical scheme of the reaction diffusion model. We also thanks Claude Husson, from the Département de la Santé des Forêts (DSF) for input on factors to take into account in the model and for help to revise the manuscript. This work was supported by grants from the DSF, French Ministry of Agriculture and Forestry and from ANSES. The UMR1136 research unit is supported by a grant managed by the French National Research Agency (ANR) as part of the ``Investissements d’Avenir'' program (ANR-11-LABX-0002-01, Laboratory of Excellence ARBRE).

\bibliographystyle{apalike}                 
\bibliography{bibliographie}

\end{document}